\documentclass[10pt,preprintnumbers,amsmath,amssymb,nofootinbib,superscriptaddress, floatfix, twocolumn]{revtex4-2}

\usepackage[english]{babel}

\usepackage[utf8]{inputenc}
\usepackage[T1]{fontenc}
\usepackage{ucs}
\usepackage{caption}
\usepackage{microtype}
\usepackage{newtxtext,newtxmath}
\usepackage{color}
\usepackage{lettrine}
\usepackage{gensymb}
\usepackage{tablefootnote} 
\usepackage{threeparttable}

\usepackage{url}
\usepackage{hyperref}

\usepackage{xspace}
\usepackage{lastpage}
\usepackage{stmaryrd}

\usepackage{graphicx}
\usepackage{empheq}
\usepackage{ifthen}
\usepackage{tensor}
\usepackage{paralist}
\usepackage{float}

\usepackage{slashed}
\usepackage{amsmath}
\usepackage{amsfonts}

\usepackage{amssymb}
\usepackage{mathrsfs}
\usepackage{extarrows}
\usepackage{verbatim}
\usepackage{upgreek}
\usepackage{wasysym}
\usepackage{multirow}

\usepackage{natbib}
\usepackage{makecell}
\usepackage{tabularx}

\newcommand{\tablefoot}[1]{\par\smallskip\noindent\textit{#1}\par}

\def \be {\begin{equation}}
\def \ee {\end{equation}}

\newcommand{\diff}{\mathrm{d}}
\newcommand{\lcdm}{$\mathrm{\Lambda CDM}$\xspace}
\newcommand{\Hc}{\ensuremath{H_0}\xspace}

\newcommand{\Ddt}{\ensuremath{D_{\Delta t}}\xspace}
\newcommand{\Om}{\ensuremath{\Omega_{\mathrm m}}\xspace}

\newcommand{\Ok}{\ensuremath{\Omega_{k}}\xspace}
\newcommand{\kext}{\ensuremath{\kappa_{\rm ext}}\xspace}

\newcommand{\lint}{\lambda_{\rm int}}

\newcommand{\Bsixteen}{B1608$+$656\xspace}
\newcommand{\RXJ}{RX\,J1131$-$1231\xspace}
\newcommand{\HEzero}{HE\,0435$-$1223\xspace}
\newcommand{\Jtwelve}{SDSS\,J1206$+$4332\xspace}
\newcommand{\WFItwenty}{WFI\,2033$-$4723\xspace}
\newcommand{\PGeleven}{PG\,1115$+$080\xspace}
\newcommand{\DESzerofour}{DES\,J0408$-$5354\xspace}
\newcommand{\WGDtwenty}{WGD\,2038$-$4008\xspace}
\newcommand{\HEeleven}{HE\,1104$-$1805\xspace}

\newcommand{\Jfourteen}{SDSS\,J1433$+$6007\xspace}
\newcommand{\Jfifteen}{J1537$-$3010\xspace}

\newcommand{\nslacsifu}{14} 
\newcommand{\nslacsifusel}{12}
\newcommand{\nslacstot}{15}
\newcommand{\nsltwossel}{8}

\def\ksmpc{${\, \mathrm{km}\, \mathrm{s}^{-1}\, \mathrm{Mpc}^{-1}}$\xspace}
\def\ks{${\, \mathrm{km}\, \mathrm{s}^{-1}}$\xspace}
\newcommand*\arcsec{\ensuremath{^{\prime\prime}}}

\newcommand{\sref}[1]{Section~\ref{#1}}
\newcommand{\ssref}[1]{Sections~\ref{#1}}
\newcommand{\aref}[1]{Appendix~\ref{#1}}
\newcommand{\fref}[1]{Fig.~\ref{#1}}
\newcommand{\fsref}[1]{Figs.~\ref{#1}}
\newcommand{\tref}[1]{Table~\ref{#1}}

\newcommand{\eref}[1]{Eq.~(\ref{#1})}

\newcommand{\hnotfiducial}{71.1^{+2.4}_{-2.3}}
\newcommand{\hnotLCDMoneb}{71.7^{+2.5}_{-2.3}}
\newcommand{\hnotLCDMfoura}{71.3^{+2.2}_{-2.1}}
\newcommand{\hnotLCDMfourb}{72.2^{+2.2}_{-2.2}}
\newcommand{\hnotULCDMone}{70.6^{+3.9}_{-3.1}}

\newcommand{\lambdaintfiducial}{0.94^{+0.05}_{-0.05}}
\newcommand{\flexshift}{\ensuremath{\Delta_3x}\xspace}

\newcommand{\glee}{\textsc{glee}\xspace}
\newcommand{\lenstro}{\textsc{lenstronomy}\xspace}
\newcommand{\jampy}{\textsc{jampy}\xspace}

\defcitealias{GalanSchweinfurth:2026}{Galan \& Schweinfurth et al.}
\newcommand{\TDCprev}{{TDCOSMO-2025}\xspace}
\newcommand{\TDCnew}{{TDCOSMO-2026}\xspace}
\newcommand{\TDCnewletter}{{TDCOSMO-2026-PRL}\xspace}


\makeatletter
\g@addto@macro\bfseries{\boldmath}
\makeatother

\begin{document}


\title{TDCOSMO 2026: A 3\% measurement of the Hubble constant from strong lensing and stellar kinematics}
  
\author{TDCOSMO Collaboration}
\noaffiliation

\author{S.~Birrer}
\affiliation{Department of Physics and Astronomy, Stony Brook University, Stony Brook, NY 11794, USA}

\author{M.~Cappellari}
\affiliation{Sub-Department of Astrophysics, Department of Physics, University of Oxford, Denys Wilkinson Building, Keble Road, Oxford OX1 3RH, UK}

\author{F.~Courbin}
\affiliation{ICC-UB Institut de Ci\`encies del Cosmos, Universitat de Barcelona, Mart\'i Franqu\`es, 1, 08028 Barcelona, Spain}
\affiliation{Institut Catal\`a de Recerca i Estudis Avan\c{c}ats (ICREA), Pg. Llu\'is Companys 23, 08010 Barcelona, Spain}
\affiliation{Institut d'Estudis Espacials de Catalunya (IEEC), Edifici RDIT, Campus UPC, Castelldefels, 08860 Barcelona, Spain}

\author{C.~D.~Fassnacht}
\affiliation{Department of Physics and Astronomy, University of California, Davis, 1 Shields Ave., Davis, CA 95616, USA}

\author{P.~Fleury}
\affiliation{Laboratoire Univers et Particules de Montpellier (LUPM), CNRS \& Universit\'{e} de Montpellier (UMR-5299), Parvis Alexander Grothendieck, F-34095 Montpellier Cedex 05, France}

\author{J.~A.~Frieman}
\affiliation{SLAC National Accelerator Laboratory, Menlo Park, CA 94025, USA}

\author{A.~Galan}
\email[Corresponding author: ]{aymeric.galan@gmail.com}
\affiliation{Department of Astronomy, University of Geneva, ch. d'Ecogia 16, 1290 Versoix, Switzerland}

\author{X.-Y.~Huang}
\affiliation{Department of Physics and Astronomy, Stony Brook University, Stony Brook, NY 11794, USA}

\author{D.~P.~Johnson}
\affiliation{Laboratoire Univers et Particules de Montpellier (LUPM), CNRS \& Universit\'{e} de Montpellier (UMR-5299), Parvis Alexander Grothendieck, F-34095 Montpellier Cedex 05, France}

\author{S.~Khanolkar}
\affiliation{Department of Physics and Astronomy, University of California, Los Angeles, CA 90095, USA}

\author{S.~Knabel}
\affiliation{Department of Physics and Astronomy, University of California, Los Angeles, CA 90095, USA}

\author{H.~Krizic}
\affiliation{D\'epartement de Physique Th\'eorique, Universit\'e de Gen\`eve, 24 quai Ernest-Ansermet, CH-1211 Gen\`eve 4, Switzerland}

\author{M.~Millon}
\email[Corresponding author: ]{martin.millon@unige.ch}
\affiliation{D\'epartement de Physique Th\'eorique, Universit\'e de Gen\`eve, 24 quai Ernest-Ansermet, CH-1211 Gen\`eve 4, Switzerland}

\author{T.~Morishita}
\affiliation{Astronomical Institute, Graduate School of Science, Tohoku University, 6–3 Aoba, Sendai 980-8578, Japan}

\author{V.~Motta}
\affiliation{Instituto de F\'{\i}sica y Astronom\'{\i}a, Universidad de Valpara\'{\i}so, Avda. Gran Breta\~na 1111, Valpara\'{\i}so, Chile}

\author{P.~Mozumdar}
\email[Corresponding author: ]{pmozumdar@astro.ucla.edu}
\affiliation{Department of Physics and Astronomy, University of California, Los Angeles, CA 90095, USA}

\author{E.~Paic}
\affiliation{Research Center for the Early Universe, Graduate School of Science, The University of Tokyo, 7-3-1 Hongo, Bunkyo-ku, Tokyo 113-0033, Japan}

\author{H.~Paugnat}
\affiliation{Department of Physics and Astronomy, University of California, Los Angeles, CA 90095, USA}

\author{A.~G.~Schweinfurth}
\affiliation{Technical University of Munich, TUM School of Natural Sciences, Department of Physics, James-Franck-Str.~1, 85748 Garching, Germany}
\affiliation{Max-Planck-Institut für Astrophysik, Karl-Schwarzschild-Str.~1, 85748 Garching, Germany}

\author{A.~J.~Shajib}
\affiliation{Department of Astronomy \& Astrophysics, University of Chicago, Chicago, IL 60637, USA}
\affiliation{Kavli Institute for Cosmological Physics, University of Chicago, Chicago, IL 60637, USA}
\affiliation{Center for Astronomy, Space Science and Astrophysics, Independent University, Bangladesh, Dhaka 1229, Bangladesh}

\author{W.~Sheu}
\affiliation{Department of Physics and Astronomy, University of California, Los Angeles, CA 90095, USA}

\author{D.~Sluse}
\affiliation{STAR Institute, University of Li{\`e}ge, Quartier Agora, All\'ee du six Ao\^ut 19c, 4000 Li\`ege, Belgium}

\author{M.~Stiavelli}
\affiliation{Space Telescope Science Institute, 3700 San Martin Dr., Baltimore, MD 21218, USA}

\author{S.~H.~Suyu} 
\affiliation{Technical University of Munich, TUM School of Natural Sciences, Department of Physics, James-Franck-Str.~1, 85748 Garching, Germany}
\affiliation{Max-Planck-Institut für Astrophysik, Karl-Schwarzschild-Str.~1, 85748 Garching, Germany}

\author{T.~Treu}
\affiliation{Department of Physics and Astronomy, University of California, Los Angeles, CA 90095, USA}

\author{F.~Urcelay}
\affiliation{Sub-Department of Astrophysics, Department of Physics, University of Oxford, Denys Wilkinson Building, Keble Road, Oxford OX1 3RH, UK}

\author{D.~M.~Williams}
\affiliation{Department of Physics and Astronomy, University of California, Los Angeles, CA 90095, USA}

\author{K.~C.~Wong}
\affiliation{Research Center for the Early Universe, Graduate School of Science, The University of Tokyo, 7-3-1 Hongo, Bunkyo-ku, Tokyo 113-0033, Japan}

\begin{abstract}
We present cosmological constraints from 11 strongly lensed quasars with measured time delays (hereafter, the TDCOSMO-2026 sample). In addition to expanding the sample, we improve many aspects of our blind analysis with respect to our 2025 analysis. Among other improvements, spatially resolved kinematics are available for nine systems, including six with JWST-NIRSpec, compared to one in 2025; for three lens models, JWST-NIRCam images now replace HST; axisymmetric dynamical modeling is applied directly to the spatially resolved data, instead of an `axisymmetric correction factor' applied to spherical models; and the line-of-sight convergence is now based on the Euclid Flagship simulations instead of the Millennium Simulation, with an improved inference methodology.

Our baseline and primary result is based on the TDCOSMO-2026 sample, in combination with a prior on $\Omega_{\rm m}$ from the DES-Dovekie supernova sample. In the flat $\Lambda$ cold dark matter ($\Lambda$CDM) cosmology, we find $H_0=\hnotfiducial$ \ksmpc, a 3.3\% uncertainty that accounts for the mass-sheet degeneracy. The SLACS and SL2S samples are in excellent agreement with the TDCOSMO-2026 sample, improving the precision on $H_0$ in flat $\Lambda$CDM to 3.0\%. Using the DESI-DR2 baryon acoustic oscillations (BAO) likelihood instead of DES-Dovekie yields fully consistent results.
We also present constraints in the open $\Lambda$CDM, $w$CDM, $w_0w_a$CDM, and $w_{\phi}$CDM cosmologies; the TDCOSMO $H_0$ inference is robust and consistent across all of them. The combination of the TDCOSMO+SLACS+SL2S likelihood with those from DESI-DR2 and DES-Dovekie disfavors $w=-1$ at the $2.3$--$2.5\sigma$ level in all three dark energy extensions that we tested. We find, e.g., $w_0=-0.88^{+0.04}_{-0.05}$ in $w_{\phi}$CDM and $w_0 = -0.85^{+0.07}_{-0.06}$ in $w_0w_a$CDM, while $w_a = -0.48^{+0.46}_{-0.44}$ remains consistent with zero.
\end{abstract}

\maketitle

\section{Introduction}

Measurements of the current expansion rate of the Universe, known as the Hubble constant ($H_0$), have increased steadily in precision over the past two decades. The most recent measurement by the H0DN Collaboration is $H_0 = 73.50 \pm 0.81$ \ksmpc \citep{H0DN:2026}. Most other measurements in the local universe agree, albeit with lower precision \citep{cosmoverse25}, with some notable exceptions \citep{Freedman24}.  

Alternatively, $H_0$ can be predicted by early-universe measurements, such as the cosmic microwave background, by assuming a cosmological model. Within a flat $\Lambda$ cold dark matter universe, \citet{Planck2020} yields $H_0 = 67.4 \pm 0.5$ \ksmpc. Most early-universe probes yield values similar to those obtained by the CMB \citep{cosmoverse25}. 

The difference between early and late universe measurements of $H_0$ is known as the Hubble tension. If the tension is real and not due to unknown biases in multiple measurements, it implies that flat $\Lambda$CDM is not the correct cosmological model. Solutions include early dark energy or new relativistic particles beyond the standard model. However, there is no natural way out from a theoretical standpoint  \citep[e.g.,][]{KnoxMillea2020,Vagnozzi2023,cosmoverse25,Turner26,Schonberg:2022,Schonberg:2026a,Schonberg:2026b}. 

Measurements of $H_0$ based on independent methods are crucial in solving the Hubble tension, provided that they can achieve a precision sufficient to distinguish the 8-9\% difference between those based on CMB and H0DN. As far as late universe probes are concerned, strong gravitational lensing time delays \citep{Refsdal1964,TreuMarshall2016,TSM22,TreuShajib_review,Birrer:2024} offer several advantages compared to traditional methods: i) they provide a one-step absolute measurement of distance; ii) they are primarily sensitive to $H_0$, with weak dependency on other cosmological parameters or the assumed background cosmology \citep{TDCOSMO2025}; iii) they are insensitive to dust\footnote{except when a standard candle is used to break the mass-sheet degeneracy as discussed below.} or photometric calibration; iv) they extend over sufficiently large distances to be insensitive to peculiar motions and local inhomogeneities \citep{Dalang2023}.

Time-delay cosmography has come a long way since it was initially proposed over 60 years ago \citep{Refsdal1964}. Hundreds of lensed quasars \citep[e.g.][]{Lemon:2023,Schmidt:2023,Napier:2023} and a handful of lensed supernovae have been discovered \citep[e.g.,][]{Kelly15,Pascale25}. Time delays have been measured with a few percent precision, using dedicated monitoring campaigns \citep[e.g.,][]{Fassnacht2002,Courbin2005,Courbin2018, Glendama2018,Millon:2020b,Millon:2020c,Dux:2025,sheu2026,Pierel:2024,Pierel:2026,Kelly2023b,Rodney:2016}. Lens modelers have taken advantage of high-resolution imaging from space- and ground-based observations to constrain the gravitational potential using the wealth of information contained in the lensed images of the host galaxy of the quasars or supernovae \citep[e.g.,][]{Suyu2010,Chen2019,Birrer2019,Shajib2019,TDCOSMO20,Grillo24}. Numerical simulations have been used to quantify the effects of over- and under-densities along the line of sight \citep[LoS;][]{Greene2013,Rusu2017,Wells2023,Wells2024}. Stellar kinematics \citep{TreuKoopmans:2002,Shajib2018,yildirim20,Shajib2023,sheu2026} and standard candles \citep{Pascale25} have been used to break residual degeneracies between models and inferred distances. Blind analysis to prevent experimenter bias has become the gold standard in the field: the analysis is kept blind to cosmological parameters until after internal review, and they are published without any changes after a formal unblinding process \citep{Suyu2013,Bonvin2017,Birrer2019,Chen2019,Rusu2020,Wong2019,Shajib2018,Wong24,Paic2026,sheu2026}.

The TDCOSMO collaboration was formed in 2020 with the goal of obtaining precise and accurate cosmological models based on time delays, building on the work of the COSMOGRAIL \citep{Courbin2005}, H0LiCOW \citep{Suyu2017}, SHARP \citep{Chen2019}, and STRIDES~\citep{Treu2018} collaborations. Since its inception, a main focus of the TDCOSMO collaboration has been to identify, evaluate, and mitigate systematic errors \citep[e.g.,][]{Millon:2020,TDCOSMO3,TDCOSMO7,TDCOSMO8,Huang2026}. 

The primary source of potential residual systematic error in time-delay cosmography is the so-called mass-sheet degeneracy \citep[MSD;][]{Falco1985,Schneider2013,Schneider2014}. Any deflector surface mass distribution that fits the observed multiple image positions can be transformed into a family of surface mass distributions that provide an equally good fit but with rescaled absolute distances, source position, and luminosity \citep{Treu10}. The simplest such transformation consists of adding an infinite sheet of mass and rescaling the surface mass distribution; hence the name. The degeneracy can be broken in a number of ways. \citet{Wong2019,Shajib2019,Millon:2020} broke it by assuming functional forms for the mass density profiles of the deflectors that go to zero at large radii, obtaining $H_0=74.2\pm1.6$\ksmpc, in agreement with the H0DN value and a 2\% uncertainty. \citet[][hereafter TDC4]{TDCOSMOIV} took an alternative approach. They explicitly accounted for the MSD by introducing a new parameter, $\lint$, that rescales the mass density profile, leaving the lensing configuration unaffected while changing $H_0$. This approach effectively maximizes the uncertainty on $H_0$, since the MSD is constrained exclusively by stellar kinematic data. For the same sample of 7 lensed quasars analyzed by \citet{Millon:2020}, the uncertainty increased from 2\% to 8\% as a result, yielding $H_0=74.5_{-6.1}^{+5.6}$\ksmpc. TDC4 also proposed to use samples of non-time-delay lenses to help constrain $\lint$ at a population level in a hierarchical analysis. Adding 33 galaxy--galaxy lenses from the external SLACS sample~\citep{Bolton08} to the time-delay lenses, they reduced the uncertainties from 8\% to 5\%, with a non-statistically significant shift to a lower value of $H_0 = 67.4_{-3.2}^{+4.1}$\ksmpc. However, the TDC4 SLACS stellar velocity dispersions were based on low signal-to-noise ratio (S/N) spectra from SDSS, which was later understood to result in underestimated uncertainties for the external sample~\citep{Knabel24}.

The TDCOSMO-2025 milestone paper \citep[][hereafter \TDCprev{}]{TDCOSMO2025} combined several improvements with respect to TDC4. First, the time-delay lens \WGDtwenty \citep{Wong24} was added, bringing the total number in the \TDCprev sample to eight. Second, the analysis used improved methods to reduce uncertainties and covariance in stellar kinematics measurements \citep{Knabel25}. Third, the analysis used JWST-NIRSpec \citep{Shajib25b} and Keck Cosmic Web Imager \citep[KCWI;][]{Shajib23} integral field spectroscopy to derive radially resolved kinematics of the time-delay lens \RXJ; JWST-NIRSpec spectra for integrated stellar velocity dispersion of 5 time-delay lenses; VLT-MUSE spectra for integrated stellar velocity dispersion of the remaining 2 time-delay lenses. Fourth, constant anisotropy replaced the previously adopted Osipkov--Merritt \citep[][hereafter OM]{Osipkov1979,Merritt1985} parameterization; this has been shown to be a better description of real systems \citep[Section~3.3]{Cappellari2025} and to provide unbiased $H_0$ when applied to simulated data \citep{V+M26}. Fifth, the SLACS external sample was based on a reanalysis of the HST images \citep{Tan24}; SDSS spectra were replaced by Keck-KCWI spectra of much higher S/N, providing radially resolved kinematic maps. Sixth, galaxy--galaxy lenses from Strong Lenses in the Legacy Survey (SL2S) sample \citep{Ruff11,Gavazzi12,Sonnenfeld13a,Sonnenfeld13b,Sheu25} were used as an additional external sample, based on remeasured stellar kinematics and updated lens models \citep{Mozumdar2025,Sheu25}. Rigorous cuts were applied to ensure that the SLACS and SL2S sample properties are well matched to those of the \TDCprev sample \citep{Mozumdar2025}. Seventh, corrections for projection effects due to departures from spherical symmetry were applied, derived from axisymmetric Jeans Anisotropic Modeling (JAM), to the spherical dynamical model predictions \citep{Huang2026}. 
Eighth, an improved treatment of the line-of-sight convergence \citep{Wells2024} was applied to the \TDCprev and external lens samples. Ninth, the surface brightness profiles of the deflector were remeasured and a multi-Gaussian expansion decomposition replaced the Hernquist models adopted in TDC4. The blind baseline result of \TDCprev was $H_0=71.6^{+3.9}_{-3.3}$\ksmpc based on 8 time-delay lenses in combination with $\Omega_{\rm m}$ constraints from the Pantheon+ Type Ia supernova dataset \citep{Scolnic2022}. The blind measurement in combination with non-time-delay lenses from the SLACS and SL2S samples was in excellent agreement $H_0=74.3^{+3.1}_{-3.7}$\ksmpc. TDCOSMO improved the precision from 8\% in TDC4 to 5\%, while resolving the (non-significant) tension between the time-delay and non-time-delay lens samples reported by TDC4.

In this 2026 milestone paper, we present an updated cosmological inference based on several significant improvements, which are briefly summarized here and described in detail in the main body. We describe the full analysis and derive \Hc and other cosmological parameters in a variety of cosmologies. A companion letter 
\citep[hereafter \TDCnewletter; ][]{TCOSMO2026_letter} focuses on the determination of \Hc in flat \lcdm and its interpretation in the context of the Hubble tension.

A summary of the sources of new information is given in Table~\ref{tab:lens_overview}. 

\begin{figure*}
    \centering
    \includegraphics[width=\textwidth]{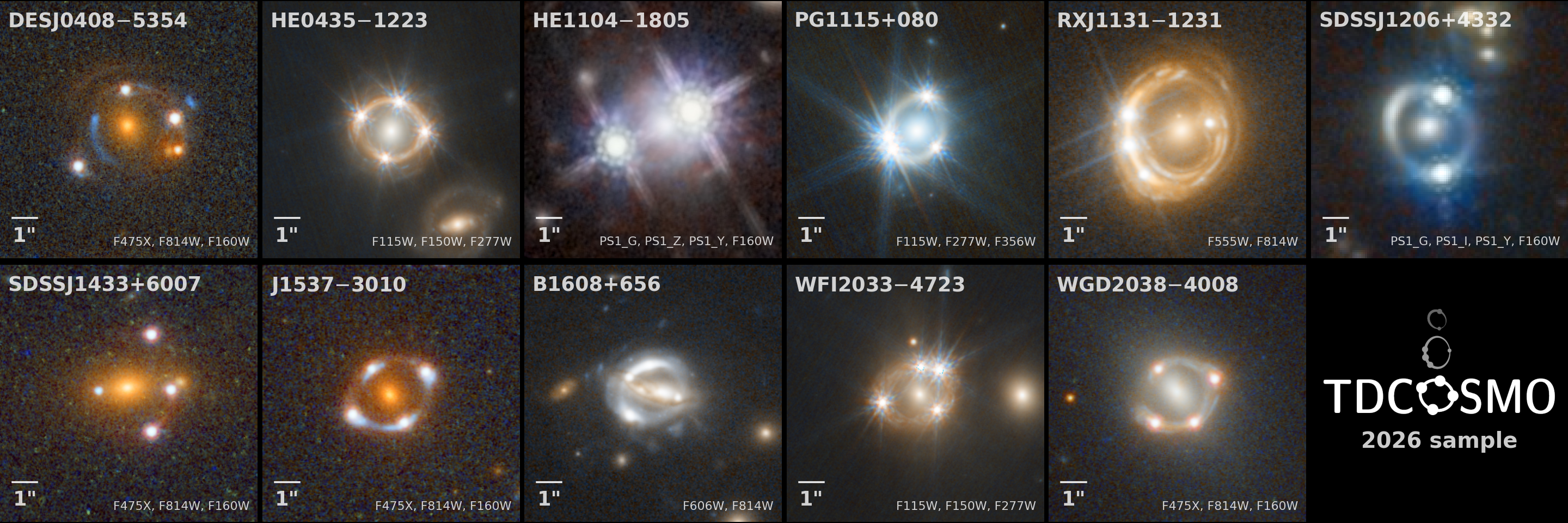}
    \caption{Montage of all the time-delay lenses in the \TDCnew sample. All cutouts cover the same angular size and are aligned North up, East left. The color composite images are made using a variety of instruments and filters from HST (F475X, F555W, F606W, F814W, F160W) and JWST-NIRCam (F115W, F150W, F277W). For \Jtwelve and \HEeleven, only one HST filter with suitable depth is available so we combined it with Pan-STARRS1 images \citep{PanstarrsData_Flewelling2020} to provide color information (PS1\_G, PS1\_I, PS1\_Z, PS1\_Y). The choice of filters shown is optimized for visual illustration and does not coincide with those used for the analysis. We refer to the modeling papers for more details on the filters used in the analysis (see Table~\ref{tab:lens_overview} for references).}
    \label{fig:montage}
\end{figure*}

\begin{enumerate}
    \item We add three new time-delay lenses: two quadruply-imaged systems (quads), \Jfourteen \citep{sheu2026} and \Jfifteen \citep{GalanSchweinfurth:2026}, and one doubly-imaged system (double), \HEeleven \citep{Paic2026}. This brings the total sample of time-delay lenses to 11, including 9 quads and 2 doubles. The \TDCnew sample is shown in Fig.~\ref{fig:montage}.
    \item For three time-delay lenses, \PGeleven, \HEzero, and \WFItwenty, we replace the HST/Keck-AO based lens models with new ones based on JWST-NIRCam imaging \citep{TDCOSMO20,Williams2026}
    \item We use spatially resolved kinematics based on JWST-NIRSpec for six lenses \citep[\RXJ;][]{Shajib25b} \citep[\HEzero, \PGeleven, and \WFItwenty;][]{Knabel2026} \citep[\Bsixteen and \Jtwelve;][]{Mozumdar2026}, compared to \TDCprev where we used a radially binned map and five integrated spectra from JWST-NIRSpec. In addition, we use ground-based spatially resolved kinematics for three of the systems \citep[\Jfourteen, \Jfifteen, \HEeleven][]{sheu2026,GalanSchweinfurth:2026,Paic2026}. For two systems, MUSE aperture stellar velocity dispersions are used~\citep[\DESzerofour, \WGDtwenty][]{TDCOSMO2025}.
    \item spatially resolved kinematics are available for \nslacsifu\ galaxy--galaxy lenses from the SLACS external sample, \nslacsifusel\, of which pass our quality cuts and are included in the analysis \citep{Knabel24,knabel26_slacs}.
    \item Integrated stellar velocity dispersion based on Keck-KCWI spectroscopy is available for three additional SLACS lenses and one additional SL2S galaxy--galaxy lens. This brings the total of SLACS and SL2S lenses used in this analysis to \nslacstot\ and \nsltwossel, respectively. 
    \item The one-dimensional posterior distribution function~(PDF) of the time-delay distance from the lens model is replaced by the joint two-dimensional PDF of time-delay distance and mass density profile slope~$\gamma_{\rm pl}$. Since $\gamma_{\rm pl}$ affects the calculation of stellar velocity dispersion, this helps reduce the degeneracy between time-delay distance and internal mass-sheet parameter~$\lint$. This was implemented only for \RXJ in \TDCprev (see Appendix to that paper), and it is now applied to all systems. 
    \item New estimates of the line-of-sight convergence are implemented, based on an improved formalism and state-of-the-art numerical cosmological simulations \citep[Euclid Flagship,][]{Johnson2026}.
    \item Instead of a statistical axisymmetric correction applied to spherical Jeans modeling \citep{Huang2026}, we use direct axisymmetric Jeans modeling \citep{Cappellari2026}.  
    
\end{enumerate}

This paper is organized as follows. In Section~\ref{sec:tdc} we briefly review the theory of time-delay cosmography. In Section~\ref{sec:new} we describe in detail the new pieces of information and methodological improvements of the present analysis compared to \TDCprev. In Section~\ref{sec:sample_selection} we apply our selection criteria to non-time-delay lenses to match the properties of the TDCOSMO sample. In Section~\ref{sec:flat} we present our results in flat $\Lambda$CDM cosmology\footnote{All likelihoods will be released upon acceptance of the paper.}. In Section~\ref{sec:alt} we present our results in alternative cosmologies. We restrict ourselves to the same set of cosmologies considered in \TDCprev to facilitate comparison. In Section~\ref{sec:analysis variation} we explore the effect of varying some of the choices of our analysis, showing the results are robust to these choices. This variational analysis includes some choices that are disfavored by physical arguments. It is meant to test the stability of the results, not as a way to quantify our systematic errors, which are discussed later. In Section~\ref{sec:literature} we compare our results with previous results based on time-delay cosmography and other approaches. In Section~\ref{sec:syst} we discuss systematic errors. In Section~\ref{sec:sum} we provide a brief summary.
  
As always, our hierarchical analysis is carried out blindly. This is achieved by removing the median of all the cosmographically relevant parameters from all the PDFs and statistics displayed by analysis codes.
The individual subcomponents of the analysis are also carried out blindly and independently of each other, as described in the referenced papers. Unblinding of the hierarchical analysis took place on September 23, 2026, after completion of the internal review, and the results were published without modification. 

Internal review was carried out by five designated reviewers (S.B., F.C., D.S., S.H.S., J.A.F.) and two code reviewers (W.S., X.-Y.H.) prior to unblinding. After unblinding, the paper was completed and further reviewed by the collaboration, including a final reader (T.T.). All explicit contributions are further referenced after the conclusions.

\section{Time-delay cosmography}
\label{sec:tdc}

Variability in strongly lensed sources, such as active galactic nuclei (AGN) or supernovae, is observed in multiple images with relative time delays due to differences in light paths and gravitational potentials.

The time delay between two images at positions $\boldsymbol{\theta}_{\rm A}$ and $\boldsymbol{\theta}_{\rm B}$, originating from a source at $\boldsymbol{\beta}$, is given by
\begin{equation}\label{eqn:time_delay}
\Delta t_{\rm AB} = \frac{D_{\Delta t}}{c} \left[\tau(\boldsymbol{\theta}_{\rm A}, \boldsymbol{\beta}) - \tau(\boldsymbol{\theta}_{\rm B}, \boldsymbol{\beta}) \right],
\end{equation}
where the Fermat potential is
\begin{equation}\label{eqn:fermat_potential}
\tau(\boldsymbol{\theta}, \boldsymbol{\beta}) = \frac{1}{2}(\boldsymbol{\theta} - \boldsymbol{\beta})^2 - \psi(\boldsymbol{\theta}),
\end{equation}
with $\psi(\boldsymbol{\theta})$ the lensing potential. The time-delay distance is defined as
\begin{equation} \label{eqn:ddt_definition}
D_{\Delta t} = (1 + z_{\rm d}) \frac{D_{\rm d} D_{\rm s}}{D_{\rm ds}},
\end{equation}
where $z_{\rm d}$ is the deflector redshift and $D_{\rm d}$, $D_{\rm s}$, and $D_{\rm ds}$ are the angular-diameter distances to the deflector galaxy, to the source galaxy, and between the deflector and the source \citep{Suyu2010}.

With a lens model and measured time delays, one can infer $D_{\Delta t}$, which scales inversely with the Hubble constant,
\begin{equation} \label{eqn:H0_ddt}
H_0 \propto D_{\Delta t}^{-1}.
\end{equation}
Thus, time-delay cosmography provides a direct constraint on $H_0$. For further details and reviews on time-delay cosmography, see \cite{TreuMarshall2016, Treu_review, TreuShajib_review, Birrer:2024}.

\subsection{The mass-sheet degeneracy}

The mass-sheet degeneracy \citep[MSD;][]{Falco1985} represents the primary source of systematic uncertainty in strong lens mass modeling and, consequently, constitutes a major hurdle to be overcome in order to measure \Hc accurately and precisely. The MSD originates from the mass-sheet transformation (MST), which rescales the convergence $\kappa(\boldsymbol{\theta}) \equiv \nabla^2 \psi(\boldsymbol{\theta})/2$ according to
\begin{equation}
\label{eq:mst}
    \kappa \to \kappa_\lambda(\boldsymbol{\theta}) = \lambda\,\kappa(\boldsymbol{\theta})
    + (1 - \lambda),
\end{equation}
alongside a corresponding rescaling of the (unobserved) source-plane coordinates
$\boldsymbol{\beta} \to \boldsymbol{\beta}_\lambda = \lambda\boldsymbol{\beta}$. From a mathematical standpoint, the MST parameter $\lambda$ is equivalent to the addition of an infinite uniform mass sheet. The MST preserves all lensing observables, namely image positions and flux ratios, while leaving only the time delays affected. Consequently, the time-delay distance (or equivalently the Fermat potential; see Eq.~\ref{eqn:fermat_potential}) transforms as $\Ddt \to \lambda^{-1}\,\Ddt$ under the MST.

The MST parameter $\lambda$ admits a decomposition into two distinct contributions,
\begin{align}
\label{eq:mst_decomp}
    \lambda \equiv (1-\kext)\,\lambda_\mathrm{int},
\end{align}
where $\lambda_\mathrm{int}$ encodes a transformation of the primary deflector's mass profile, and \kext denotes the convergence due to matter along the LoS. \kext can be constrained for a given LoS using photometric and spectroscopic observables of the LoS environment \citep[e.g][]{Greene2013, Rusu2017, Wells2023} or weak lensing \citep{Tihhonova2018, Tihhonova2020}. 
The quantity $\kext$ can itself be expressed as
\begin{align}
\label{eq:los_breakdown}
    1-\kext = \frac{(1-\kappa_\mathrm{d})(1-\kappa_\mathrm{s})}{1-\kappa_\mathrm{ds}},
\end{align}
where $\kappa_\mathrm{d}$, $\kappa_\mathrm{s}$, and $\kappa_\mathrm{ds}$ are the weak lensing convergence contributions from the observer--lens, observer--source, and lens--source planes, respectively, computed in the absence of the primary deflector \citep{TDCOSMO20, Fleury_2021, Johnson2024}. In previous TDCOSMO work, the approximation $\kext \approx \kappa_\mathrm{s}$ was made, as it simplifies the galaxy number counts analysis. In this work, for the first time, we relax this assumption and compute the exact $\kappa_\mathrm{d}$, $\kappa_\mathrm{s}$, and $\kappa_\mathrm{ds}$ contributions as detailed by \citet{Johnson2026} and summarized by \sref{subsec:kappa_ext}.

The internal part of the MSD, parametrized by $\lambda_\mathrm{int}$, can be mitigated through stellar kinematic measurements of the primary deflector, as shown in \ssref{subsec:Kinematic_modeling}. 

\subsection{Lens modeling}

Our baseline model is a power-law elliptical mass distribution \citep[PEMD;][]{Kormann:1994, Barkana:1998}, with an additional external shear component. The convergence of the mass profile is parameterized as 
\begin{equation} \label{eqn:pl_profile}
\kappa(\theta_1, \theta_2) = \frac{3-\gamma_{\rm pl}}{2} \left[\frac{\theta_{\rm E}}{\sqrt{q_{\rm m}\theta_1^2 + \theta_2^2/q_{\rm m}}} \right]^{\gamma_{\rm pl}-1},
\end{equation}
where $\gamma_{\rm pl}$ is the logarithmic slope, $q_{\rm m}$ the axis ratio of the elliptical profile, and $\theta_{\rm E}$ the Einstein radius. This profile is fit to each individual system using high-resolution imaging data. Nearby perturber galaxies close to the main lens are explicitly added to the model as additional explicit mass profile if the flexion shift $\flexshift$ that they induce at the position of the main deflector is larger than $10^{-4}$ \citep{McCully2017}.  The details of this modeling can be found for each system in the lens modeling references listed in \tref{tab:lens_overview}. 

Since the PEMD profile implicitly breaks the MSD, its parameters can be tightly constrained (to within a few percent) using high-quality imaging. To capture the full flexibility in the radial mass profile of the lens galaxy, we include an internal MSD parameter, $\lambda_\mathrm{int}$, which provides an additional degree of freedom that is exactly degenerate with \Hc. This parameter describes deviations from an elliptical power-law profile and is constrained at the population level using kinematic data (see \sref{subsec:Kinematic_modeling}). 
A value of $\lambda_\mathrm{int} = 1$ corresponds to the PEMD profile, with no deviation from a power law. Values of $\lambda_\mathrm{int} < 1$ imply that, on average, a positive mass-sheet transform is present in the sample. We note that this does not imply that the galaxies contain an infinite mass sheet. Instead, it can be physically interpreted as  modification of the power-law mass density profile. This modification can be ascribed, for example, to the presence of a dark matter core in the galaxy mass profile extending well beyond the Einstein radius of the lens \citep{Blum2020}, which locally acts as a mass sheet. Conversely, $\lambda_\mathrm{int} > 1$ indicates the presence of a negative mass sheet, which can be physically interpreted as a deviation from a power-law profile corresponding to an inner mass profile that is steeper than the slope, $\gamma_{\rm pl}$, measured at the Einstein radius. Regardless of the interpretation, the internal mass-sheet parameter enables a modification of the PEMD profile that is maximally degenerate with $H_0$ and thus allows us to take into account this source of uncertainty in our analysis.

$\gamma_{\rm pl}$ is indeed a local quantity inferred at the Einstein radius and constrained by the modeling of the quasar host galaxy, lensed in gravitational arcs \citep{Kochanek2002}. Since the Einstein radius depends on both the lens mass and the distances to the source and to the lens, the physical scale at which $\gamma_{\rm pl}$ is measured varies across the lens sample. If massive elliptical galaxies deviate from a PEMD through an MST in a way that depends on the projected radius, applying a single global MST correction to all individually inferred PEMD profiles may introduce biases. To account for this possibility, we allow the MST parameter to vary linearly with the ratio of the deflector's effective radius to its Einstein radius, $R_{\rm eff}/\theta_{\rm E}$,
\begin{equation}
\label{eqn:lambda_int}
\lambda_{\rm int}(R_{\rm eff}/ \theta_{\rm E}) = \lambda_{\rm int, 0} + \alpha_{\lambda}\left( \frac{R_{\rm eff}}{\theta_{\rm E}}-1 \right),
\end{equation}
where $\lambda_{\rm int,0}$ is the MST parameter for lenses with $R_{\rm eff}/\theta_{\rm E}=1$, and $\alpha_{\lambda}$ describes its radial dependence \citep{TDCOSMOIV}. This parameterization assumes self-similarity of the galaxy mass profiles when scaled by the half-light radius. We further include a Gaussian intrinsic scatter with standard deviation $\sigma(\lambda_{\rm int})$ at fixed $R_{\rm eff}/\theta_{\rm E}$ to account for expected lens-to-lens variations at the population level.

\subsection{Kinematic modeling}
\label{subsec:Kinematic_modeling}
Single-aperture velocity dispersions remain susceptible to the mass-anisotropy degeneracy~\citep[MAD;][]{Binney1982}, whereas spatially resolved kinematic data further constrain the radial mass-profile slope, thereby alleviating both the internal MSD and the MAD simultaneously \citep[e.g.,][]{Yildirim+2023,Shajib2023,Wang2025}. Here, we recall the most important steps of this procedure, but more details can be found in \citep[e.g.][]{knabel26_slacs}.

The line-of-sight velocity dispersion predicted by a lens model can be separated into cosmology-dependent and model-dependent terms \citep[see e.g.,][]{Birrer2016, Birrer2019}:
\begin{equation}
\label{eq:sigma_los}
\sigma_{\rm los}^2 = \frac{1-\kappa_{\rm s}}{1 - \kappa_{\rm ds}} \frac{D_{\rm s}}{D_{\rm ds}}c^2 J(\boldsymbol{\xi}_{\rm lens}, \beta_{\rm ani},\lambda_{\rm int}),
\end{equation}
where $J$ is dimensionless and depends on the lens model parameters $\boldsymbol{\xi}_{\rm lens} = \{\boldsymbol{\xi}_{\rm mass}, \boldsymbol{\xi}_{\rm lens \ light}\}$, describing the mass and light profiles of the deflector, the stellar orbital anisotropy $\beta_{\rm ani}$, the internal mass component $\lambda_{\rm int}$, and the observational setup, including the PSF and the spectroscopic aperture \citep[e.g.,][]{Binney1982, TreuKoopmans2004, Suyu2010}. The quantities $\kappa_{\rm s}$ and $\kappa_{\rm ds}$ describe the LoS convergence between the observer and the source, and between the deflector and the source, respectively.

The parameter $\lambda_{\rm int}$ accounts for an internal mass component affecting the stellar dynamics and is therefore included in $J$~\citep{Teodori2022}. In contrast, the external component, characterized by $\kext$, does not affect the stellar kinematics of the main deflector. This distinction motivates treating the internal and external contributions separately. For a sheet-like component extending well beyond the effective radius of the deflector, its effect can be approximated as
\begin{equation}\label{eq:J}
J(\boldsymbol{\xi}_{\rm lens}, \beta_{\rm ani},\lambda_{\rm int}) \approx \lambda_{\rm int} J(\boldsymbol{\xi}_{\rm lens}, \beta_{\rm ani}).
\end{equation}

In \eref{eq:sigma_los}, the external convergence can be constrained from an analysis of the lens environment along the LoS, while the ratio $D_{\rm s}/D_{\rm ds}$ can either be inferred by assuming a cosmological model, jointly fitted with the time-delay distance $D_{\Delta t}$, or constrained using probes of the relative expansion history. Comparing the observed line-of-sight velocity dispersion with the model prediction then provides a constraint on $\lambda_{\rm int}$. Residual degeneracies between $J(\boldsymbol{\xi}_{\rm lens}, \beta_{\rm ani})$ and $\lambda_{\rm int}$ can be further reduced using spatially resolved kinematics, with velocity-dispersion measurements at multiple radii from the galaxy center. These measurements provide additional leverage on the stellar orbital anisotropy $\beta_{\rm ani}$, the mass-profile slope $\gamma_{\rm pl}$, and $\lambda_{\rm int}$. 

In practice, $J(\boldsymbol{\xi}_{\rm lens}, \beta_{\rm ani})$ is precomputed for each lens model on a three-dimensional parameter grid spanning $\beta_{\rm ani}$, $\gamma_{\rm pl}$, and the intrinsic three-dimensional axis ratio of the mass distribution, $q_{\rm int}$, using axisymmetric Jeans modeling (see \ssref{ssec:axisymmetric_modelling} for details). The dependence on $\lambda_{\rm int}$ is then applied linearly using \eref{eq:J}. This enables a fast evaluation of the kinematic likelihood without solving the axisymmetric Jeans equations at each sampling step.

\subsection{Hierarchical Bayesian inference}
\label{ssec:hierarc}

Our hierarchical analysis follows the framework introduced by \cite{TDCOSMOIV} and was subsequently used in \cite{TDCOSMO2025}. We briefly summarize it below, with particular emphasis on the parameterizations, assumptions, and priors that govern our inference. We refer the reader to~\citet{TDCOSMOIV} for a thorough description of the method and to \citet{TDCOSMO8} for tests of the framework on simulated data.

Our goal is to determine the posterior distribution of the cosmological parameters $\boldsymbol{\pi}$ given a collection of strong-lensing observations $p(\boldsymbol{\pi} | \{\mathcal{D}_{i} \}_{N})$, where $\mathcal{D}_i$ denotes the dataset associated with each individual lens system $i$ (imaging data, time-delay measurements, stellar kinematics, and line-of-sight characterization) and $N$ is the total number of lenses.

We use $\boldsymbol{\xi}$ to collectively denote all model parameters, both at the level of individual lenses and at the population level, in addition to the cosmological parameters $\boldsymbol{\pi}$. Applying Bayes' theorem and assuming statistical independence between individual lens datasets, the posterior can be written as
\begin{align} \label{eqn:full_inference}
    p(\boldsymbol{\pi} \mid \{\mathcal{D}_{i} \}_{N})
    &\propto \mathcal{L}(\{\mathcal{D}_{i} \}_{N} \mid \boldsymbol{\pi})\, p(\boldsymbol{\pi}) \nonumber \\
    &= \int \mathcal{L}(\{\mathcal{D}_{i} \}_{N} \mid \boldsymbol{\pi}, \boldsymbol{\xi})\, p(\boldsymbol{\pi}, \boldsymbol{\xi})\, \diff\boldsymbol{\xi} \nonumber \\
    &= \int \prod_i^N \mathcal{L}(\mathcal{D}_{i} \mid \boldsymbol{\pi}, \boldsymbol{\xi})\, p(\boldsymbol{\pi}, \boldsymbol{\xi})\, \diff\boldsymbol{\xi}.
\end{align}
We further partition the nuisance parameters $\boldsymbol{\xi}$ into two subsets: lens-specific parameters $\boldsymbol{\xi}_i$, constrained separately for each system, and population-level hyperparameters $\boldsymbol{\xi}_{\rm pop}$, which are inferred jointly across the full lens sample. The per-lens parameters $\boldsymbol{\xi}_i$ encompass the lens mass model, the source and lens light distributions, and any additional model quantities required to reproduce the observed imaging data. The population parameters $\boldsymbol{\xi}_{\rm pop}$ include $\lambda_{\rm int,0}$, $\alpha_{\lambda}$ (as described in \eref{eqn:lambda_int}), the mean and scatter of the ratio of tangential to radial velocity dispersions, $\langle \sigma_{\rm t}/\sigma_{\rm r} \rangle$ and $\sigma(\sigma_{\rm t}/\sigma_{\rm r})$, and the mean and scatter of the mass axis ratio of the deflector galaxies, $\langle q_{\rm int} \rangle$ and $\sigma( q_{\rm int} )$. The ratio of tangential to radial velocity dispersions relates to the anisotropy parameter $\beta_{\rm ani} \equiv 1-\sigma^2_{\rm t}/\sigma^2_{\rm r}$, where $\beta_{\rm ani}=0$ corresponds to the isotropic case. $q_{\rm int}$ relates to the mass profile axis ratio (\eref{eqn:pl_profile}) through the inclination angle $i$ between the symmetry axis and the line of sight (\eref{eqn:inclination_angle}).

The hierarchical posterior (\eref{eqn:full_inference}) can then be recast as
\begin{multline} \label{eqn:full_inference_extended}
    p(\boldsymbol{\pi} \mid \{\mathcal{D}_{i} \}_{N})
     \propto \int \prod_i \left[ \mathcal{L}(\mathcal{D}_i \mid 
     D_{\rm d, s, ds}(\boldsymbol{\pi}), \boldsymbol{\xi}_i,  
     \boldsymbol{\xi}_{\rm pop})
     p(\boldsymbol{\xi}_i) \right]
     \\ \times
     \frac{p(\boldsymbol{\pi}, \{\boldsymbol{\xi}_i \}_{N}, 
     \boldsymbol{\xi}_{\rm pop})}{\prod_i p(\boldsymbol{\xi}_i)} 
     \diff \boldsymbol{\xi}_{\{i\}} \diff \boldsymbol{\xi}_{\rm pop},
\end{multline}
where $\{\boldsymbol{\xi}_i \}_{N} = \{\boldsymbol{\xi}_1, \boldsymbol{\xi}_2, ..., \boldsymbol{\xi}_N \}$ collects the parameters of all individual lenses, and $p(\boldsymbol{\xi}_i)$ are the interim priors adopted during the single-lens inference stage. Note that these interim priors are divided out to obtain the hierarchical posterior. The dependence on cosmological parameters $\boldsymbol{\pi}$ enters exclusively through the set of angular-diameter distances $\{D_{\rm d}, D_{\rm s}, D_{\rm ds} \} \equiv D_{\rm d, s, ds}$. Therefore, we substitute $D_{\rm d, s, ds}(\boldsymbol{\pi})$ for $\boldsymbol{\pi}$ in \eqref{eqn:full_inference_extended}. 

Constraining $H_0$ from a lens ensemble requires specifying priors on the population hyperparameters, as detailed in \tref{tab:param_prior}. The hierarchical Bayesian framework also allows us to include lenses without measurable time delays. Galaxy--galaxy lenses, which lack a variable AGN and therefore do not provide a $D_{\Delta t}$ measurement, can still constrain the population parameters~$\boldsymbol{\xi}_{\rm pop}$ when lensing and kinematic data are available. Their kinematic likelihood is evaluated by comparing the observed and modeled kinematics through \eref{eq:J}, while the distance ratio $D_{\rm s}/D_{\rm ds}$ is determined by the cosmological parameters $\boldsymbol{\pi}$. Including galaxy--galaxy lenses, therefore, relies on the assumption that the time-delay (or TDCOSMO) and non-time-delay lens samples are described by the same population and cosmological models.

\begin{table*}[]
\caption{\label{tab:param_prior}
Summary of the parameters and priors for the baseline hierarchical model.}
\renewcommand{\arraystretch}{1.2}
\centering
\begin{tabular}{l l l}
    \hline \hline
    Parameter & Prior & Description \\
    \hline
    \textbf{Cosmology (flat $\Lambda$CDM)} \\
    $H_0$ [\ksmpc] & $\mathcal{U}\,(0, 150)$  & Hubble constant \\
    $\Omega_{\rm m}$ & $\mathcal{U}\,(0.05, 0.5)$ & Matter density \\
    \hline
    \textbf{Mass profile} \\
    $\lambda_{\rm int, 0}$ & $\mathcal{U}\,(0.5, 1.5)$ & Internal MST population mean \\
    $\alpha_{\lambda}$ & $\mathcal{U}\,(-1, 1)$  & Slope of $\lambda_{\rm int}$ with $R_{\rm eff}/\theta_{\rm E}$ of the deflector \\
    $\sigma(\lambda_{\rm int})$ & $\textrm{Loguniform}\,(0.001, 0.5)$ & 1$\sigma$ Gaussian scatter in the internal MST \\
    $\gamma_{\mathrm{pl}, i}$ & $p\,(\gamma_{\mathrm{pl}, i})$ of individual lenses  & Lens model-only prior on the power-law slope (see \sref{sec:ddt_gamma_conditioning}) \\
    \hline
    \textbf{Stellar kinematics} \\
    $\langle \sigma_{\rm t}/\sigma_{\rm r} \rangle$ & $\mathcal{U}\,(0.87,1.12)$ & Mean anisotropy parameter for a spatially constant anisotropy \\
    $\sigma(\sigma_{\rm t}/\sigma_{\rm r})$ & $\textrm{Loguniform}\,(0.03, 1)$  & 1$\sigma$ Gaussian scatter in $\sigma_{\rm t}/\sigma_{\rm r} $\\
    $\langle q_{\rm int} \rangle$ & $\mathcal{U} (0.2, q_{\rm max})$  & Mean intrinsic mass axis ratio. $q_{\rm max}$ is the maximal observed axis ratio in the sample. \\
    $\sigma(q_{\rm int})$ & $\mathcal{U}$ (0.012, 0.3) & 1$\sigma$ Gaussian scatter in $q_{\rm int}$\\
    
    \hline
    \textbf{LoS} \\
    \kext & $p\,(\kext)$ of individual lenses & External convergence of lenses \\
    \hline
\end{tabular}
\end{table*}

\section{Improvements compared to previous TDCOSMO works}
\label{sec:new}

Several key methodological improvements and updated datasets are introduced in this paper compared to our \TDCprev analysis. The main modifications of our analysis are listed below.

\subsection{New systems}
We include 3 new lensed quasars in our sample, in addition to the 8 previously analyzed in \TDCprev. The analysis of these systems, namely \HEeleven, \Jfourteen, and \Jfifteen, is presented at the individual level in the accompanying papers \citep{Paic2026, sheu2026, GalanSchweinfurth:2026}, and their HST images are shown in \fref{fig:montage}. The key findings from these papers are summarized here. 
\paragraph{\HEeleven: } This system is doubly imaged, making \HEeleven the only double of the sample with \Jtwelve. The time delay is measured at 5.8\% thanks to three sets of data spanning 16 years from the COSMOGRAIL program (PI: F. Courbin). Analyzing MUSE data (PI: A. Agnello), the resolved velocity dispersion was measured in 3 concentric bins with precision from 5\% to 13\%. Lens modeling was performed with \lenstro using HST/F160W observations (HST-GO 12889, PI: S.~Suyu). For this system, the kinematic data were not of sufficient quality to constrain both $\lint$ and cosmography. Therefore, $\lint=1$ was adopted. Also, since this system was completed before the update in the calculation of $\kext$ by \citet{Johnson2026},  $\kext = \kappa_{\rm s}$ was adopted. Under these assumptions, the cosmographic analysis of this system carried out in \citet{Paic2026}, yielded $H_0=64.2^{+5.8}_{-5.0}$ \ksmpc. Naturally, for the analysis in this milestone paper, $\lint$ is free to vary and is constrained hierarchically, while $\kext$ has been updated to the new methodology. 
\paragraph{\Jfourteen: } This system was analyzed by \citet{sheu2026}, where the authors used \textit{HST} imaging in the F160W, F814W, and F475X filters (PID: 15320, PI: T. Treu) to construct a lens model using \lenstro.  A time-delay measurement with precision $2.6\%$ was achieved by combining monitoring campaigns from the Maidanak and Wendelstein Observatories \citep{q25, sheu2026}.  \Jfourteen was observed for 8 hours with KCWI/KCRM to construct a detailed spatially resolved kinematic map. The time-delay cosmography analysis by \citet{sheu2026} yields $H_0=73.2^{+4.8}_{-4.7}$ \ksmpc and $\lambda_{\rm int}=1.12^{+0.05}_{-0.06}$ for this individual system.

\paragraph{\Jfifteen:} Lens models of this system based on the HST imaging data (PI: Treu; PID: 15652) were presented by Galan, Schweinfurth et al.~[\citenum{GalanSchweinfurth:2026}] and obtained using both \lenstro and \glee modeling codes using time delays measured with $\sim2\%$ precision. From that single system, they measured $H_0=75.5^{+9.3}_{-5.8}$ \ksmpc and $\lambda_{\rm int}=0.89^{+0.09}_{-0.07}$, combining constraints from both codes with equal weights. Galan, Schweinfurth et al.~[\citenum{GalanSchweinfurth:2026}] also concluded that while the two codes returned mutually consistent results, the \lenstro model of \Jfifteen provides slightly better fits to the imaging data, time delays, and stellar kinematics. For the present hierarchical analysis, we therefore use the \lenstro lensing constraints while using the \glee constraints as a systematic check.

The new sample is described in \tref{tab:lens_overview}.
\begin{table*}[]
    \renewcommand{\arraystretch}{1.2}
    \caption{\label{tab:lens_overview}
    Overview of the TDCOSMO-2026 sample of time-delay lenses (strongly lensed quasars).
    }
    \centering
    \begin{tabular}{l l l l l l}
        \hline \hline
        Lens & Discovery \& Redshifts & Kinematics & Time delays & Lens modeling & Environment \\
        \hline
        \DESzerofour  
        & \cite{Lin2017} 
        & \cite{TDCOSMO2025} (MUSE, aperture) 
        & \cite{Courbin2018} 
        & \cite{Shajib2019} (HST)
        & \cite{Buckley-Geer2020,Johnson2026} \\
        
        \HEzero       
        & \cite{Wisotzki2002, Morgan2005, Eigenbrod2006, Sluse2012} 
        & \cite{Knabel2026} (NIRSpec, resolved)    
        & \cite{Bonvin2017} 
        & \cite{Wong2017, Chen2019,Williams2026} (JWST)
        & \cite{Sluse2017, Rusu2017,Johnson2026} \\

        \HEeleven
        & \cite{Wisotzki1993,Lidman2000}
        & \cite{Paic2026} (MUSE, radial bins)
        & \cite{Paic2026}
        & \cite{Paic2026} (HST)
        & \cite{Paic2026,Johnson2026} \\

        
        \PGeleven     
        & \cite{Weymann1980, Tonry1998} 
        & \cite{Knabel2026} (NIRSpec, resolved)   
        & \cite{Bonvin2018} 
        & \cite{Chen2019, Williams2026} (JWST) 
        & \cite{Chen2019,Johnson2026} \\
        
        \RXJ          
        & \cite{Sluse2003, Sluse2007} 
        & \cite{Shajib25b} (KCWI+NIRSpec, resolved)  
        & \cite{Tewes2013} 
        & \cite{Suyu2014, Chen2019} (HST) 
        & \cite{Suyu2013,Johnson2026}\\
        
        \Jtwelve      
        & \cite{Oguri2005, Agnello2016}
        & \citep{Mozumdar2026} (NIRSpec, resolved)   
        & \cite{Eulaers2013, Birrer2019} 
        & \cite{Birrer2019} (HST)
        & \cite{Birrer2019,Johnson2026} \\

        \Jfourteen
        & \cite{Agnello_2017}
        & \cite{sheu2026} (KCWI, resolved)
        & \cite{q25, sheu2026}
        & \cite{sheu2026} (HST)
        & \cite{sheu2026,Johnson2026} \\

        \Jfifteen
        & \cite{Lemon:2019,Delchambre:2019,GalanSchweinfurth:2026}
        & \cite{GalanSchweinfurth:2026} (MUSE, resolved)
        & \cite{Dux:2025,GalanSchweinfurth:2026}
        & \cite{GalanSchweinfurth:2026} (HST)
        & \cite{GalanSchweinfurth:2026,Johnson2026} \\
        
        \Bsixteen     
        & \cite{Myers1995, Fassnacht1996} 
        & \cite{Mozumdar2026} (NIRSpec, resolved)  
        & \cite{Fassnacht2002} 
        & \cite{Suyu2009, Jee2019} (HST)
        & \cite{Suyu2010,Johnson2026}\\
        
        \WFItwenty    
        & \cite{Morgan2004, Sluse2019, Sluse2012}
        & \cite{Knabel2026} (NIRSpec, resolved)  
        & \cite{Bonvin2019} 
        & \cite{Rusu2020,Williams2026} (JWST)
        & \cite{Sluse2019,Johnson2026} \\
        
        \WGDtwenty    
        & \cite{Agnello2018, Buckley-Geer2020} 
        & \cite{TDCOSMO2025} (MUSE, aperture) 
        & \cite{Wong24} 
        & \cite{Shajib2022} (HST)
        & \cite{Buckley-Geer2020,Johnson2026} \\
        \hline
    \end{tabular}

    \tablefoot{The table lists the references where the data products can be found.}
\end{table*}

\begin{table*}[!th]
\renewcommand{\arraystretch}{1.2}
\centering
\begin{threeparttable}
\setlength{\tabcolsep}{10pt}
\caption{Key properties of the TDCOSMO-2026 time-delay lenses. \label{tab:tdcosmo_sample_props}}
\begin{tabular}{l c c c c c c c}
\hline\hline
Lens & $z_{\rm d}$ & $z_{\rm s}$ & $\theta_{\rm E}$\ ["] & $\gamma_{\rm pl}$ & $R_{\rm eff}$\ ["] & $q_{\rm mass}$\tnote{\dag} & $q_{\rm light}$\tnote{\ddag} \\
\hline
\DESzerofour & $0.597$ & $2.375$ & $1.920 \pm 0.010$ & $1.90 \pm 0.03$ & $1.94 \pm 0.05$ & $0.720 \pm 0.020$ & $0.800$ \\
\HEzero & $0.455$ & $1.693$ & $1.134 \pm 0.003$ & $1.97 \pm 0.01$ & $2.25 \pm 0.01$ & $0.830 \pm 0.003$ & $0.787$ \\
\HEeleven & $0.729$ & $2.319$ & $1.359 \pm 0.130$ & $1.91 \pm 0.02$ & $1.17 \pm 0.08$ & $0.843 \pm 0.024$ & $0.794$ \\
\PGeleven & $0.310$ & $1.727$ & $0.950 \pm 0.030$ & $2.16 \pm 0.09$ & $1.20 \pm 0.00$ & $0.930 \pm 0.009$ & $0.914$ \\
\RXJ & $0.295$ & $0.657$ & $1.630 \pm 0.020$ & $1.95 \pm 0.05$ & $1.51 \pm 0.05$ & $0.762 \pm 0.006$ & $0.865$ \\
\Jtwelve & $0.745$ & $1.789$ & $1.250 \pm 0.010$ & $1.95 \pm 0.05$ & $0.53 \pm 0.05$ & $0.874 \pm 0.017$ & $0.808$ \\
\Jfourteen & $0.407$ & $2.737$ & $1.538 \pm 0.006$ & $1.95 \pm 0.01$ & $2.18 \pm 0.03$ & $0.754 \pm 0.006$ & $0.692$ \\
\Jfifteen & $0.590$ & $1.721$ & $1.360 \pm 0.005$ & $2.02 \pm 0.02$ & $1.57 \pm 0.05$ & $0.818 \pm 0.023$ & $0.784$ \\
\Bsixteen & $0.630$ & $1.394$ & $0.810 \pm 0.020$ & $2.08 \pm 0.03$ & $0.65 \pm 0.09$ & $0.604$ & $0.411$ \\
\WFItwenty & $0.657$ & $1.662$ & $0.987 \pm 0.003$ & $1.85 \pm 0.01$ & $1.55 \pm 0.01$ & $0.782 \pm 0.009$ & $0.686$ \\
\WGDtwenty & $0.228$ & $0.777$ & $1.380 \pm 0.001$ & $2.26 \pm 0.05$ & $2.57 \pm 0.05$ & $0.626 \pm 0.005$ & $0.702$ \\
\hline
\end{tabular}
\begin{tablenotes}
\small
\item[\dag] Axis ratio of the surface mass density model. Note that we ignore uncertainties in $q_{\rm mass}$ in this analysis. The value for \Bsixteen is taken from \citet[][]{Koopmans2003}, in which no uncertainty is quoted.
\item[\ddag] Axis ratio of the surface brightness model. Note that we ignore uncertainties in $q_{\rm light}$ in this analysis (uncertainties $\sim10^{-2}$).
\end{tablenotes}
\end{threeparttable}
\end{table*}

\subsection{Conditioning $D_{\Delta t}$ on the mass-density slope}
\label{sec:ddt_gamma_conditioning}

The lens model of each system delivers a joint posterior probability distribution function of the time-delay distance and the logarithmic slope of the power-law mass profile, $P(D_{\Delta t}, \gamma_{\mathrm{pl}} \mid \mathbf{d}_{\rm img}, \Delta t)$. 
However, $\gamma_{\rm pl}$ is not a nuisance parameter that needs to be integrated out: it is sampled explicitly when combining lensing and kinematic data in the hierarchical inference because the velocity dispersion measurements further constrain the slope of the mass profile.

This step needs to be handled with care. In fact, feeding the marginal posterior $P(D_{\Delta t})$ into such a model would count the slope uncertainty twice: once through the marginalization already performed on the imaging chain and once more through the $\gamma_{\rm pl}$ prior that is re-sampled at the hierarchical level. It would also discard the correlation between $\gamma_{\rm pl}$ and $D_{\Delta t}$, which is precisely the information the kinematics are meant to exploit. A steeper profile lowers the inferred $D_{\Delta t}$ and raises the predicted velocity dispersion.

Therefore, we proceed as follows. We replace the \emph{marginal} posterior with a \emph{conditional} posterior evaluated at a fixed pivot slope, taken to be the mean $\gamma_{\rm pl}$, and use a linear model that restores the correlation at sampling time. Over our ranges of interest, the relationship between $D_{\Delta t}$ and $\gamma_{\rm pl}$ is well approximated by a line. Therefore, this model is sufficient to restore the information obtained during the lens modeling procedure. We define the quantities
\begin{equation}
    \gamma_{\rm piv} = \langle \gamma_{\rm pl} \rangle,
    \quad {\rm and} \quad
    S \equiv \left. \frac{\partial D_{\Delta t}}{\partial \gamma_{\rm pl}} \right|_{\gamma_{\rm piv}}, 
\end{equation}
which are measured on the individual posterior of each lens. These quantities are used at the hierarchical sampling stage to displace the model prediction back along the measured degeneracy before it is compared with the conditional posterior: 
\begin{equation}
  D_{\Delta t}^{\rm mod} \;\longrightarrow\;
  D_{\Delta t}^{\rm mod} \;-\; \left( \gamma_{\rm pl} - \gamma_{\rm piv} \right) S ,
  \label{eq:displace}
\end{equation}
with $\gamma_{\rm pl}$ a sampled parameter for each lens. The likelihood for each lens, entering \eref{eqn:full_inference}, becomes
\begin{align}
\mathcal{L} &=
  P\!\left( D_{\Delta t}^{\rm mod} - (\gamma_{\rm pl} - \gamma_{\rm piv})\,S
     \;\middle|\; \gamma_{\rm piv} , \boldsymbol{\pi} \right) \nonumber \\
  &\quad \times
  \mathcal{L}_{\rm kin}\!\left(\sigma_{\rm los} | \gamma_{\rm pl}, \beta_{\rm ani}, \lambda_{\rm int}, q_{\rm int}, \boldsymbol{\pi} \right) \nonumber \\
  &\quad\times
  \mathcal{N}\!\left( \gamma_{\rm pl} \mid \langle \gamma_{\rm pl} \rangle, \sigma_{\gamma}^2 \right).
\end{align}

Note that, as desired, the slope uncertainty enters only once through the explicit $\gamma_{\rm pl}$ normal prior, centered on $\langle \gamma_{\rm pl} \rangle$ and with standard deviation $\sigma_{\gamma}$.

For example, for \Jtwelve, this procedure reduces the $\Ddt$ posterior at fixed $\gamma_{\rm pl}$ to $9.2\%$ (excluding kinematics) from $10.6\%$ when marginalized over $\gamma_{\rm pl}$. This approach was only applied to \RXJ in our \TDCprev analysis. In this paper, we take advantage of the improvement in data quality by extending this treatment to all the lenses that have spatially resolved velocity dispersion measurements.

\subsection{Axisymmetric Jeans modeling}
\label{ssec:axisymmetric_modelling}

In \TDCprev, deviations from spherical symmetry were incorporated through the projection-effect correction of \citet{Huang2026}. Axisymmetric JAM models were used to construct a correction to the spherical dynamical model, marginalizing over the intrinsic shape and inclination of the deflectors. This approach captures the effect of projection on the predicted kinematics, but does not jointly model inclination and orbital anisotropy, thereby neglecting potential correlations between them. Moreover, the spatially resolved kinematics were compressed into a single aperture or radial annuli rather than modeled as full two-dimensional maps.

In this work, we model the deflectors as axisymmetric systems and fit the two-dimensional kinematic information, jointly sampling the intrinsic shape, inclination, and anisotropy. 
The lens imaging constrains the projected mass distribution, which we deproject into an axisymmetric three-dimensional density profile for the dynamical modeling. Lensing and dynamics are combined by using the lens-model posterior as a prior for the kinematic inference. For the 2 systems with unresolved kinematics, the logarithmic slope, $\gamma_{\rm pl}$, of the power-law mass density is constrained only through lensing and marginalized over the lensing prior; for the 9 systems with resolved kinematics, $\gamma_{\rm pl}$ is sampled jointly with that prior.

We compute the predicted line-of-sight second velocity moment, $v_{\rm rms}\equiv\sqrt{v_{\rm los}^2+\sigma_{\rm los}^2}$, with the \jampy implementation of JAM \citep{Cappellari2008,Cappellari2020,Cappellari2026}, where $v_{\rm los}$ and $\sigma_{\rm los}$ are the line-of-sight mean velocity and velocity dispersion, respectively. Following the usual JAM framework, we consider both spherical and cylindrical alignment of the velocity ellipsoids \citep{Cappellari2020}. For spherically aligned ellipsoids, $\sigma_\theta=\sigma_\phi \equiv \sigma_{\rm t}$ and $\beta_r\equiv1-\sigma_{\rm t}^2/\sigma_r^2$ is the spherical anisotropy parameter, which measures the relative amount of radial and tangential motion. For cylindrically aligned ellipsoids, $\sigma_R=\sigma_\phi$ and $\beta_z\equiv1-\sigma_z^2/\sigma_R^2$ is the cylindrical anisotropy parameter, which measures the relative amount of vertical and radial motion. Here, the $\sigma$ terms denote velocity dispersions along the corresponding spherical $(r,\theta,\phi)$ or cylindrical $(R,\phi,z)$ coordinates. In the spherical coordinates, the prior that we adopt $\mathcal{U}(0.87, 1.12)$ on $\langle \sigma_{\rm t}/\sigma_{\rm r} \rangle$ corresponds to $-0.25 \lesssim \beta \lesssim 0.24$, consistent with what was found for local early-type galaxies \citep[][Section~3.3]{Cappellari2025}.

Strong lenses are expected to be predominantly slow rotators owing to the lensing cross section's strong preference for high velocity dispersion galaxies \citep[e.g.,][and references therein]{Treu10}; although slow rotators can be mildly triaxial, their intrinsic shapes are close to nearly oblate and a nearly spherical population \citep{Weijmans2014,Li2018}. As shown in Appendix~\ref{app:lambdaR}, the \TDCnew sample includes 2 fast rotators out of 11 galaxies. Given the prevalence of slow rotators in our sample, we adopt the spherically aligned anisotropy as the baseline. However, in Section~\ref{sec:analysis variation}, we show that adopting cylindrically aligned anisotropy changes our inference of $H_0$ by much less than our statistical errors. 

Axisymmetric deprojection requires the inclination angle $i$ between the symmetry axis and the line of sight, related to the observed axis ratio $q_{\rm obs}$ and the intrinsic axis ratio $q_{\rm int}$ by
\begin{equation}
\label{eqn:inclination_angle}
\cos^2 i = \frac{q_{\rm obs}^2-q_{\rm int}^2}{1-q_{\rm int}^2}.
\end{equation}
We treat the intrinsic axis-ratio distribution as a population-level parameter inferred hierarchically (see \ssref{ssec:hierarc}), so that the uncertainty in galaxy shape is marginalized and propagated to the cosmological inference. We adopt the lens-model mass axis ratio, $q_{\rm m}$ as $q_{\rm obs}$ in Eq.~\eqref{eqn:inclination_angle}, so that the intrinsic shape inferred by our model corresponds to that of the total mass distribution. In a variant of the analysis presented in \sref{sec:analysis variation}, we adopt the lens light-profile axis ratio $q_{\rm light}$ as $q_{\rm obs}$, defining the inclination relative to the observed light profile of the lensing galaxy. This axisymmetric treatment is applied to all lenses, including systems with unresolved single-aperture kinematics, for which the intrinsic shape is not individually constrained but is marginalized at the population level.

\subsection{New JWST-NIRCam Lens Modeling}

\begin{figure*}
    \centering
    \includegraphics[width=\textwidth]{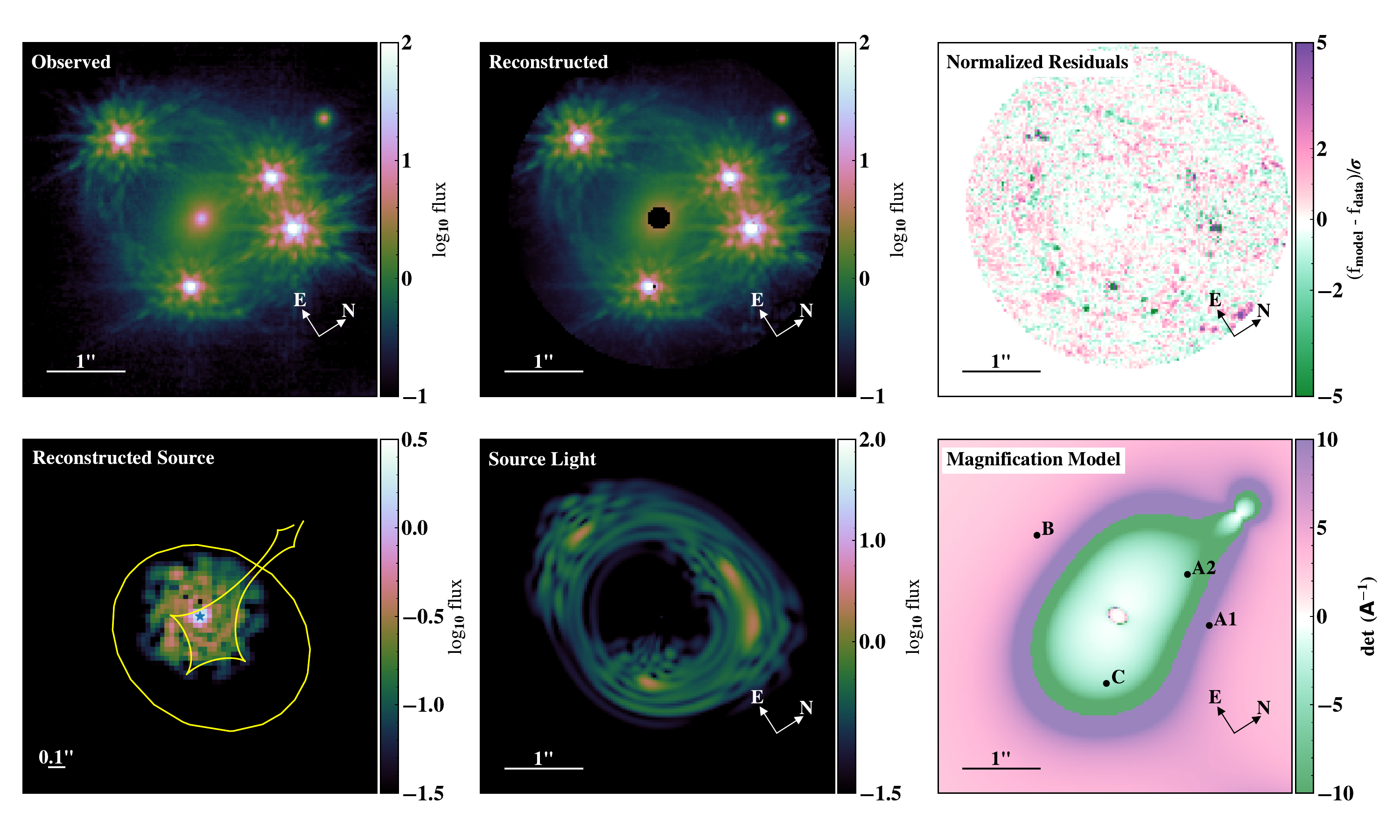}
    \caption{Example JWST model fit of \WFItwenty from \citet{Williams:2025}. The top-left panel shows the observed image and the top-middle panel shows the model-predicted reconstruction of the image. The top-right panel shows the fit's residuals which are normalized by the estimated uncertainty of each pixel, taking into account additional uncertainty of the PSF. The bottom-left panel shows the reconstructed quasar host galaxy, with the star symbol denoting the location of the quasar within the host galaxy. The bottom-middle panel shows this extended quasar host galaxy light when it is lensed and unconvolved, and the bottom-right panel shows the magnification map of the system.}
    \label{fig:example_Nircam}
\end{figure*}

We include in our analysis updated models based on JWST-NIRCam images for three systems, namely \HEzero, \PGeleven and \WFItwenty \citep{Williams:2025, Williams2026}. In addition to the higher-resolution and deeper imaging that JWST provides relative to HST and Keck adaptive-optics (AO) data, these models incorporate significant advancements in PSF reconstruction techniques, enabling high-precision lens models even when the point sources are saturated. Based on recent studies exploring the impact of PSF choice on lens models \citep{Shajib2022, Williams:2025, Krizic2026}, the choice of PSF reconstruction method is now an explicit systematic test that is marginalized over. An example of such model of the JWST images is shown in \fref{fig:example_Nircam} in the case of \WFItwenty. Improvements in aperture-integrated kinematic measurements for the main lensing galaxies were incorporated from JWST-NIRSpec \citep{Knabel2026}. The new JWST-based lens models remain consistent with those based on HST and Keck-AO data despite differing instruments, software, PSF reconstruction techniques, and external convergence estimates. The final distributions for the three systems are in excellent mutual agreement, yielding, for a fixed value of $\lambda_{\rm int} = 1$,  $H_0 = 73.5^{+2.7}_{-2.8} \text{km s}^{-1}\text{Mpc}^{-1}$. In this present analysis, $\lint$ is allowed to be free and hierarchically constrained  with the information obtained from the spatially resolved kinematic data \citep{Knabel2026}, replacing the aperture-integrated kinematics used by \citet{Williams2026}.

\subsection{New JWST kinematic data for the lensed quasar sample}

New spatially resolved kinematic maps have been produced for five galaxies from the lensed quasar sample using JWST-NIRSpec IFU spectroscopy from Cycle 1 program JWST-GTO-1198 (PI: M.~Stiavelli, CoI: T.~Morishita) and Cycle 2 program JWST-GO-2974 (PI: T.~Treu, CoPI: A.~J.~Shajib). These data were presented for the first time as spatially integrated single-velocity dispersions by \cite{TDCOSMO2025}. Since then, the reduction and analysis have been improved to produce spatially resolved kinematics. Details regarding data reduction and kinematic extraction are presented for \PGeleven, \HEzero, and \WFItwenty by \cite{Knabel2026} and for \Bsixteen and \Jtwelve by \cite{Mozumdar2026}, building upon the methods described by \cite{Shajib25b}. Kinematics were extracted using the \textsc{squirrel}\footnote{\url{https://github.com/ajshajib/squirrel}} pipeline \citep{Shajib25b} built on the penalized pixel-fitting (\textsc{pPXF}) software program \citep{Cappellari04, cappellari17, cappellari22_ppxf}. Uncertainties and covariances are calculated following the methodology of \citet{Knabel25}.

Instead of annular bins, as utilized for \RXJ in \cite{TDCOSMO2025} in the context of their spherical dynamical models, we took spatial bins from the IFU datacubes (two spatial axes and one wavelength axis) using the PowerBin method~\citep{cappellari_powerbin}, an improvement over the Voronoi tessellation method used in earlier papers \citep{Shajib2023}. The spaxels (spatial pixels) were binned according to a target ``specific'' S/N (sS/N) \citep[see, e.g.,][]{Shajib25b} that was individually determined for each object through experimentation to sufficiently cover a useful range of radii from the center of the galaxy and achieve stable kinematic measurements without spurious extreme velocity dispersions that are indicative of systematic effects standing out above the optimal sS/N (around 30--40). Contamination from the background quasar and host galaxy light was explicitly handled as a fit component utilizing spatial masks to isolate those spectra and fit them before using them in the fitting of the deflector \citep[see][for details]{Knabel2026,Mozumdar2026}. 

\begin{figure*}[htbp]
    \includegraphics[width=\textwidth]{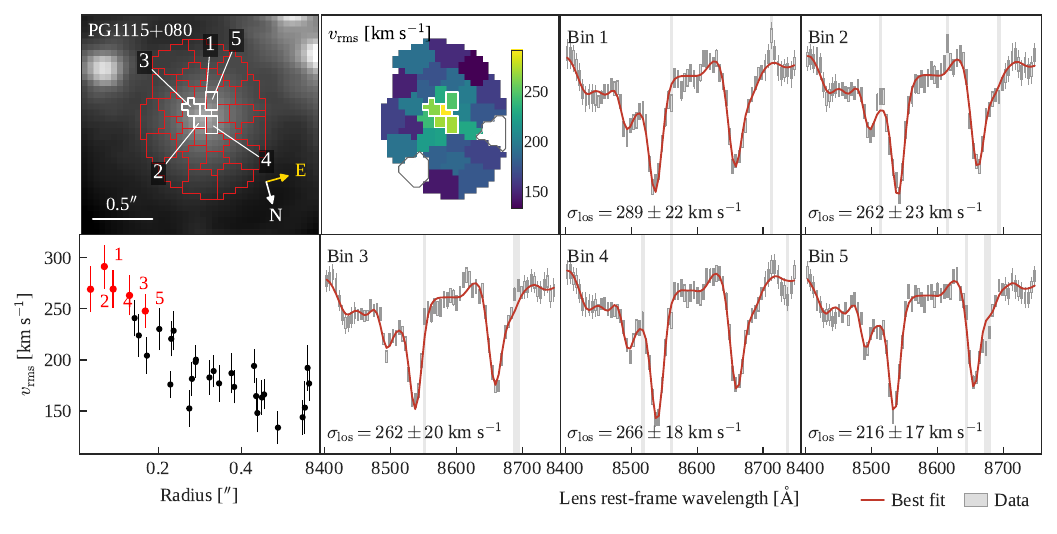}
    \caption{Example \textit{JWST}-NIRSpec IFU kinematics of \PGeleven from \cite{Knabel2026}. The top-left panel shows the white-light image of the data cube summed over wavelengths with the spatial binning overlaid. The $v_{\rm rms}$ map is shown in the top-middle-left, and the same profile is plotted against the radius of the luminosity-weighted bin center in the lower left. The other five panels show the spectra extracted from the corresponding bins identified in the white-light image, the kinematic map, and the radial profile. The wavelength range shown here is the one used for kinematic fitting.}
    \label{fig:example_model_fit}
\end{figure*}

One example of the reconstructed kinematic maps and the associated spectral fit in five bins is shown in \fref{fig:example_model_fit} for the lensed quasar \PGeleven. 

\subsection{New KCWI kinematic data for the SLACS sample}
\label{sec:kcwi_slacs}

spatially resolved kinematics for the non-time-delay lens SLACS sample \citep{Bolton06} were extracted from optical KCWI integral-field spectroscopy on the W.~M.~Keck Observatory \citep{morrissey12_kcwi,morrissey18_kcwi}. The data and kinematics were presented by \cite{Knabel24}, and joint lensing and dynamical models were produced by \cite{knabel26_slacs}. As for the time-delay lenses, mass density models were power-law profiles with a free MSD parameter $\lambda_\mathrm{int}$ for each individual object, and the dynamical models were solved for spherical and axisymmetric geometry using the JAM method~\citep{cappellari08,Cappellari2020}. The dynamical models preferred nearly isothermal power-law mass density profiles with a sample mean density slope of $\gamma_\mathrm{pl} = 2.04\pm0.02$ with intrinsic scatter of $0.08^{+0.03}_{-0.02}$ \citep[consistent with earlier studies, e.g.,][]{Koopmans2009,Auger2010}. The sample mean of the MSD parameter was found to be $\lambda_\mathrm{int}=1.01\pm0.03$, consistent with no significant deviation from unity at the sample level. The intrinsic scatter on $\lambda_\mathrm{int}$ is $0.11\pm0.03$. 

In addition to the SLACS KCWI observations presented by \citet{Knabel24}, we have new KCWI/KCRM \citep{KCRM} observations for three SLACS systems (SLACJ0959+0410, SLACJ1218+0830, and SLACJ0728+3835) and one SL2S system (SL2SJ0226-0420).  While the data are spatially resolved and will be used in the future to produce kinematic maps, only high-quality aperture-integrated kinematic measurements have been performed for these four systems at the moment. The spectra and kinematic fits are presented in Appendix~\ref{sec:unpub_spec}.

\subsection{External convergence}
\label{subsec:kappa_ext}

\begin{figure*}[!ht]
    \centering
    \includegraphics[width=\linewidth]{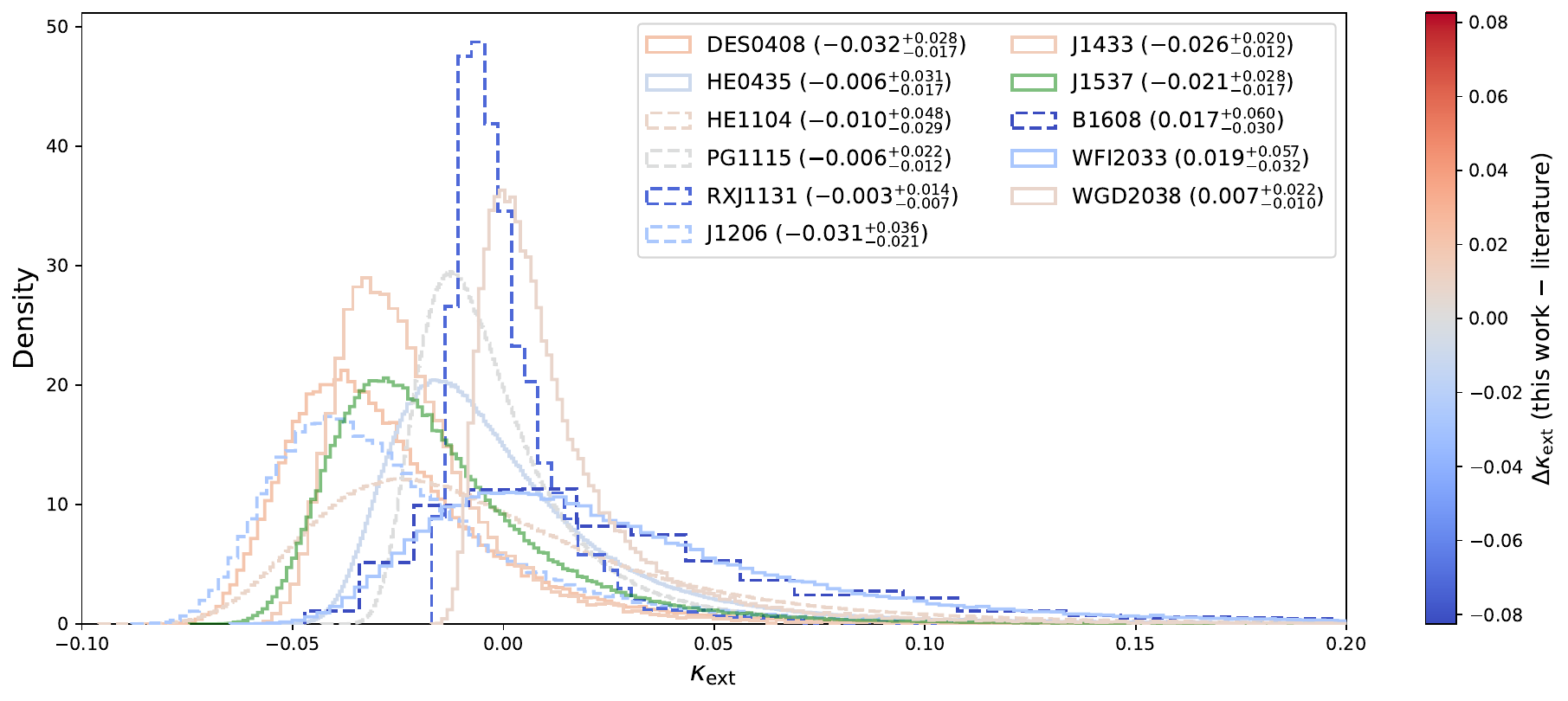}
    \caption{External convergence ($\kext$) estimates used in this work, from \citet{Johnson2026}. Solid lines correspond to systems for which photometric redshifts were used to weight the number counts, dotted lines for systems for which these were not available. The line colors correspond to the difference between the values used in this work and those estimated with previous techniques in the literature, with the exception of \Jfifteen, for which no such estimate had previously been carried out.}
    \label{fig:delta_kappa_ext_vs_lit}
\end{figure*}

Following previous TDCOSMO analyses, the external convergence of each lens system was estimated using the weighted number counts method, introduced by \cite{Suyu2010,Fassnacht_2010} and further developed by \cite{Suyu2013,Greene2013,Rusu2017,TDCOSMOIV,Wells2023,Wells2024}. We re-applied this method to each of our 11 lens systems, with three key updates with respect to previous analyses. Firstly, we make use of the \textit{Euclid} flagship simulation galaxy catalogs and convergence maps \cite{Castander_2025}, which offer a massive upgrade in simulation size compared to the Millennium simulation used historically, improved angular resolution, and a carefully calibrated relationship between the simulated galaxy catalog and the weak lensing convergence maps, which is ideal for our purposes. Secondly, using the method introduced by \cite{Johnson_2025}, we no longer approximate $\kext \approx \kappa_\mathrm{s}$, and explicitly include the contributions from the foreground and background terms $\kappa_\mathrm{d}$ and $\kappa_\mathrm{ds}$ (\eref{eq:los_breakdown}). Finally, given the challenge in identifying the origin of external shear \citep[see][]{Etherington2024}, we follow \cite{Buckley-Geer2020, Paic2026, sheu2026} and other recent analyses in omitting $\gamma_\mathrm{ext}$ as a weight by which simulated lines of sight are selected. The full details of the analysis choices and their application to this lens sample can be found in the paper by \citet{Johnson2026}. A summary of their results and a comparison to previous estimates found in the literature \citep[see][]{Suyu2010,Suyu2013,Rusu2017,Birrer2019,Chen2019,Rusu2020,Buckley-Geer2020,Paic2026,sheu2026} is shown in \fref{fig:delta_kappa_ext_vs_lit}. 

\section{Lens selection} \label{sec:sample_selection}

As in \TDCprev, we use the SLACS and SL2S galaxy--galaxy lenses to constrain the population-level parameters of the deflectors in the hierarchical analysis. Since these parameters are shared with the time-delay lenses and therefore indirectly affect cosmological inference, the external samples must be representative of the TDCOSMO lenses and have lensing and kinematic data of sufficient quality. We therefore apply a set of selection criteria designed to match the properties of the \TDCnew sample and retain only systems with ``cosmography-grade'' measurements. All \TDCnew lenses satisfy the quality criteria described below.

We first require the data products needed for the joint analysis: stellar kinematics, lens models, and a characterization of the LoS. For SLACS, we use the KCWI kinematics from \citet{Knabel24}, supplemented by new KCWI observations of SDSSJ0959+0410, SDSSJ1218+0830, and SDSSJ0728+3835; lens models from \citet{Tan24} and \citet{Shajib21}; and LoS measurements from \citet{TDCOSMOIV}. For SL2S, we use the LRIS, DEIMOS, and X-Shooter kinematics presented by \citet{Mozumdar2025}, supplemented by new KCWI observations of SL2SJ0226$-$0420; lens models from \citet{Tan24} and \citet{Sheu25}; and the LoS analysis of \citet{Wells2024}. After this pre-selection, the parent samples contain 16 SLACS and 27 SL2S lenses.

\paragraph{Lens morphology selection}
We require the projected light axis ratio of the deflector to satisfy $q_{\rm light} > 0.41$, matching the range spanned by the TDCOSMO sample. This criterion is less restrictive than the $q_{\rm light}>0.7$ cut adopted in \TDCprev. The previous threshold also served to remove flattened systems that could not be adequately described by the spherical dynamical models used in our previous analysis. In the present analysis, the dynamics are modeled explicitly under axisymmetry, allowing us to include flatter galaxies and to account for both ordered rotation and velocity dispersion. This criterion excludes 4 out of 27 SL2S lenses and no SLACS lenses.

\paragraph{Lensing information selection}
We follow the lensing-information selection of \TDCprev. Briefly, we use the lensing-information metric $\mathcal{I}$ introduced by \citet{Tan24}, which quantifies the information in the extended lensed arcs and correlates with the precision with which the mass-density slope can be inferred from imaging \citep{Sheu25}. For systems with unresolved kinematics, we require $\mathcal{I} > 150$.
This cut is not applied to systems with spatially resolved kinematics, since this information provides an additional direct constraint on the radial mass profile. This selection excludes 9 out of 27 SL2S systems, one of which is already excluded by the morphology criterion, corresponding to eight additional exclusions.

\paragraph{Kinematic data quality selection}
We adopt the same kinematic quality requirements as in \TDCprev, retaining only measurements for which the statistical and systematic uncertainties are sufficiently controlled for cosmological constraints. In particular, for the unresolved SL2S measurements, we require a spectral ${\rm S/N}>13\,\text{\AA}^{-1}$, for which the systematic uncertainty in the velocity dispersion is below $2\%$ \citep{Mozumdar2025}. 
In total, this criterion excludes 9 SL2S systems, four of which are already excluded by the preceding criteria, corresponding to five additional exclusions.

\paragraph{Velocity-dispersion selection}
As in \TDCprev, we use the lensing velocity dispersion, $\sigma_{\rm SIS}$, as a precisely measured proxy for deflector mass \citep[e.g.,][]{Treu2006} and restrict the external samples to the range covered by the TDCOSMO lenses,
$150 < \sigma_{\rm SIS}/({\rm km\,s^{-1}}) < 350$.
This ensures that the hierarchical inference is not informed by external lenses substantially outside the mass range of the time-delay deflectors. This cut excludes one SL2S system that is not already excluded by any of the preceding criteria.

\paragraph{Mass-density slope selection}
To exclude systems with mass profiles that differ substantially from those of TDCOSMO deflectors, we require that the power-law slope be within $1.5 < \gamma_{\rm pl} < 2.5$. This range is broader than that spanned by the TDCOSMO lenses and is intended to remove only significant outliers. The cut excludes one SLACS and two SL2S lenses, with one of the latter already excluded by the lensing-information criterion.

\paragraph{Environment selection}
We follow the environment selection of \TDCprev, designed to exclude systems whose local environment cannot be adequately described by the main lens model together with an external LoS contribution. In particular, we exclude lenses in strongly overdense environments or with significant nearby perturbers, for which the assumption that the remaining LoS contribution can be treated linearly may break down.
We characterize the local environment through the inverse-distance-weighted relative overdensity, $\zeta_{1/R}$, which compares the density of galaxies around each lens with that measured along random LoSs, giving greater weight to galaxies at smaller projected separations~\citep{Suyu2010,Greene2013}. We use the measurements of \citet{TDCOSMOIV} for SLACS and the LoS characterization of \citet{Wells2024} for SL2S, and apply the same overdensity criterion, $\zeta_{1/R}<2.10$, to both samples.

We additionally inspect for nearby perturbers that are not explicitly included in the lens model. No additional lenses were excluded by this criterion.

\begin{figure*}[htbp]
    \centering
    \includegraphics[width=0.85\linewidth]{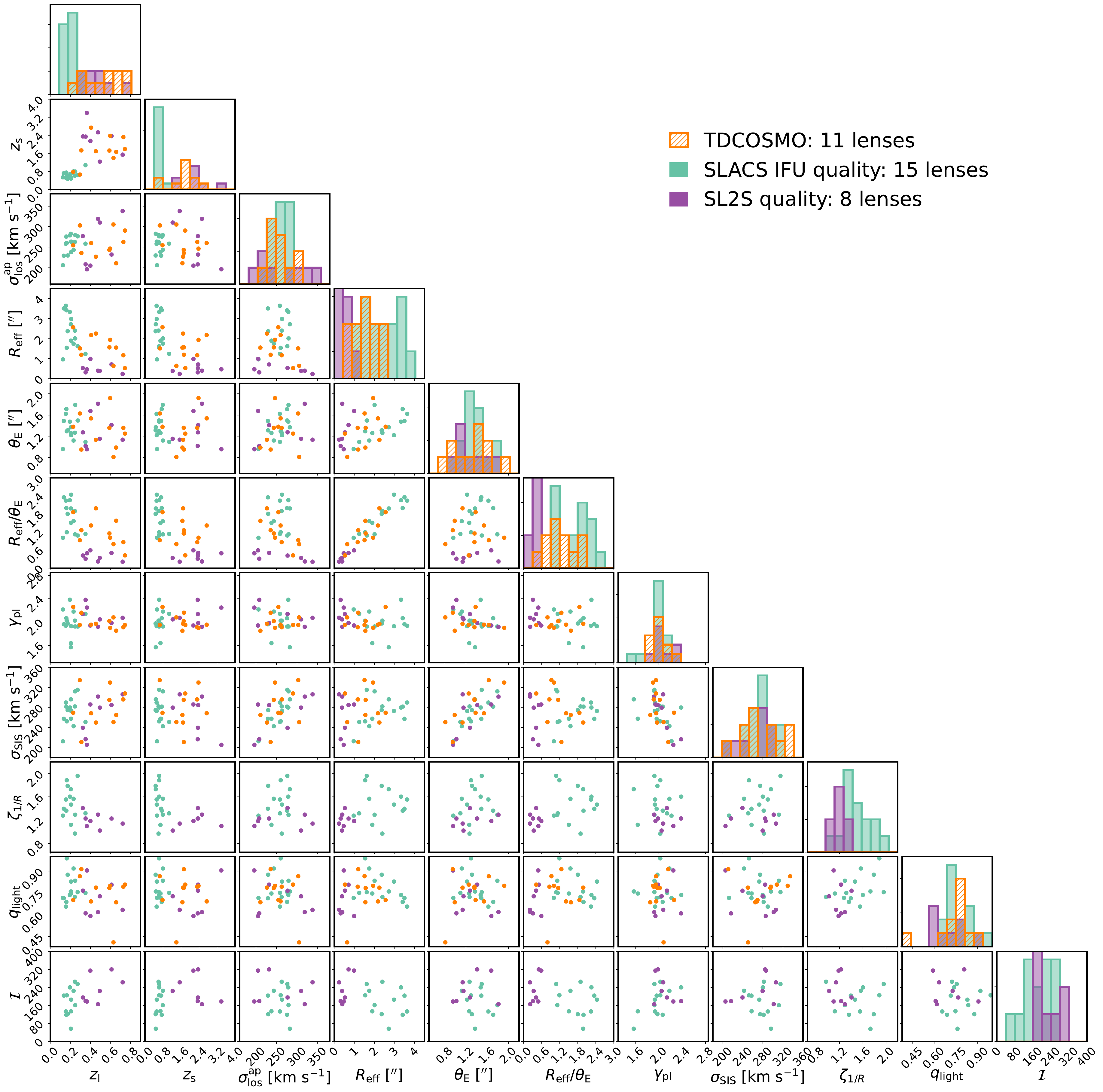}
    \caption{Properties of selected lens samples. The TDCOSMO, SLACS, and SL2S lenses shown here satisfy all the selection criteria described in Section~\ref{sec:sample_selection}. 
    The selected systems have the lensing, stellar-kinematic, and LoS information required for the joint analysis and satisfy the morphology, mass range, mass-density slope, and environment described therein.
    From left to right, we show the lens redshift, source redshift, aperture stellar velocity dispersion $\sigma^{\rm ap}_{\rm los}$, deflector half-light radius $R_{\rm eff}$, Einstein radius~$\theta_{\rm E}$, effective-to-Einstein radius ratio $R_{\rm eff}/\theta_{\rm E}$, power-law mass-density slope $\gamma_{\rm pl}$, SIS-equivalent velocity dispersion inferred from the Einstein radius $\sigma_{\rm SIS}$, relative LoS overdensity $\zeta_{1/R}$, projected light axis ratio $q_{\rm light}$, and lensing information $\mathcal{I}$, computed using the scaling parameters of \citet{Sheu25}.}
    \label{fig:sample_selection}
\end{figure*}

\subsection{Selected SLACS and SL2S lenses}

After applying the above criteria, our external sample consists of \nslacstot\ SLACS and \nsltwossel\ SL2S lenses. Their main properties are compared with those of the TDCOSMO lenses in \fref{fig:sample_selection}. Overall, the three samples span similar ranges in the quantities most relevant to our lensing and dynamical analysis. The main differences are in redshift and effective radius: the SLACS lenses and sources lie at systematically lower redshifts, while the effective radii follow the approximate ordering $R_{\rm eff}^{\rm SL2S}<R_{\rm eff}^{\rm TDCOSMO} <R_{\rm eff}^{\rm SLACS}$. The larger sizes of the lower-redshift SLACS galaxies are qualitatively consistent with the redshift evolution of the galaxy size--mass relation \citep{Vd10}. However, this does not fully explain the difference between SL2S and TDCOSMO, which occupy similar redshift ranges and follow the same Fundamental Plane within the uncertainties \citep{Mozumdar2025}. Differences in sample selection or in the photometric measurements may contribute to this offset. In particular, the SL2S light profiles used here are measured from Canada-France-Hawaii Telescope $r$-band imaging \citep{Sheu25}. Despite these differences, the distributions of the remaining observables show good agreement among the three samples.

Compared with \TDCprev, both the size and information content of the external samples have increased. The SLACS sample grows from 11 to \nslacstot\ lenses, of which \nslacsifusel\ have spatially resolved KCWI kinematics, and three have single-aperture measurements. Rather than compressing the IFU measurements into 6--7 radial annuli as in \TDCprev, we model the two-dimensional kinematic information with 25--114 spatial bins per lens. The SL2S sample increases from four to \nsltwossel\ lenses, primarily because the axisymmetric dynamical modeling allows us to relax the morphology cut adopted in \TDCprev. All selected SL2S lenses have the updated lens models of \citet{Sheu25}.

At the same time, the kinematic information available for the TDCOSMO lenses has substantially increased relative to the external samples. Nine out of 11 systems now have spatially resolved kinematics. Thus, while the external samples remain important for constraining the population-level properties of the deflectors, our hierarchical inference is increasingly informed by detailed kinematic information from the time-delay lenses themselves.

\section{Cosmological results in flat \lcdm}
\label{sec:flat}

Our baseline analysis choices were decided by the collaboration before unblinding. A consensus described our baseline analysis to include only the TDCOSMO lenses, with the DES-Dovekie supernova sample relative distance measurement, effectively serving as an $\Omega_{\rm m}$ prior \cite{DESDovekie}. We adopt axisymmetric JAM with spherically aligned anisotropy. We assume $q_{\rm obs} = q_{\rm m}$, i.e. adopting the mass axis ratio from the lens models as our measure of flattening. We also adopt a log-uniform prior on the scatter parameters $\sigma(\lint)$ and $\sigma(\sigma_{\rm t}/\sigma_{\rm r})$. We do not apply any automatic masking strategy to the kinematic maps, except for \Bsixteen, where we adopt $T(\boldsymbol{x})=20\%$ for the tidal disruption tolerance parameter defined by \citet{Mozumdar2026}. This lens has a close perturber that is included in the lens model but affects the dynamics of the main deflector. As a result, the outer bins cannot be properly modeled with JAM and are therefore masked out. We refer the reader to \citet{Mozumdar2026} for further details on this point. 

Under these assumptions, our baseline results yield
\begin{equation}
    H_0 = \hnotfiducial \, \mathrm{km}\, \mathrm{s}^{-1}\, \mathrm{Mpc}^{-1}, 
\end{equation}
at 3.3\% precision, improving our 5.0\% precision constraints from \TDCprev. This represents a 35\% reduction in the uncertainties, mainly due to the new lenses and the new kinematic data included in the analysis. In combination with the SLACS and SL2S lens samples, we obtain 
\begin{equation}
    H_0 = \hnotLCDMfoura \, \mathrm{km}\, \mathrm{s}^{-1}\, \mathrm{Mpc}^{-1}, 
\end{equation}
at 3.0\% precision. Replacing the DES-Dovekie $\Omega_{\rm m}$ prior with DESI-DR2 BAO, we obtain 
\begin{equation}
    H_0 = \hnotLCDMoneb \, \mathrm{km}\, \mathrm{s}^{-1}\, \mathrm{Mpc}^{-1},
\end{equation}
without the SLACS and SL2S lenses and
\begin{equation}
    H_0 = \hnotLCDMfourb \, \mathrm{km}\, \mathrm{s}^{-1}\, \mathrm{Mpc}^{-1},
\end{equation}
when including them. 

We emphasize that when combining TDCOSMO and BAO data, the time-delay distance serves as an anchor to calibrate the BAO distance, bypassing the need for CMB data or other constraints, as discussed below. To do this, we use the 13 relative DESI DR2 distance measurements and their covariances\footnote{\url{https://github.com/CobayaSampler/bao_data/tree/master/desi_bao_dr2}} across seven redshift bins spanning $z=0.295$--2.33. These distances constrain the transverse comoving and Hubble distances, $D_{\rm M}(z)$ and $D_{\rm H}(z)=c/H(z)$, relative to the sound horizon $r_{\rm d}$ through $D_{\rm M}/r_{\rm d}$ and $D_{\rm H}/r_{\rm d}$. To break the resulting $r_{\rm d}$--\Hc degeneracy, DESI adopts a prior on $r_{\rm d}$ from either Big Bang nucleosynthesis (BBN) or \textit{Planck} CMB data. Here, we do not use any informative prior on the sound horizon $r_{\rm d}$. Instead, it is inferred from the time-delay distances, and we obtain $r_{\rm d}=140.7 \pm 4.3$ Mpc, in $\sim 1.5\sigma$ tension with the \textit{Planck} value,  $r_{\rm d}=147.05 \pm 0.3$ Mpc \citep{Planck2020}. 

The constraints from strong lensing alone, without any external information on \Om, are listed as models U$\Lambda$CDM1--4 in \tref{tab:results_LCDM}, where \Om is sampled with a uniform prior, $\mathcal{U}(0.05, 0.5)$. For the \TDCnew sample, we find:
\begin{equation}
    H_0 = \hnotULCDMone \, \mathrm{km}\, \mathrm{s}^{-1}\, \mathrm{Mpc}^{-1},
\end{equation}
at 4.9\% precision. Since \Om is only weakly constrained by the lenses ($\Om = 0.356^{+0.103}_{-0.135}$, spanning most of the prior range), the external datasets mainly reduce the uncertainty in \Hc through the mild \Hc--\Om degeneracy of the time-delay distances, without shifting its central value. This lensing-only measurement is completely independent of the supernovae, BAO, and CMB data, and of the local distance ladder.

\begin{figure*}[h!]
    \centering
    \includegraphics[width=0.9\linewidth]{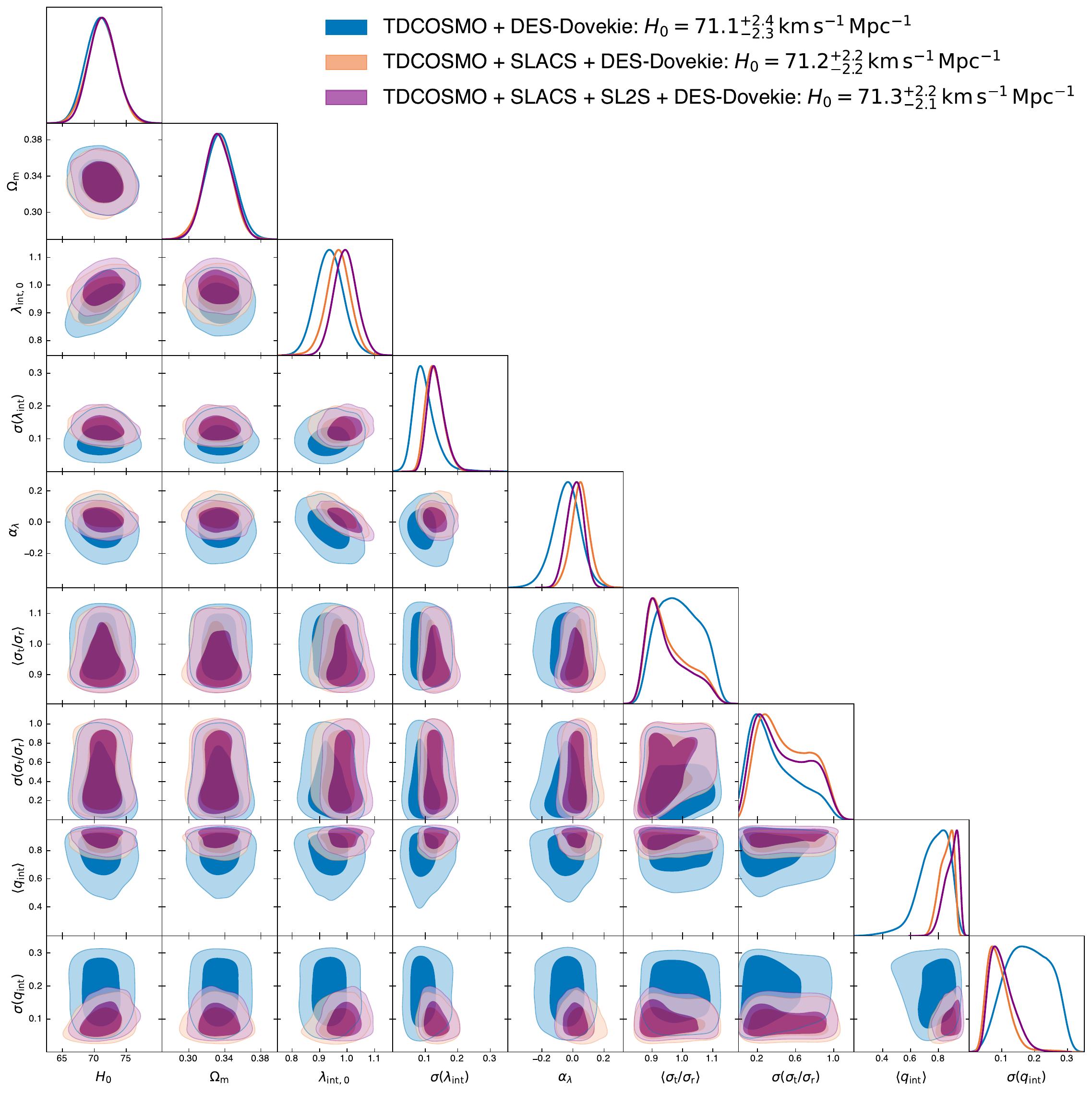}
    \caption{Posterior distributions for the population and cosmological parameters,  defined with their prior in \tref{tab:param_prior}. Individual lens power-law slopes $\gamma_{\mathrm{PL}, i}$ are also sampled but not shown on this plot. Blue: constraints from the 11 \TDCnew lenses. Orange: constraints from the \TDCnew sample + \nslacstot ~SLACS lenses, including \nslacsifusel ~with spatially resolved kinematic from Keck/KCWI. Purple: constraints from \TDCnew + SLACS + SL2S sample, including \nsltwossel\ additional lenses with aperture kinematics. All contours include \Om constraint from the DES-Dovekie sample \citep{DESDovekie}.}
    \label{fig:cornerplot_main_result}
\end{figure*}

The posterior distributions of all cosmological and population-level parameters are shown in \fref{fig:cornerplot_main_result} for the TDCOSMO, TDCOSMO + SLACS, and TDCOSMO + SLACS + SL2S data sets. Although the individual power-law slope, $\gamma_{\mathrm{pl}, i}$, for each lens is sampled for the TDCOSMO and SLACS lenses, they are not shown in this figure. The median, 16\textsuperscript{th}, and 84\textsuperscript{th} percentiles of the posterior distribution for \Hc, \Om and $\lint$ are listed in \tref{tab:results_LCDM}. The rest of the population parameters are presented in \aref{app:all_posteriors} and listed in \tref{tab:results_LCDM_all_parameters}.

\renewcommand*\arraystretch{1.5}
\setlength{\tabcolsep}{2pt}
\begin{table*}[]
\caption{\label{tab:results_LCDM}
Cosmological parameters in flat \lcdm.}
\begin{minipage}{\linewidth}
\centering
\begin{tabular}{lllccc}
\hline \hline
Model ID &
Experiments/datasets & Lensing datasets &
$H_{0}$ [km s$^{-1}$ Mpc$^{-1}$] &
$\Om$ & $\lambda_{\rm int}$ \\
\hline
U$\Lambda$CDM1 & TDCOSMO & TDCOSMO-2026 & $70.6^{+3.9}_{-3.1}$ & $0.356^{+0.103}_{-0.135}$ & $0.935^{+0.060}_{-0.055}$ \\
U$\Lambda$CDM2 & TDCOSMO & TDCOSMO-2026 + SLACS & $71.9^{+4.3}_{-3.4}$ & $0.306^{+0.133}_{-0.157}$ & $0.976^{+0.058}_{-0.050}$ \\
U$\Lambda$CDM3 & TDCOSMO & TDCOSMO-2026 + SL2S & $71.0^{+3.7}_{-3.0}$ & $0.374^{+0.090}_{-0.135}$ & $0.953^{+0.055}_{-0.048}$ \\
U$\Lambda$CDM4 & TDCOSMO & TDCOSMO-2026 + SLACS + SL2S & $72.1^{+4.1}_{-3.4}$ & $0.323^{+0.118}_{-0.144}$ & $1.002^{+0.050}_{-0.046}$ \\ \hline
$\Lambda$CDM1a & TDCOSMO + DES-Dovekie & TDCOSMO-2026 & $71.1^{+2.4}_{-2.3}$ & $0.335^{+0.016}_{-0.016}$ & $0.936^{+0.050}_{-0.049}$ \\
$\Lambda$CDM2 & TDCOSMO + DES-Dovekie & TDCOSMO-2026 + SLACS & $71.2^{+2.2}_{-2.2}$ & $0.332^{+0.016}_{-0.015}$ & $0.969^{+0.042}_{-0.043}$ \\
$\Lambda$CDM3 & TDCOSMO + DES-Dovekie & TDCOSMO-2026 + SL2S & $71.5^{+2.4}_{-2.3}$ & $0.334^{+0.015}_{-0.016}$ & $0.956^{+0.046}_{-0.042}$ \\
$\Lambda$CDM4a & TDCOSMO + DES-Dovekie & TDCOSMO-2026 + SLACS + SL2S & $71.3^{+2.2}_{-2.1}$ & $0.333^{+0.016}_{-0.015}$ & $0.993^{+0.042}_{-0.041}$ \\ \hline
$\Lambda$CDM1b & TDCOSMO + DESI BAO & TDCOSMO-2026 & $71.7^{+2.5}_{-2.3}$ & $0.298^{+0.009}_{-0.009}$ & $0.942^{+0.051}_{-0.047}$ \\
$\Lambda$CDM4b & TDCOSMO + DESI BAO & TDCOSMO-2026 + SLACS + SL2S & $72.2^{+2.2}_{-2.2}$ & $0.298^{+0.009}_{-0.008}$ & $0.999^{+0.038}_{-0.037}$ \\

\hline
\end{tabular}
\end{minipage}
\tablefoot{Reported values are medians, with errors corresponding to the 16\textsuperscript{th} and 84\textsuperscript{th} percentiles.}
\end{table*}
\setlength{\tabcolsep}{6pt}
\renewcommand*\arraystretch{1.0}

\renewcommand*\arraystretch{1.5}
\setlength{\tabcolsep}{2pt}
\begin{table*}[hbt!]
\caption{Cosmological parameters for extensions to the \lcdm model, from time-delay cosmography alone, or combined with other probes. \label{tab:results_LCDM_extension}}
\begin{minipage}{\linewidth}
\centering
\begin{tabular}{llcccccccc}
\hline \hline
Model &
Experiments/datasets &
$H_{0}$ [km s$^{-1}$ Mpc$^{-1}$] &
$\Om$ &
$\Ok$ &
$w$ or $w_{0}$ &
$w_{a}$
\\
\hline
Uo$\Lambda$CDM & TDCOSMO & $72.3^{+3.7}_{-3.7}$ & $0.339^{+0.113}_{-0.146}$ & $-0.015^{+0.323}_{-0.309}$ & -- & -- \\
o$\Lambda$CDM & TDCOSMO + \textit{Planck} & $66.5^{+1.6}_{-1.5}$ & $0.324^{+0.017}_{-0.013}$ & $-0.003^{+0.004}_{-0.004}$ & -- & -- \\ \hline
U$w$CDM & TDCOSMO & $67.9^{+6.1}_{-6.0}$ & $0.285^{+0.151}_{-0.157}$ & -- & $-0.62^{+0.50}_{-0.53}$ & -- \\
$w$CDM1 & TDCOSMO + \textit{Planck} & $78.0^{+3.4}_{-4.5}$ & $0.237^{+0.031}_{-0.022}$ & -- & $-1.33^{+0.15}_{-0.12}$ & -- \\
$w$CDM2 & TDCOSMO + DESI BAO & $71.3^{+2.4}_{-2.3}$ & $0.297^{+0.009}_{-0.009}$ & -- & $-0.91^{+0.07}_{-0.08}$ & -- \\
$w$CDM3 & TDCOSMO + DES-Dovekie & $71.4^{+2.2}_{-2.1}$ & $0.266^{+0.065}_{-0.088}$ & -- & $-0.84^{+0.15}_{-0.15}$ & -- \\ 
$w$CDM4 & TDCOSMO + DESI BAO + DES-Dovekie & $71.6^{+2.1}_{-2.1}$ & $0.297^{+0.009}_{-0.008}$ & -- & $-0.91^{+0.04}_{-0.04}$ & -- \\ \hline
U$w_0w_a$CDM & TDCOSMO & $69.8^{+5.9}_{-6.1}$ & $0.347^{+0.103}_{-0.151}$ & -- & $-0.63^{+0.61}_{-0.58}$ & $-1.93^{+3.05}_{-4.35}$ \\
$w_0w_a$CDM1 & TDCOSMO + \textit{Planck} + BAO$^\dagger$& $65.2^{+2.7}_{-2.4}$ & $0.338^{+0.026}_{-0.028}$ & -- & $-0.56^{+0.26}_{-0.31}$ & $-1.38^{+0.85}_{-0.78}$ \\
$w_0w_a$CDM2 & TDCOSMO + DESI BAO & $66.1^{+4.0}_{-3.0}$ & $0.380^{+0.037}_{-0.052}$ & -- & $-0.26^{+0.36}_{-0.45}$ & $-2.46^{+1.57}_{-1.27}$ \\
$w_0w_a$CDM3 & TDCOSMO + DES-Dovekie & $70.6^{+2.1}_{-2.1}$ & $0.454^{+0.029}_{-0.062}$ & -- & $-0.64^{+0.22}_{-0.19}$ & $-5.52^{+3.09}_{-2.57}$ \\ 
$w_0w_a$CDM4 & TDCOSMO + DESI BAO + DES-Dovekie & $70.6^{+2.2}_{-2.1}$ & $0.312^{+0.013}_{-0.016}$ & -- & $-0.85^{+0.07}_{-0.06}$ & $-0.48^{+0.46}_{-0.44}$ \\ \hline
U$w_\phi$CDM & TDCOSMO & $66.5^{+6.3}_{-5.6}$ & $0.316^{+0.125}_{-0.142}$ & -- & $-0.41^{+0.55}_{-0.66}$ & -- \\
$w_\phi$CDM1 & TDCOSMO + \textit{Planck} & $72.7^{+2.8}_{-3.9}$ & $0.274^{+0.032}_{-0.022}$ & -- & $-1.31^{+0.23}_{-0.14}$ & -- \\
$w_\phi$CDM2 & TDCOSMO + DESI BAO & $71.1^{+2.3}_{-2.4}$ & $0.307^{+0.012}_{-0.011}$ & -- & $-0.85^{+0.11}_{-0.12}$ & -- \\
$w_\phi$CDM3 & TDCOSMO + DES-Dovekie & $71.1^{+2.2}_{-2.1}$ & $0.286^{+0.043}_{-0.044}$ & -- & $-0.84^{+0.12}_{-0.13}$ & -- \\
$w_\phi$CDM4 & TDCOSMO + DESI BAO + DES-Dovekie & $71.1^{+2.3}_{-2.0}$ & $0.304^{+0.007}_{-0.008}$ & -- & $-0.88^{+0.04}_{-0.05}$ & -- \\ \hline
\hline
\end{tabular}

\tablefoot{Reported values are medians, with errors corresponding to the 16\textsuperscript{th} and 84\textsuperscript{th} percentiles. In this table, we utilized all lensing datasets, i.e. the TDCOSMO-2026, SLACS and SL2S lens samples. $^\dagger$ Planck chains in $w_0w_a$ cosmology from \cite{Planck2020} include BAO data from 6dFGS, SDSS-MGS, BOSS LOWZ and CMASS-DR11. }

\end{minipage}
\end{table*}
\setlength{\tabcolsep}{6pt}
\renewcommand*\arraystretch{1.0}

\section{Alternative cosmological models}
\label{sec:alt}

The measurement presented so far assumes a flat $\Lambda$CDM cosmology. This assumption enters the inference through the distance--redshift relation used to convert the measured time-delay and angular-diameter distances into cosmological parameters; the lens modeling, kinematic modeling, and hierarchical treatment of the deflector population are otherwise independent of the choice of cosmological model.\footnote{Two notable exceptions are the use of the Euclid Flagship simulation for our \kext estimate, which assumes a flat-\lcdm Universe, and the explicit treatment of multi-plane perturbers in some of our lens systems, which requires distance ratios involving the main lens plane. For the latter, \citet{Rusu2020} showed for \WFItwenty, and \citet{Wong2017} for \HEzero, that letting $\Omega_{\rm m}$ and $w$ vary in the multi-plane ray tracing changes the effective \Ddt by less than 1\%, well below our per-lens uncertainties. We expect the effect on \kext to be similarly small.}

We can therefore repeat the inference under a different expansion history at no additional modeling cost. We consider four extensions: non-flat $\Lambda$CDM model, dark energy with a constant equation of state, dark energy with standard $(w_0, w_a)$ parameterization, and dark energy scalar-field model. In every case, the deflector population parameters, their priors, and the lens sample are exactly those of the baseline analysis, so that the differences below are entirely due to the cosmological model.

As time-delay cosmography provides a direct measurement of an absolute distance scale, it constrains $H_0$ far better than it constrains the shape of the expansion history. Therefore, the parameters describing the dark-energy sector are less well informed by our data. However, the information content of time-delay cosmography is complementary to the relative-distance probes, e.g. the BAO and type-Ia supernovae \citep{Linder11}, which constrain the expansion history well but calibrate the absolute scale only through additional assumptions. Combinations are therefore considerably more informative than either alone. We combine our sample with the \textit{Planck} CMB likelihood \citep{Planck2020}, the DESI-DR2 BAO \citep{DESIDR2} and the DES-Dovekie supernova sample~\citep{DESDovekie}.

The results are presented in \tref{tab:results_LCDM_extension}.

\subsection{Non-flat \lcdm}
\label{sec:oLCDM}

The first extension relaxes flatness while keeping dark energy as a cosmological constant, adding the curvature density $\Omega_k$ as a free parameter with $\Omega_k \sim \mathcal{U}(-0.5, 0.5)$ and requiring $\Omega_\Lambda = 1 - \Omega_{\rm m} - \Omega_k > 0$. 

From the lensing sample alone, we find $H_0 = 72.3^{+3.7}_{-3.7}$ \ksmpc and $\Omega_k = -0.015^{+0.323}_{-0.309}$.  Curvature is, for time-delay cosmography alone, almost unconstrained, as our lenses sit at low enough redshift that $\Omega_k$ barely alters the measured \Ddt. Therefore, $\Omega_k$ itself is left essentially unmeasured, with a posterior spanning most of the prior range, but \Hc remains well measured with 5.1\% precision.

\begin{figure}[h!]
    \centering
    \includegraphics[width=\linewidth]{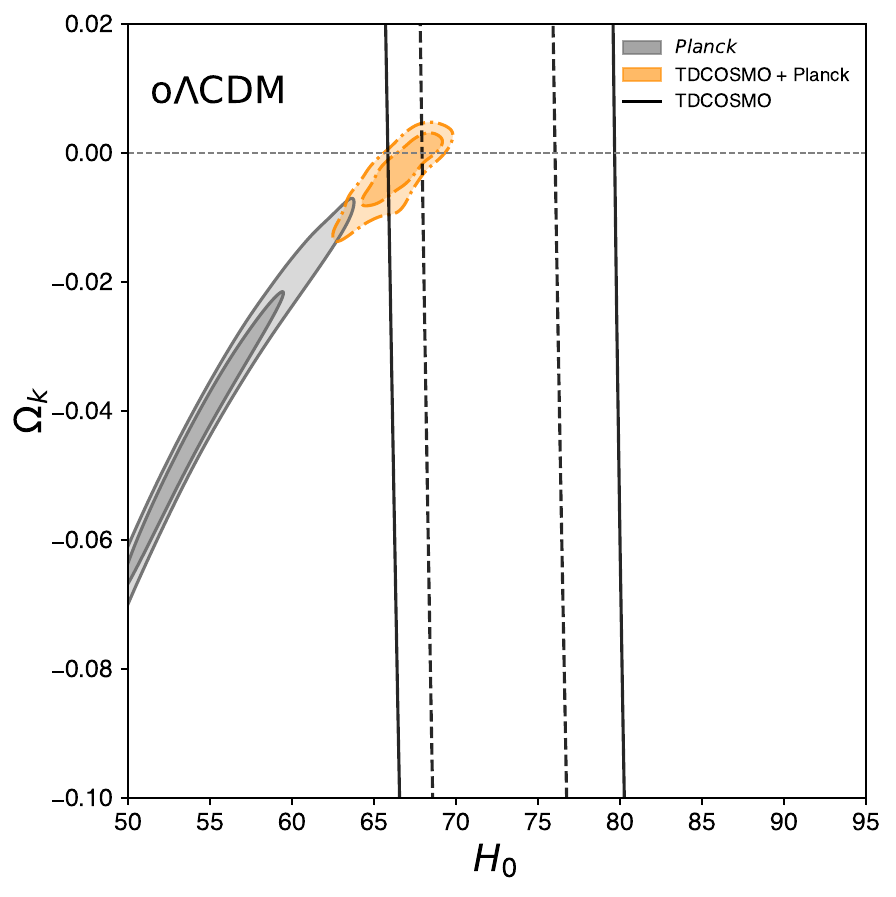}
    \caption{Constraints on \Hc and $\Omega_k$ in non-flat \lcdm. Black unfilled contours show the time-delay lenses alone (Uo\lcdm; dashed and solid lines enclose the 68\% and 95\% credible regions), gray filled contours \textit{Planck} alone \citep{Planck2020}, and orange filled contours their combination. The dashed horizontal line marks a flat universe, $\Omega_k = 0$. \textit{Planck} constrains a narrow, nearly degenerate combination of \Hc and $\Omega_k$, whereas the lensing constraint on \Hc is almost independent of $\Omega_k$. The two datasets are in $\sim 3.4\sigma$ tension in \Hc (\sref{sec:oLCDM}), so the joint constraint should be interpreted with caution.}
    \label{fig:oLCDM}
\end{figure}

The independence of the time-delay-inferred $H_0$ on $\Omega_k$ makes the combination with the CMB interesting. The CMB constrains a tight, nearly degenerate combination of $H_0$ and $\Omega_k$: the geometric degeneracy, which is visible as the narrow gray band in \fref{fig:oLCDM}. An independent \Hc measurement breaks that degeneracy, and the joint TDCOSMO + \textit{Planck} constraint yields $H_0 = 66.5^{+1.6}_{-1.5}$ \ksmpc with a tight bound on the curvature parameter: $\Omega_k = -0.003^{+0.004}_{-0.004}$. These results should, however, be read while keeping in mind that combining two datasets is only meaningful when they are statistically consistent within the model being fitted. In contrast, the two datasets are in significant tension: \textit{Planck} alone prefers a closed universe with $H_0 = 54.2^{+3.8}_{-3.4}$\ksmpc and $\Omega_k = -0.043^{+0.016}_{-0.018}$, while our lensing-only constraint is $H_0 = 72.3^{+3.7}_{-3.7}$\ksmpc. The tension between the two posteriors is $3.4\sigma$ in $H_0$ alone and $\sim 4\sigma$ in the full $(H_0, \Om, \Omega_k)$ space. This is, in some sense, a manifestation of the Hubble tension, extended to non-flat \lcdm.

\subsection{Flat $w$CDM}

We next restore flatness and let the dark-energy equation of state take a constant value $w \neq -1$, with $w \sim \mathcal{U}(-1.5, 0.5)$ and the baseline priors on $H_0$ and $\Omega_{\rm m}$ unchanged. The cosmological constant corresponds to $w = -1$.

Freeing $w$ creates a degeneracy with $H_0$, since a more negative equation of state can be compensated by a higher expansion rate over the redshift range spanned by our lenses. Our lensing-only constraint, $H_0 = 67.9^{+6.1}_{-6.0}$ \ksmpc, is weaker than in flat $\Lambda$CDM. The fractional error bar grows from $5.2\%$ to $8.9\%$ and $w$ itself is poorly determined, $w = -0.62^{+0.50}_{-0.53}$.

\begin{figure}[h!]
    \centering
    \includegraphics[width=\linewidth]{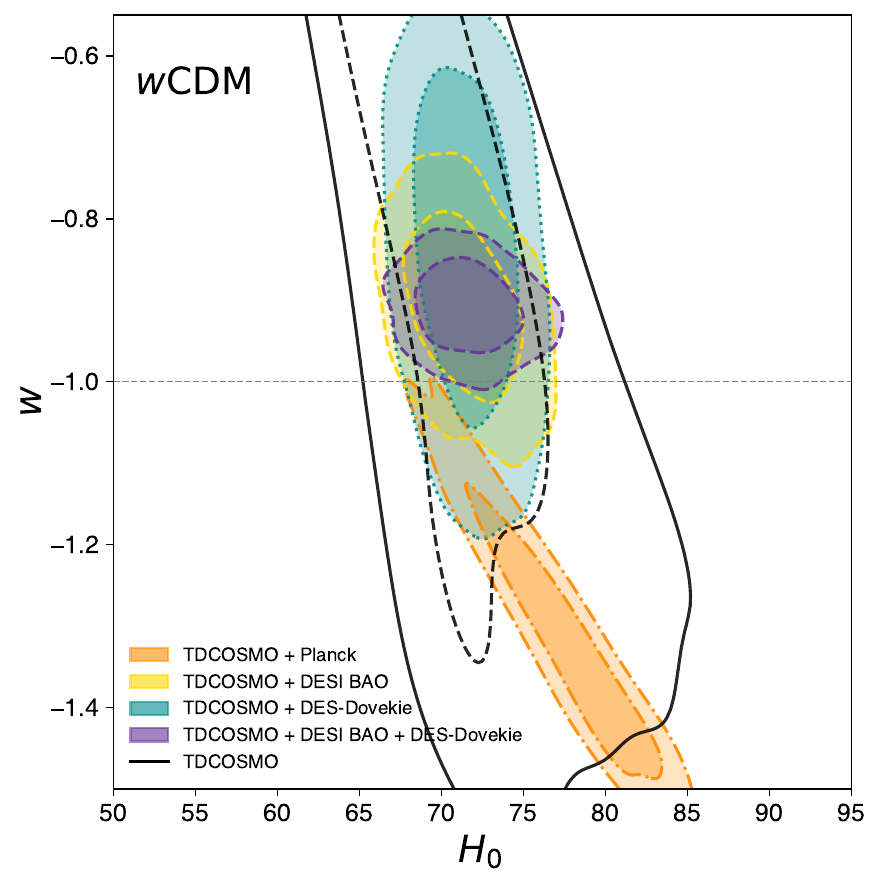}
    \caption{Posterior distributions of $H_0$ and $w$ in a flat $w$CDM cosmology. The constraints are obtained by combining the \TDCnew, SLACS, and SL2S lens samples, with i) \textit{Planck}, ii) DESI-DR2 BAO, iii) DES-Dovekie, and iv) the combination of those two datasets. The corresponding contours enclose the 1$\sigma$ and 2$\sigma$ credible regions. The combination with BAO and SN data disfavor a cosmological constant at $2.3\sigma$. }
    \label{fig:wCDM}
\end{figure}

Adding relative distance measurements, such as the BAO or the supernova data, restores the precision on $H_0$ and returns an equation of state compatible with a cosmological constant: $H_0 = 71.3^{+2.4}_{-2.3}$ \ksmpc with $w = -0.91^{+0.07}_{-0.08}$ for DESI BAO, and $H_0 = 71.4^{+2.2}_{-2.1}$ \ksmpc with $w = -0.84^{+0.15}_{-0.15}$ for DES-Dovekie. Both sit within about $1\sigma$ of $w = -1$. The fractional error on \Hc is
$3.3\%$ and $3.0\%$ respectively, close to the precision achieved in \lcdm despite the extra degree of freedom. Combining both datasets with TDCOSMO yields $H_0 = 71.6^{+2.1}_{-2.1}$ \ksmpc and $w = -0.91^{+0.04}_{-0.04}$, disfavoring $w = -1$ at $2.3\sigma$. All posterior distribution functions of \Hc and $w$ in $w$CDM cosmology are displayed in \fref{fig:wCDM}. 

The combination with {\it Planck} behaves differently. It prefers $w = -1.33^{+0.15}_{-0.12}$, in $2.3\sigma$ tension with a cosmological constant and a Hubble constant of $H_0 = 78.0^{+3.4}_{-4.5}$ \ksmpc. However, this result is rather another manifestation of the $H_0$ tension than independent evidence for phantom dark energy. Within $w$CDM, a phantom equation of state is one way to reconcile the acoustic scale with a high local expansion rate. In fact, the fit moves along that known degeneracy. We report this combination with \textit{Planck} as a diagnostic of the tension. We do not regard it as a measurement of $w$.

\subsection{Flat $w_0w_a$CDM}

A more flexible description lets the equation of state vary with redshift. We adopt the standard two-parameter form \citep{Chevallier2001,Linder2003}
\begin{equation}
  w(z) = w_0 + w_a \, \frac{z}{1+z},
  \label{eq:cpl}
\end{equation}
with uniform priors $w_0 \sim \mathcal{U}(-1.5, 0.5)$ and $w_a \sim \mathcal{U}(-10,10)$. The parameterization is phenomenological: it is a convenient expansion rather than a physical model, and it does not enforce any energy condition on the dark-energy
component.

On its own, our sample does not provide useful constraints: we obtain $w_0 = -0.63^{+0.61}_{-0.58}$ and $w_a = -1.93^{+3.05}_{-4.35}$, both dominated by the priors, while $H_0$ remains measurable at $H_0 = 69.8^{+5.9}_{-6.1}$ \ksmpc, though with an uncertainty inflated by the extra freedom in the cosmological model.

\begin{figure}[h!]
    \centering
    \includegraphics[width=\linewidth]{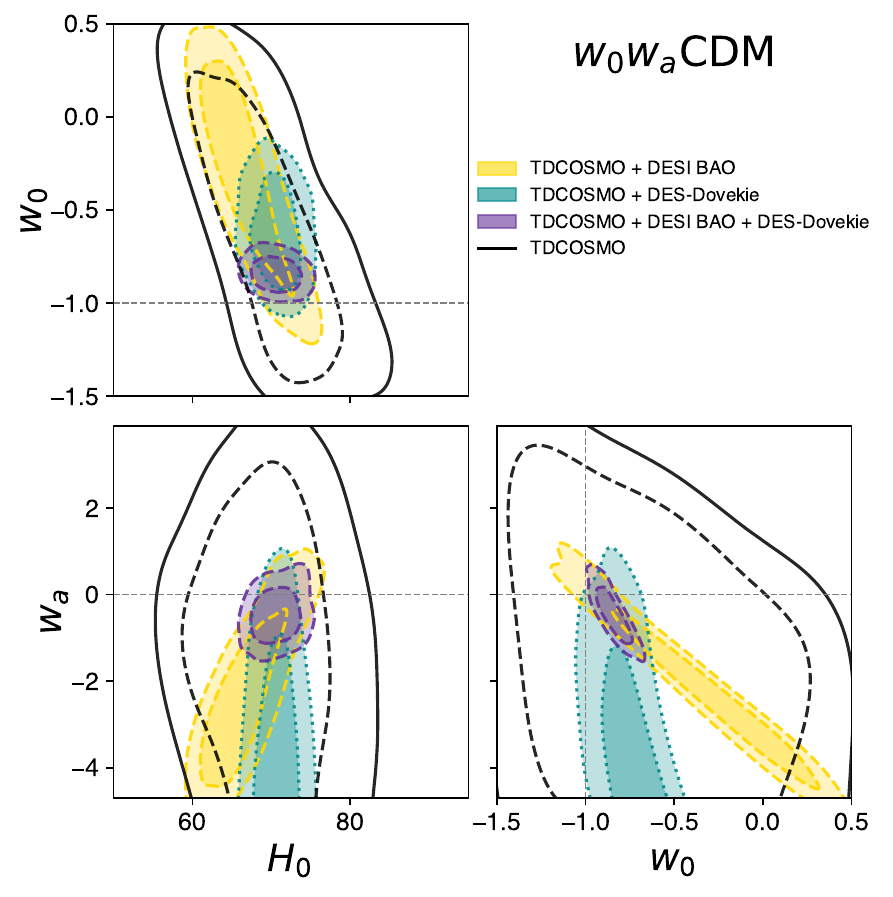}
    \caption{Posterior distributions of $H_0$, $w_0$, and $w_a$ in a flat $w_0w_a$CDM cosmology. The constraints are obtained by combining the \TDCnew, SLACS, and SL2S lens samples, with DESI-DR2 BAO, DES-Dovekie and the combination of those two datasets. The corresponding contours are indicated by the colors specified in the legend and enclose the 1$\sigma$ and 2$\sigma$ credible regions.}
    \label{fig:w0waCDM}
\end{figure}

Combined individually with the CMB, the BAO or the supernova data, the results stay consistent with a cosmological constant within the quoted intervals: $w_0 = -0.56^{+0.26}_{-0.31}$, $w_a = -1.38^{+0.85}_{-0.78}$ with \textit{Planck} (whose public $w_0w_a$CDM chains include earlier BAO data, see \tref{tab:results_LCDM_extension}); $w_0 = -0.26^{+0.36}_{-0.45}$, $w_a = -2.46^{+1.57}_{-1.27}$ with DESI BAO; and $w_0 = -0.64^{+0.22}_{-0.19}$, $w_a = -5.52^{+3.09}_{-2.57}$ with DES-Dovekie. The supernova combination drifts furthest from $\Lambda$CDM ($2.0\sigma$ in the $w_0$--$w_a$ plane), and does so in a suspicious direction as it pairs its strongly negative $w_a$ with a very large $\Omega_{\rm m} = 0.454^{+0.029}_{-0.062}$, well above its expected value from BAO and CMB. This solution is a compromise between a high matter density and a rapidly evolving equation of state. The posterior distribution for \Hc, $w_0$, and $w_a$ in these different combinations of probes is shown in \fref{fig:w0waCDM}. 

Including both DES-Dovekie and DESI BAO data, we find $w_0 = -0.85^{+0.07}_{-0.06}$ and $w_a = -0.48^{+0.46}_{-0.44}$, with $\Omega_{\rm m} = 0.312^{+0.013}_{-0.016}$, in $2.4\sigma$ tension with \lcdm in the $w_0$--$w_a$ plane. Here $w_0$ is significantly above $-1$ ($w_0 < -1$ is excluded at $2.5\sigma$), while $w_a$ remains consistent with zero at about $1\sigma$. The reported preference for dynamical dark energy in recent analyses emerges only when several relative-distance probes are combined. Without CMB data, we find no evidence for $w_a \neq 0$, consistent with the finding of \citet{DESIDR2} that the preference for dynamical dark energy strengthens when CMB data are included. However, our time-delay cosmography lens sample is not large enough to significantly constrain the expansion history $H(z)$. The real gain of combining our data with relative distance probes in $w_0w_a$CDM cosmology is that it provides an absolute calibration that is independent of the sound horizon, and in that combination, we find $H_0 = 70.6^{+2.2}_{-2.1}$ \ksmpc, a $3.0\%$ measurement. Remarkably, the precision is similar to that achieved in $\Lambda$CDM even though there are more free parameters.

\begin{figure}
    \centering
    \includegraphics[width=\linewidth]{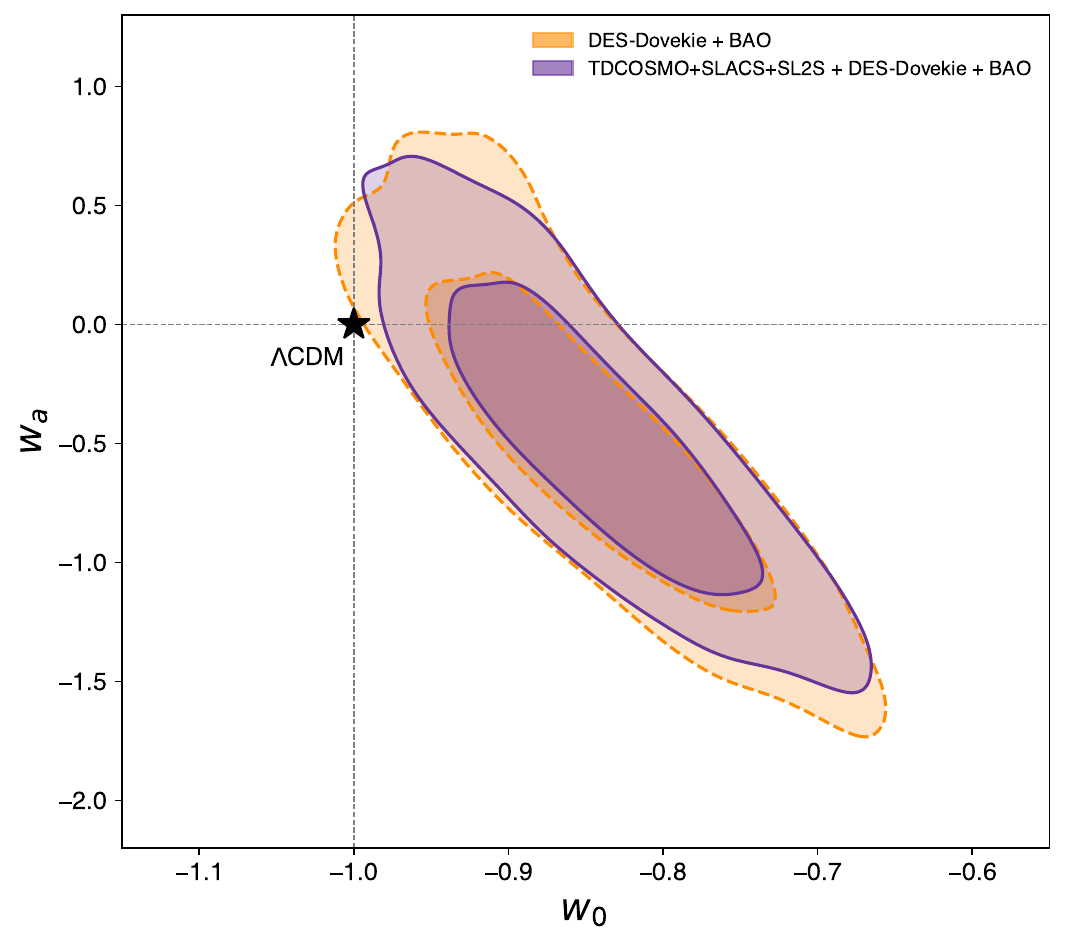}
    \caption{Constraints in the $w_0$--$w_a$ plane of flat $w_0w_a$CDM from DES-Dovekie and DESI-DR2 BAO, without (orange, dashed) and with (purple, solid) the \TDCnew, SLACS and SL2S lens samples. Contours enclose the 68\% and 95\% credible regions. Although time-delay cosmography is mainly sensitive to \Hc, adding the lens samples  tightens the constraint along the $w_0$--$w_a$ degeneracy, increasing the figure of merit from 51 to 60. Without any CMB data, the tension with \lcdm is at $2.4\sigma$ in the $w_0$--$w_a$ plane.}
    \label{fig:w0wa_information gain}
\end{figure}

A competitive measurement of the dark-energy equation of state and its redshift evolution with time-delay cosmography will only come with a larger sample of lenses, e.g. from \textit{Euclid}, Rubin/LSST and Roman. The larger samples will allow a much finer sampling of the \Ddt -- ($z_{\rm d}$, $z_{\rm s}$) relation \citep{Erickson26}. For now, with 11 time-delay lenses and 23 non-time-delay lenses, the dark-energy figure of merit, ${\rm FoM} = 1/ \sqrt{\det {\rm Cov}(w_0, w_a)}$, with and without the lenses, improves marginally from 51.2 to 59.8 (+17\%) when the lenses are combined with DES-Dovekie and DESI DR2, compared to DES-Dovekie and DESI DR2 alone (see \fref{fig:w0wa_information gain}).

\subsection{Flat $w_\phi$CDM}

Finally, we consider a physically motivated alternative in which dark energy is a slowly rolling scalar field, such as the pseudo-Nambu-Goldstone bosons, rather than a free function of redshift \citep{Frieman95, Shajib25c}. For the class of potentials considered there, the equation of state is described by
\begin{equation}
  w(z) = -1 + (1 + w_0)\,\mathrm{e}^{-\alpha z},
  \label{eq:wphi}
\end{equation}
where $w_0$ is its present-day value and $\alpha$ is set by the shape of the potential. Since $\alpha$ is confined to a narrow range by the underlying physics, we sample it with $\alpha \sim \mathcal{U}(1.35, 1.55)$ and $w_0 \sim \mathcal{U}(-1.5, 0.5)$. No current dataset is sensitive enough to constrain $\alpha$ within that interval, so in practice, this model behaves as a single-parameter extension of $\Lambda$CDM, with the important difference that its parameter space respects the null energy condition by construction.

\begin{figure}[h!]
    \centering
    \includegraphics[width=\linewidth]{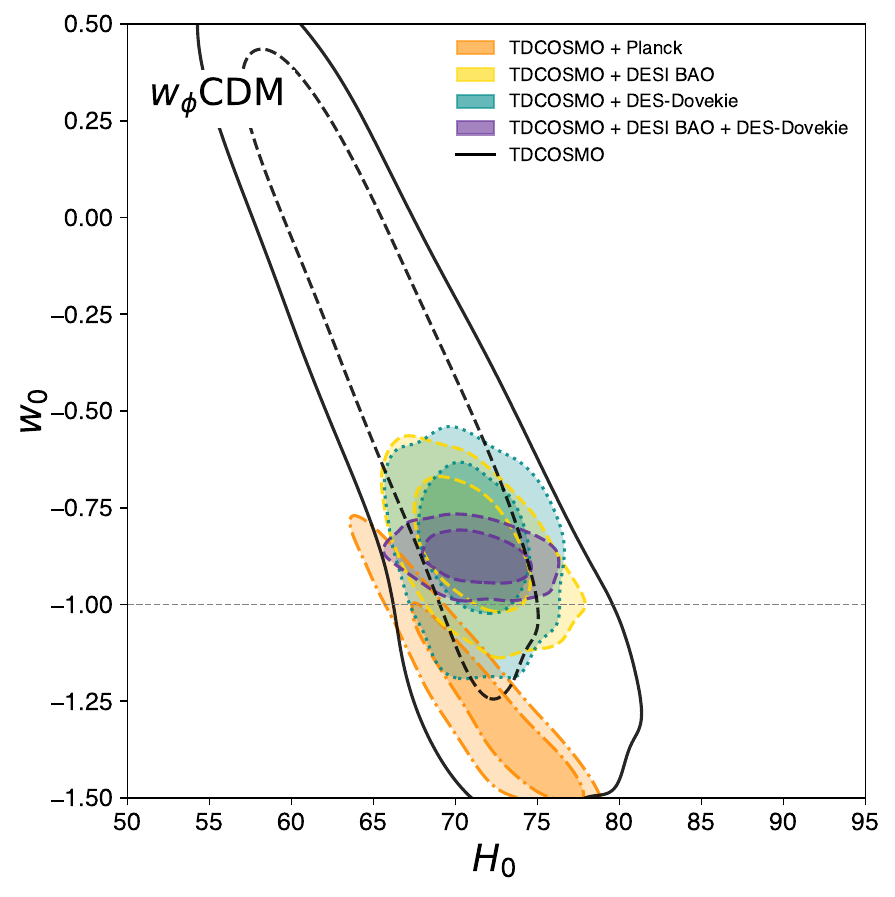}
    \caption{Posterior distributions of $H_0$ and $w_0$ in a flat $w_{\phi}$CDM cosmology. The constraints are obtained by combining the \TDCnew, SLACS, and SL2S lens samples, with \textit{Planck}, DESI-DR2 BAO, DES-Dovekie, and the combination of those last two datasets. The corresponding contours corresponds to the 1$\sigma$ and 2$\sigma$ credible regions.}
    \label{fig:wphiCDM}
\end{figure}

The lensing-only constraint is $H_0 = 66.5^{+6.3}_{-5.6}$ \ksmpc with
$w_0 = -0.41^{+0.55}_{-0.66}$, again prior-dominated. The combinations with BAO and the supernova data give $w_0 = -0.85^{+0.11}_{-0.12}$ and $w_0 = -0.84^{+0.12}_{-0.13}$, respectively. One feature of this model is its ability to cross-check the results found in the phenomenological CPL parameterization. Combining DESI BAO and DES-Dovekie, we find $w_0 = -0.88^{+0.04}_{-0.05}$ ($w_0 < -1$ excluded at $2.4\sigma$) and $H_0 = 71.1^{+2.3}_{-2.0}$ \ksmpc, statistically indistinguishable from $w_0 = -0.85^{+0.07}_{-0.06}$ with the CPL parameterization using the same data. We see that there is a slight preference for dynamical dark energy, regardless of the adopted parameterization.  

The combination with the CMB is again the exception and resembles the $w$CDM case: it returns $w_0 = -1.31^{+0.23}_{-0.14}$, favoring $w_0 < -1$ at $1.4\sigma$. A present-day value $w_0 < -1$ in \eqref{eq:wphi} implies $w(z) < -1$ at all redshifts, which no minimally coupled scalar field can produce. This combination should therefore be read as the same statement of tension seen in $w$CDM, and not taken to mean that our data favor phantom dark energy.

\section{Analysis variations}
\label{sec:analysis variation}

\begin{figure*}
    \centering
    \includegraphics[width=0.8\linewidth]{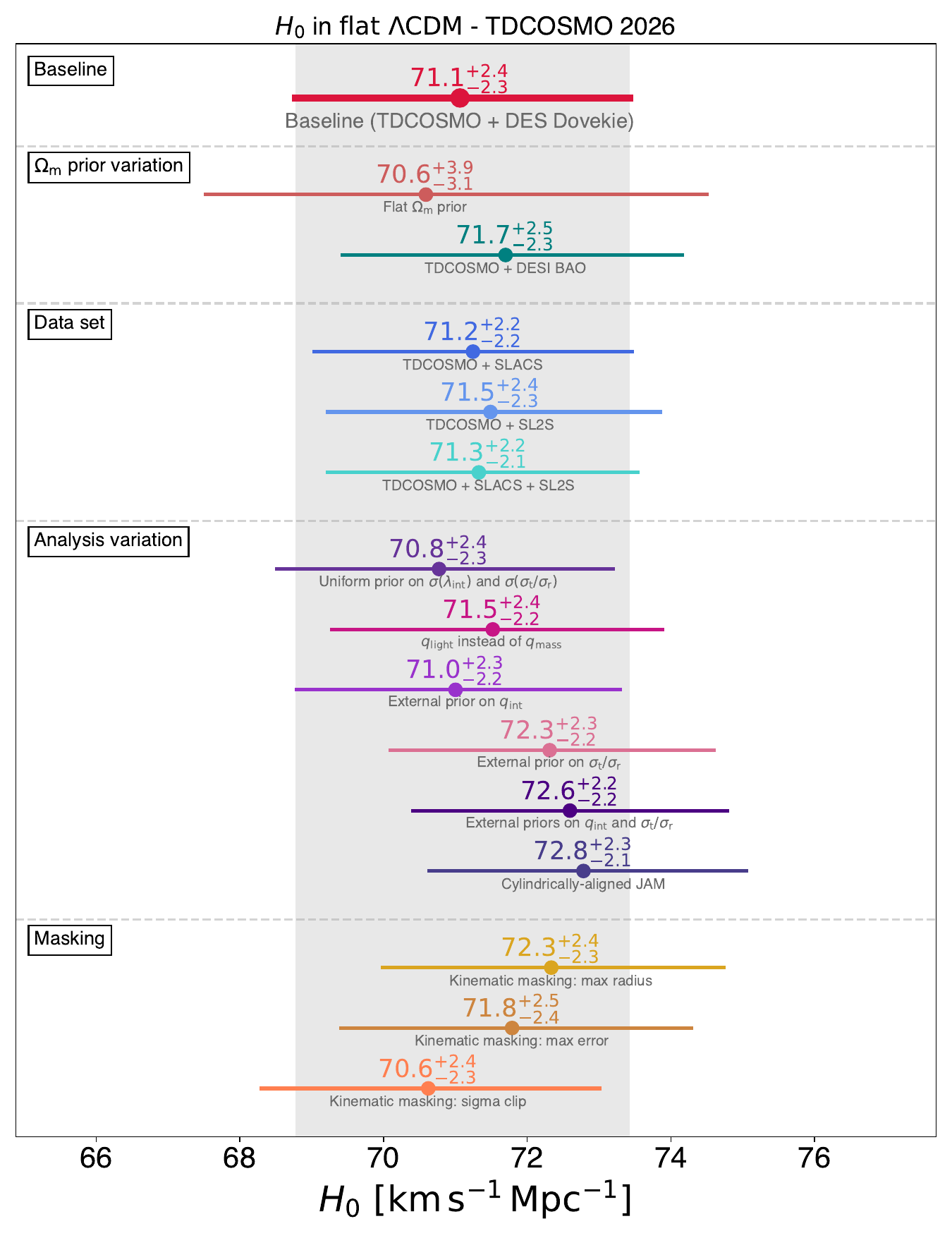}
    \caption{\Hc in flat \lcdm for each variation of the analysis, changing one ingredient at a time relative to the baseline. Points show the median and error bars the 68\% credible interval, while the gray band marks the 68\% interval of the baseline result. We emphasize that the baseline result represents the collaboration consensus, established before unblinding, on the preferred set of analysis choices. All alternative analyses should be interpreted as robustness tests rather than as a quantification of the systematic uncertainties, as some of them are disfavored by physical arguments. They use the same data and are therefore strongly correlated, and their shifts should not be added in quadrature.}
    \label{fig:H0variations}
\end{figure*}

In this section, we present alternative analysis choices, differing from our baseline results, in order to investigate the impact of these modeling choices in our cosmological inference. We re-ran the full hierarchical inference for each of these choices in turn, changing one ingredient at a time with respect to the baseline and leaving everything else fixed. The resulting shifts are summarized in \fref{fig:H0variations}, which shows, for every configuration, the inferred $H_0$ together with its $68\%$ interval.

We quote each shift in units of the baseline uncertainty,  $\sigma_{\rm base} = 2.3$\ksmpc, and we stress that the runs share the same data: their shifts are strongly correlated and must not be added in quadrature, nor interpreted as independent systematic error terms. Also, some analysis choices are preferred to others for physical reasons. Therefore, they should not be regarded as equally viable alternatives.

Overall, the inferred \Hc is robust to all the variations we considered. The shifts with respect to the baseline range from $-0.20\,\sigma_{\rm base}$ to $+0.7\,\sigma_{\rm base}$, and none exceeds $1\,\sigma_{\rm base}$. The precision is also stable, between 3.0\% and 3.4\%, with the only exception of the uniform \Om prior, which, as expected, lowers the precision. Most variations move \Hc slightly upward, by $+0.2\,\sigma_{\rm base}$ on average, so that our baseline lies at the lower end of the range spanned by the alternative choices. 

\paragraph{Prior on $\Omega_{\rm m}$.}
Replacing the supernova constraint by a flat prior on $\Omega_{\rm m} \in [0.05, 0.5]$ isolates how much of the inference is carried by the external data set rather than by the lenses. Since time-delay cosmography constrains \Hc far better than it constrains $\Omega_{\rm m}$, this variation mainly widens the $H_0$ uncertainty from 3.3\% to 4.9\%, instead of displacing the median value. Replacing the $\Omega_{\rm m}$ prior from DES-Dovekie with the DESI BAO distances changes \Hc by +0.6 \ksmpc ($+0.3\,\sigma_{\rm base}$) with similar precision. 

\paragraph{External lens samples.}
The baseline uses the time-delay lenses alone. Adding the SLACS lenses with KCWI kinematics, the SL2S lenses, and both together, relies on the assumption that the mass profiles of the lens population are shared across samples. In turn, the combined analysis allows the external lenses to constrain $\lambda_{\rm int}$ and the anisotropy without contributing any time delay of their own. This is a more precise result at the cost of an additional assumption. 

\paragraph{Priors on the population scatter.} 
In the baseline, the width of some population distribution, i.e., the scatter of $\lambda_{\rm int}$, and of the anisotropy parameter carries a scale-invariant $1/\sigma$ (log-uniform) prior. The variation replaces these with uniform priors on the population scatter parameters, i.e. $\sigma(\lambda_{\rm int}) \in [0.001 - 0.5]$ and $\sigma(\sigma_{\rm t}/\sigma_{\rm r}) \in [0.03 - 1]$). The two choices differ in how much weight is given to small scatter, so this variation probes the sensitivity of the population widths to a prior choice that is only weakly informed by the data. The effect on \Hc is negligible ($-0.1\,\sigma_{\rm base}$, same precision).

\paragraph{Flattening of the mass model.}
The axisymmetric kinematic model requires the observed axis ratio $q_{\rm obs}$, which determines the inclination and thus the projection of the dynamical model. In our baseline analysis, we use the axis ratio of the total mass distribution inferred from the lens model, while the variation uses the axis ratio of the lens light. These two quantities differ systematically, with the mass generally being rounder than the light distribution (with the exception of \RXJ and \WGDtwenty). We consider the mass axis ratio $q_{\rm m}$ to be a more appropriate tracer of the potential shape, as it accounts for both baryonic and dark matter. However, the ellipticity-shear degeneracy can bias $q_{\rm m}$ in systems with strong external shear. We therefore test $q_{\rm light}$ as an alternative choice and find that it has only a minor impact on our results, shifting \Hc by $+0.2\,\sigma_{\rm base}$.

\paragraph{Informative priors on the intrinsic axis ratio and anisotropies.}
Two population parameters are only weakly constrained by the lensing and kinematic data themselves: the mean intrinsic axis ratio $\langle q_{\rm int}\rangle$ and the mean anisotropy $\langle\sigma_{\rm t}/\sigma_{\rm r}\rangle$. The baseline samples both, together with their population widths. For our three variations, we instead fix them to an external prior $q_{\rm int} = 0.73 \pm 0.08$ and $\sigma_{\rm t}/\sigma_{\rm r} = 1.00 \pm 0.07$, taken from the slow rotators of the SDSS-IV MaNGA sample \citep{Li2018}. We test these two priors separately and both at once. Each replaces a sampled mean and scatter by a fixed distribution, so the variation measures how much of the inference rests on the internal calibration of these parameters and how much external information would change it. The prior on $q_{\rm int}$ has no effect on \Hc. The anisotropy prior instead shifts \Hc by $+0.5\,\sigma_{\rm base}$. This is the expected consequence of the mass--anisotropy degeneracy: the prior centered on isotropic orbits raises $\lambda_{\rm int,0}$ from 0.936 to 0.954, and \Hc by the same fraction. The precision on \Hc improves only marginally, to 3.1\%.

\paragraph{Cylindrically aligned JAM modeling.}
In general, the velocity ellipsoid in early-type galaxies is expected to be spherically aligned \citep[see discussion in Sec.~2.1 of][]{Cappellari2026}. However, evidence shows that in fast rotators the velocity ellipsoid flattening varies with polar angle in such a way that the overall flattening is approximately cylindrically aligned \citep[Fig.~2]{Cappellari2007}. In our sample, 9/11 time-delay deflectors are slow rotators for which the cylindrical alignment is not expected to be the best choice. However, assuming a cylindrical alignment represents a useful extreme assumption to test as a worst-case scenario. In practice, the cylindrical alignment is implemented by using the spherical alignment with the angular variation of \citet[Eq.~37]{Cappellari2026}. The anisotropy is then parameterized by $\sigma_z/\sigma_R$ rather than by $\sigma_{\rm t}/\sigma_{\rm r}$. In this variant, we adopt a uniform prior on $\langle \sigma_z/\sigma_R \rangle \in [0.7, 1]$ and a log-uniform prior on $\sigma( \sigma_z/\sigma_R)$ identical to the spherically aligned case. This is the largest shift among our variations, $+0.7\,\sigma_{\rm base}$ ($\lambda_{\rm int,0}$ rising from 0.936 to 0.963), which remains well within the statistical uncertainty.

\paragraph{Kinematic bin masking.}
As an alternative to our baseline analysis, which does not apply any masking to the kinematic data except in cases of an obvious failure of the kinematic fit, we apply here three different strategies to restrict which bins of the resolved kinematic maps enter the likelihood, each targeting a potential failure mode of our kinematic extraction pipeline:
\begin{itemize}
  \item \emph{maximum radius:} bins beyond a projected radius of 0.8 $\theta_{\rm E}$, are dropped. These are the bins where the surface brightness of the lens galaxy is the lowest and where the light from the quasar and its host galaxy could contaminate the spectrum and potentially affect the velocity dispersion measurement. This choice masks 17 out of the 229 kinematic bins of the \TDCnew sample (7\%). 
  \item \emph{maximum error:} bins whose velocity dispersion is measured with an uncertainty above a 30 \ks are dropped. This indicates a poorly constrained \textsc{pPXF} fit, which might potentially be inaccurate. This removes 28 kinematic bins out of the \TDCnew sample (12\%). 
  \item \emph{sigma clipping:} a smooth polynomial trend in $\log_{10} r$ is fitted to the map and bins departing from it by more than $3\sigma$ of their own uncertainty are dropped. This removes only 4 out of the 229 kinematic bins (2\%). 
\end{itemize}

We note that the ``maximum radius'' and ``maximum error'' often remove the same low-SNR outer bins, hence explaining the similarity of the +0.6$\sigma_{\rm base}$ and +0.3$\sigma_{\rm base}$ shifts on \Hc that those two runs yield. Each mask changes the number of constraints, but it also modifies the radial leverage on the anisotropy. Removing some of the outer bins tends to reduce the scatter on the anisotropy parameter $\sigma_{\rm t}/\sigma_{\rm r}$, resulting in slightly tighter constraints on its mean $\langle \sigma_{\rm t}/\sigma_{\rm r} \rangle$ and in a small shift on \Hc.

\section{Comparison with the literature}
\label{sec:literature}
In this section, we compare our results with the ones obtained previously by the TDCOSMO collaboration in \ssref{ssec:comparison_TDCOSMO}, before comparing them with other \Hc measurements from time-delay cosmography in \ssref{ssec:comparison_others}. A comparison with independent probes in the context of the Hubble tension is presented in a companion letter (\TDCnewletter). 

\subsection{Comparison with \TDCprev}
\label{ssec:comparison_TDCOSMO}

\begin{figure*}
    \includegraphics[width = 0.8\textwidth]{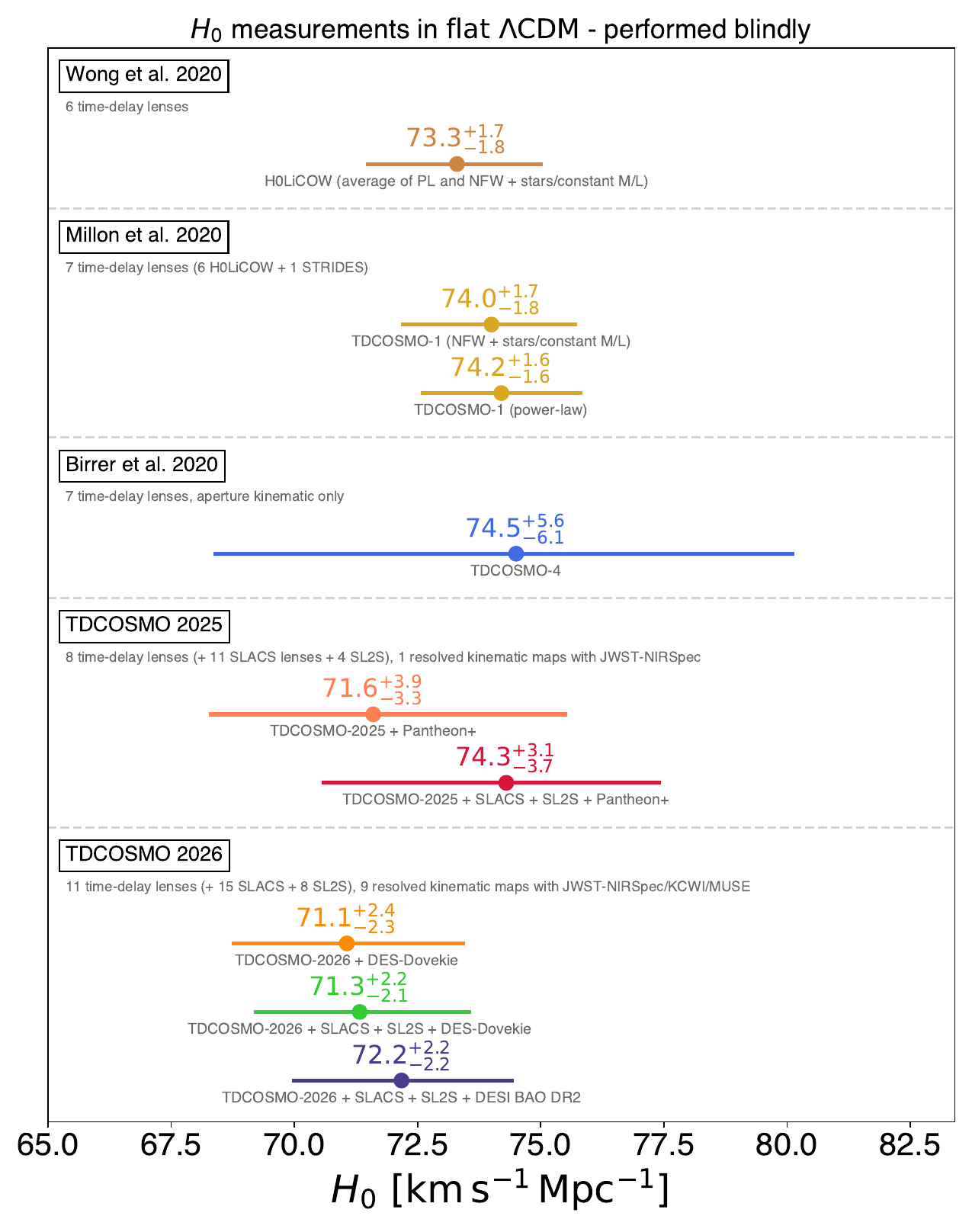}
    \caption{Summary of the \Hc{} measurements of the TDCOSMO collaboration (and its predecessor H0LiCOW) in flat $\Lambda$CDM, in chronological order. The measurements by \citet{Wong2019} and \citet[][TDCOSMO-1]{Millon:2020} are ``assertive'' in the definition of \citet{Treu_review}: they break the mass-sheet degeneracy (MSD) by assuming a power-law (or NFW+stars) radial mass profile. All subsequent measurements are ``conservative'': they sample the internal MSD parameter $\lambda_{\rm int}$ and constrain it with stellar kinematics. The \TDCprev points \citep{TDCOSMO2025} are shown with the Pantheon+ prior on $\Omega_{\rm m}$ with which they were published; the \TDCnew points (this work) use DES-Dovekie, or DESI-DR2 BAO where stated. All the analyses were performed blindly.\label{fig:h0comparison}}
\end{figure*}

Figure~\ref{fig:h0comparison} places our results in the sequence of \Hc measurements by the TDCOSMO collaboration and its predecessor H0LiCOW. The most visible change with respect to \TDCprev is the precision. For the TDCOSMO sample combined with supernovae, the uncertainty decreases from 5.0\% ($\Hc = 71.6^{+3.9}_{-3.3}$\ksmpc, with Pantheon+) to 3.3\% ($\Hc = \hnotfiducial$\ksmpc, with DES-Dovekie), and with the addition of the SLACS and SL2S lenses from 4.6\% ($74.3^{+3.1}_{-3.7}$\ksmpc) to 3.0\% ($\hnotLCDMfoura$\ksmpc), a reduction of about 35\% in both cases. This gain results from the larger sample, the new high-quality kinematic data and several methodological improvements. Our most precise result, which combines \TDCnew, SLACS and SL2S with DESI BAO, $\Hc = \hnotLCDMfourb$\ksmpc (3.0\%), approaches the precision of the H0LiCOW ($73.3^{+1.7}_{-1.8}$\ksmpc, 2.4\%) and TDCOSMO-1 power-law ($74.2 \pm 1.6$\ksmpc, 2.2\%) measurements. Those results, however, relied on an assumption about the radial mass profile that is equivalent to fixing $\lint = 1$ in the present analysis. TDCOSMO-4 relaxed this assumption on almost the same data, and the uncertainty grew from 2.2\% to 7.9\% ($\Hc = 74.5^{+5.6}_{-6.1}$\ksmpc), because the single-aperture velocity
dispersions available at the time could not constrain the internal MST at a competitive level. \TDCprev brought this uncertainty back to 5\% with much better kinematic data. With the present work, a measurement that makes
minimal assumptions on the radial mass profile reaches 3\%. This precision is close to what was previously obtained under much more assertive assumptions.

Our central value is fully
consistent with that of \TDCprev. For the \TDCnew sample with supernovae, it moves by only 0.5\ksmpc, about $0.15\sigma$. The \TDCprev baseline used a single-aperture velocity dispersion for seven of its eight lenses, plus 6 and 8 radial bins from JWST/NIRSpec and Keck/KCWI for \RXJ. These are now replaced by 227 kinematic bins for the nine lenses with spatially resolved kinematics from several instruments, plus aperture measurements for \DESzerofour and \WGDtwenty. Interestingly, this extra information does not translate into a much tighter population mean of the internal MST: we find $\lint = \lambdaintfiducial$, compared with $\lint = 0.97^{+0.06}_{-0.05}$ in \TDCprev, i.e.\ a precision of 5.3\% instead of 5.5\%. The gain in \Hc instead comes from the resolved kinematics constraining the MST of each time-delay lens individually, which loosens the coupling between \Hc and the population mean. Better kinematic data informs $\lint$ for individual lenses, decreasing the relative weight of the population prior in the inference. 

The population mean does not tighten further because the data now are good enough to detect intrinsic scatter in $\lint$ among the time-delay lenses. This scatter was undetected in \TDCprev, where we found $\sigma(\lint) = 0.045^{+0.066}_{-0.040}$, consistent with zero. It is now $\sigma(\lint) = 0.093^{+0.038}_{-0.026}$, statistically consistent with the previous value, but significantly non-zero. This sets a floor on how well the population mean can be determined from only eleven lenses within the current hierarchical analysis.  When the external lenses are included, the data prefer a somewhat larger scatter, $\sigma(\lint) =  0.131^{+0.029}_{-0.022}$. The same trend was present in \TDCprev, where TDCOSMO+SLACS+SL2S gave $\sigma(\lint) = 0.127^{+0.039}_{-0.028}$, consistent with the value found by \citet{knabel26_slacs} for the SLACS sample alone. It may indicate that the SLACS deflectors differ slightly from the time-delay deflectors \citep{knabel26_slacs}, which would challenge the assumption that the two samples are drawn from the same population. In \TDCprev, it was hypothesized that the scatter partly absorbed the limitations in the modeling of the SLACS kinematics, which were modeled with spherical JAM. We now use axisymmetric JAM and still observe a similar scatter. This might be an indication that the measured scatter is indeed the true intrinsic scatter of the population. Fitting independent $\lint$ for each lens, rather than using it as a population parameter, may be necessary to reduce the scatter.

The population-level stellar anisotropy, in contrast, remains only weakly constrained. The mean of $\sigma_{\rm t}/\sigma_{\rm r}$ is $0.987^{+0.082}_{-0.073}$ for the \TDCnew sample, compared with $1.023^{+0.069}_{-0.092}$ in \TDCprev. Both are largely set by the $\mathcal{U}(0.87, 1.12)$ prior. Adding the SLACS and SL2S posterior shifts the mean slightly toward radial orbits, $0.928^{+0.100}_{-0.044}$, without narrowing it significantly.

Interestingly, the \TDCnew sample alone shows a mild preference for a negative slope of the MST parameter with the ratio of effective to Einstein radius, $\alpha_\lambda = -0.038^{+0.082}_{-0.088}$, whereas \TDCprev found $\alpha_\lambda = -0.003^{+0.052}_{-0.051}$. As can be seen in \fref{fig:cornerplot_main_result}, $\alpha_\lambda$ is anti-correlated with the mean $\lint$, so the low value of $\lint$ and the negative $\alpha_\lambda$ are partly two expressions of the same feature of the data. Adding the \nslacstot ~ SLACS and SL2S lenses, which contribute excellent spatially resolved kinematics, removes the trend completely, $\alpha_\lambda = 0.016^{+0.051}_{-0.056},$ and $\lint = 0.993^{+0.042}_{-0.041}$. This is remarkably consistent with no departure from a power law. A larger sample of time-delay lenses with resolved kinematics will show whether the time-delay deflectors differ from the SLACS and SL2S deflectors, or whether this is a statistical fluctuation of the present sample.

These results also explain the comparison with the earlier assertive measurements. The \TDCnew time-delay lenses deviate, on average, from a pure power law by about $1\sigma$, with $\lint$ between 0.935 and 0.942 depending on the prior on \Om. This deviation is not statistically significant and accounts for most of the shift with respect to the power-law result of TDCOSMO-1. With the same uniform prior on \Om, we find $\Hc = 70.6^{+3.9}_{-3.1}$\ksmpc, $0.9\sigma$ below TDCOSMO-1. Since $\Hc \propto \lint$ to first order, rescaling our value to $\lint = 1$ gives $\approx 75.5$\ksmpc, in agreement with TDCOSMO-1. The trend disappears once the SLACS and SL2S lenses are included: the population mean is then
$\lint = 1.002^{+0.050}_{-0.046}$ (uniform \Om), indistinguishable from a power law, and $\Hc = 72.1^{+4.1}_{-3.4}$\ksmpc agrees with TDCOSMO-1 within $0.5\sigma$. 
The consistency is very re-assuring considering that many things have changed since TDCOSMO-1: the sample has expanded, all the kinematic data are new, external convergences are new and NIRCam-based models have replaced older models for three systems.

\subsection{Comparison with other works}
\label{ssec:comparison_others}

We limit this section to a comparison with other time-delay cosmography estimates, referring to the companion Letter for a more general overview of the Hubble tension.

The main recent development in time-delay cosmography has been the accelerated pace of discovery and exploitation of multiply-imaged supernovae. Four such measurements have been published to date: supernova Refsdal \citep{Kelly2023b,Grillo24}; supernova H0pe \citep{Pascale25}; supernova Encore \citep{Pierel:2026,Suyu:2026}; and supernova Winny \citep{Tautenberger:2026,Winny5,Johansson2025}. The first three are lensed by a cluster-scale deflector, the last one by a galaxy-scale deflector. The four measurements have precision ranging from $\sim$6--10\%, and they are all in statistical agreement with the \TDCnew result. This is encouraging, considering that the potential sources of systematics are very different:  galaxy-scale deflectors vs cluster-scale deflectors; time delay measurements in quasars vs supernovae light curves. As the number of supernovae with measured time delays increases, it will be beneficial to carry out a hierarchical analysis of the results to infer a combined estimate accounting for correlated uncertainties. This will provide a very important independent way to measure $H_0$. 

When more supernovae lensed by galaxies are available, one can imagine carrying out a very similar analysis to the one presented here, fully accounting for the mass-sheet degeneracy. This will yield yet another important test of systematics. 

There are also efforts under way to use quasars lensed by clusters to infer $H_0$ \citep[e.g.,][]{Napier:2023}. Those are promising but have not yet reached the precision needed to put our results through a stringent test.

\section{Systematics}
\label{sec:syst}

In this section, we discuss residual systematic errors and our most recent efforts to quantify and mitigate them (Section~\ref{ssec:modelass}). We also briefly revisit the issue of potential selection effects (Section~\ref{ssec:sel}). These sections were written prior to unblinding, except for the few words presenting unblinded results. As the statistical precision improves, we are continuing to investigate systematic errors, building upon previous work. We focus here on the more recent work on the topic, referring to \TDCprev for a discussion of the pre-2025 work. 

\subsection{Model assumptions}
\label{ssec:modelass}

The main advances in understanding, mitigating, and correcting potential systematics errors since the 2025 milestone are:

\begin{itemize}
    \item \textit{Uncertainties on the lens models:} Several studies have shown that the modeled Fermat potential differences are robust to the choice of lens modeling code and data. \citet{TDCOSMO20} modeled the lens \WFItwenty based on JWST-NIRCam data and using \lenstro. They found Fermat potential differences that are more precise and consistent within the errors of the estimates based on HST data and the \glee code presented by \citet{Rusu2020}. The same data were analyzed by an independent team \citep{Li2026} with the independent lensing code Herculens \citep{Galan:2022}. They found consistent results and comparable uncertainties. \citet{GalanSchweinfurth:2026} analyzed the lens system \Jfifteen\ using the same data and two independent codes led by the two lead authors (\glee and \lenstro) and found overall agreement, although the \lenstro one better fit the data with a more physical model. \citet{Williams2026} modeled with \lenstro three quads that are part of the \TDCnew sample combining JWST-NIRCam imaging, JWST-NIRSpec spectroscopy, and new estimates of the external convergence.   They compared their results with the previous analysis based on HST imaging, ground-based spectroscopy, and older estimates of \kext, finding consistent results and reduced scatter. 
    \item \textit{A central supermassive black hole} could affect the interpretation of the stellar velocity dispersion at the JWST-NIRSpec angular resolution \citep{Wang2025}. This is now accounted for in our most recent dynamical models. For the systems for which a black hole has been included, we find that it makes no significant difference given the current data \citep[see, e.g.,][]{sheu2026}.
    \item \textit{Departures from spherical symmetry in dynamical models:} we have replaced the correction to spherical models \citep{Huang2026}, with axisymmetric models. Impacts of potential triaxiality are below the 1\% level on average~\citep{Huang2026}. 
    \item \textit{Parameterization of the orbital anisotropy:} \citet{V+M26} show using cosmological numerical simulations that the assumption of constant anisotropy adopted in this paper yields unbiased measurements of the $H_0$.
    
\end{itemize}

In summary, the work done in the past year strengthens the conclusions of \TDCprev that the model assumption uncertainties are smaller than the statistical uncertainties associated with our measurement of $H_0$, and are uncorrelated. Therefore, they are subdominant in our error budget, and can be neglected for now. 

As the statistical uncertainties continue to improve, we will extend this investigation to ensure that systematic errors remain subdominant. This implies updating some of our analysis techniques to explicitly capture additional model degeneracies \citep{Krizic2026} and establish a detailed error budget per source of uncertainties, either in a data-driven way or based on controlled simulations.

\subsection{Selection effects}
\label{ssec:sel}

As discussed extensively in \TDCprev, selection effects could potentially bias our result in combination with external datasets. 

One potential source of bias is informative priors on population-level hyperparameters taken from external samples. To avoid this potential source of bias we have adopted non-informative priors on the distribution of intrinsic flattening in 3D, rather than relying on the much better characterized distribution estimated from much larger samples of non-lens galaxies~\citep{Li2018}. This comes at the cost of two additional population parameters, $\langle q_{\rm int} \rangle$ and $\sigma(q_{\rm int})$, which need to be constrained by the data of our 11 time-delay lenses. We show in \sref{sec:analysis variation} that adopting a prior from the MaNGA sample would only improve our precision from 3.3\% to 3.2\%, with no change in the central value of \Hc.

Another potential source of bias is the combination with external samples of non-time-delay lenses. If, for example, the properties of the SLACS and SL2S sample were intrinsically different than those of the time-delay lenses, the inference of $\lint$ and therefore of $H_0$ could be biased. It is for this reason that our primary result is that based exclusively on the time-delay lenses, and we report the one in combination as an extension of this work. Reassuringly, the results with and without the external samples are in excellent statistical agreement, providing no evidence of measurable bias.

\section{Summary and Conclusions}
\label{sec:sum}

We have presented the blind hierarchical analysis of the \TDCnew dataset, comprising 11 lensed quasars with measured time delays. We have also considered the combination of \TDCnew with expanded samples of non-time-delay lenses, used to constrain the internal mass-sheet parameter $\lint$. The dataset and analysis methods have substantially changed and improved with respect to our previous \TDCprev milestone. The most important changes are: we have expanded the sample with three new systems; we have improved the kinematic data using JWST-NIRSpec spatially resolved kinematics for six out of the 11 systems; we have replaced inclination-corrected spherical models with axisymmetric ones; we have replaced estimates of the line-of-sight external convergence based on the Millennium simulations with ones based on Euclid Flagship simulations; we have used JWST-NIRCam lens models for three lenses, replacing previous ones based on HST imaging data. The main results can be summarized as follows:

\begin{enumerate}
    \item In flat-$\Lambda$CDM we find $H_0 = \hnotfiducial$ \ksmpc for the \TDCnew sample in our baseline model that contains solely time-delay lenses. The result is robust to variations in analysis choices and cosmological priors.
    \item In Flat $\Lambda$CDM we find $H_0 = \hnotLCDMfoura$ \ksmpc  for the \TDCnew+SLACS+SL2S sample that also contains non-time-delay lenses.
    \item With a 3\% precision, TDCOSMO is among the most precise late-Universe probes, second only to the local distance ladder. 
    \item Our inferred $H_0$ in flat $\Lambda$CDM is within 1$\sigma$ of measurements based on the local distance ladder \citep{Freedman24,H0DN:2026}.  
    \item Our baseline $H_0$ lies $1.6\sigma$ above the \textit{Planck} value. Combining the time-delay lenses with the SLACS and SL2S samples raises this to $1.9\sigma$ with DES-Dovekie and $2.2\sigma$ with DESI-DR2 BAO, the largest difference among our \lcdm results.
    \item Combining TDCOSMO with other late-universe probes yields a tension with Planck at the $\sim2.7\sigma$ level or more, even when excluding the local distance ladder, as discussed in our companion article (\TDCnewletter).
    \item The Hubble constant is robust to the choice of background cosmology. We find $H_0 = 72.3^{+3.7}_{-3.7}$ \ksmpc in open $\Lambda$CDM.  In combination with our DES-Dovekie baseline, we find $H_0 = 71.4^{+2.2}_{-2.1}$ \ksmpc in $w$CDM, $H_0 = 70.6^{+2.1}_{-2.1}$ \ksmpc in $w_0w_a$CDM, and $H_0 = 71.1^{+2.2}_{-2.1}$ \ksmpc in $w_\phi$CDM.  
    \item Without CMB data, the combination of TDCOSMO with DESI BAO and DES-Dovekie favors a present-day equation of state $w_0 > -1$ in every extension of \lcdm we consider: $w = -0.91^{+0.04}_{-0.04}$ in $w$CDM, $w_0 = -0.85^{+0.07}_{-0.06}$ in $w_0w_a$CDM and $w_0 = -0.88^{+0.04}_{-0.05}$ in $w_\phi$CDM, disfavoring $w = -1$ at $2.3$--$2.5\sigma$.
    \item Our blind results are in excellent agreement with our previous blind determinations of the Hubble constant. This was not a foregone conclusion considering the blindness of the analysis and the many changes in the data, methods, and individual investigators involved in the analysis.
\end{enumerate}

\section{Author contributions and Acknowledgments}
\paragraph{Author contributions:}
The authors of this paper are listed in alphabetical order. S.B. provided guidance on the hierarchical model and on the lens modeling, is the original developer of the hierarchical analysis framework, and served as internal reviewer.
M.C. provided guidance on the kinematic extraction pipeline and on the kinematic modeling.
F.C. led the monitoring campaign of the latest time delays, contributed to the analysis of \HEeleven, served as internal reviewer, and wrote parts of the paper.
C.D.F. created the catalogs for the external convergence analysis of several of the systems.
P.F. developed the methodology of the external convergence estimates.
J.A.F. contributed to the cosmological analysis and served as an internal reviewer.
A.G. led the \Jfifteen analysis, provided feedback in lens modeling efforts, implemented parts of the code, wrote parts of the paper.
X.-Y.H. reviewed the code and contributed to the selection of the external lens samples.
D.P.J. contributed to the development of and carried out the external convergence analysis, and wrote parts of the paper.
S.Kh. reduced the KCWI/KCRM data of the SLACS and SL2S lenses.
S.K. reduced the raw JWST-NIRSpec data, extracted the stellar kinematics, and wrote parts of the paper.
H.K. refitted the lens light of \Bsixteen.
M.M. led the analysis, implemented key improvements in the cosmological inference and the combinations with external probes, measured the time delays of \Jfourteen, provided feedback on the lens modeling, designed and ran the analysis variation tests, and wrote parts of the paper.
T.M. prepared the JWST GTO program.
V.M. contributed to the light-curve data collection and participated in observing proposals and coordination efforts.
P.M. contributed to the reduction of the raw JWST-NIRSpec data, extracted stellar kinematics, and wrote parts of the paper.
E.P. carried out the full analysis of \HEeleven, and reviewed the code and the paper.
H.P. acquired the KCWI/KCRM data.
A.G.S. co-led the \Jfifteen analysis, producing the \glee lens model and catalogs.
A.J.S. co-led the JWST proposal, and did crucial work on the data reduction, calibration, and kinematic measurements.
W.S. reviewed the code, carried out the data reduction and analysis of \Jfourteen, and reduced the KCWI/KCRM data of the external lenses.
D.S. contributed to the data and environment analysis of several targets, to coordination efforts within TDCOSMO, and to proposals, and served as internal reviewer.
S.H.S. supported the distance measurements of \Jfifteen and the data acquisition for \HEeleven,  was the PI of the JWST-NIRSpec of \RXJ,  provided feedback and support on the analysis, and served as an internal reviewer of the paper.
M.S. is the PI of the JWST GTO program and prepared the program for execution.
T.T. was PI of the JWST-NIRSpec, HST, and Keck proposals, worked closely with S.K., P.M., W.S., and D.M.W. on data reduction, methods, and manuscripts, wrote parts of the paper, and served as final reader.
F.U. implemented the axisymmetric dynamical modeling, and contributed to the selection of the external samples and to the preprocessing of the kinematic data.
D.M.W. produced the JWST lens models and their analysis, and contributed to the external convergence analysis.
K.C.W. helped with the lens modeling, and reviewed the code and the paper.
\paragraph{Acknowledgments:}
MM and HK acknowledge support by the Swiss National Science Foundation (SNSF) through Ambizione grant PZ00P2\_223738. AG acknowledges support by the SNSF through PostDoc Mobility Return grant P5R5-2\_235360. The UCLA group acknowledges support by NSF through grants NSF-AST-2407277, by NASA through grant JWST-GO-2974, and from the Moore Foundation through grant 8548.AJS acknowledges support by NASA through grant JWST-GO-2974.KCW is supported by JSPS KAKENHI Grant Numbers JP24K07089, JP24H00221.E.P. is supported by JSPS KAKENHI Grant Number JP24H00221. DS acknowledges the support of the Fonds de la Recherche Scientifique-FNRS, Belgium, under grant No. 4.4503.1. CDF acknowledges support for this work from the US National Science Foundation under Grant No. AST-2407278. DJ acknowledges support by the First Rand Foundation, South Africa, and the Centre National de la Recherche Scientifique of France. PF acknowledges support from the French Agence Nationale de la Recherche through the ELROND project (ANR-23-CE31-0002). This work was supported by the Agencia Estatal de Investigación (AEI), Ministerio de Ciencia, Innovación y Universidades, Spain, under the project PID2024‑155455NB‑I00, within the 2024 Call for Knowledge Generation Projects. FC has also received support through ERC Synergy RedH0T under grant agreement 101225120. V.M. acknowledges support from ANID FONDECYT Regular grant number 1231418, and Centro de Astrof\'{\i}sica de Valpara\'{\i}so CIDI 21. SB and XH are supported by DoE Grant DE-SC0026113. The research reported here is supported by Stanford University under Contract No. DE-AC02-76SF00515 with the U.S. Department of Energy, Office of Science, Office of High Energy and Nuclear Physics. T.M. acknowledge support by JSPS through KAKENHI Grant Number 25K24918. MS acknowledge support under NASA grant 80NSSC22K1294.
This research made use of
\textsc{hierArc} \citep{TDCOSMOIV},
\lenstro \citep{Birrer2018, Birrer2021lenstronomy},
\jampy \citep{Cappellari2008, Cappellari2020}, \glee \citep{SuyuHalkola2010},
\textsc{emcee} \citep{Foreman2013},
\textsc{GetDist} \citep{GetDist},
\textsc{squirrel} \citep{Shajib25b} and \textsc{pPXF} \citep{Cappellari04},
\textsc{STARRED} \citep{Michalewicz2023, Millon2024},
\textsc{Astropy} \citep{Astropy2013, Astropy2018},
\textsc{matplotlib} \citep{Hunter2007}, and
\texttt{STCI}\footnote{\url{https://pypi.org/project/STCI/}} for the color composite images.

\appendix

\section{Kinematic classification of the TDCOSMO deflectors}
\label{app:lambdaR}

\begin{figure}
    \includegraphics[width = \linewidth]{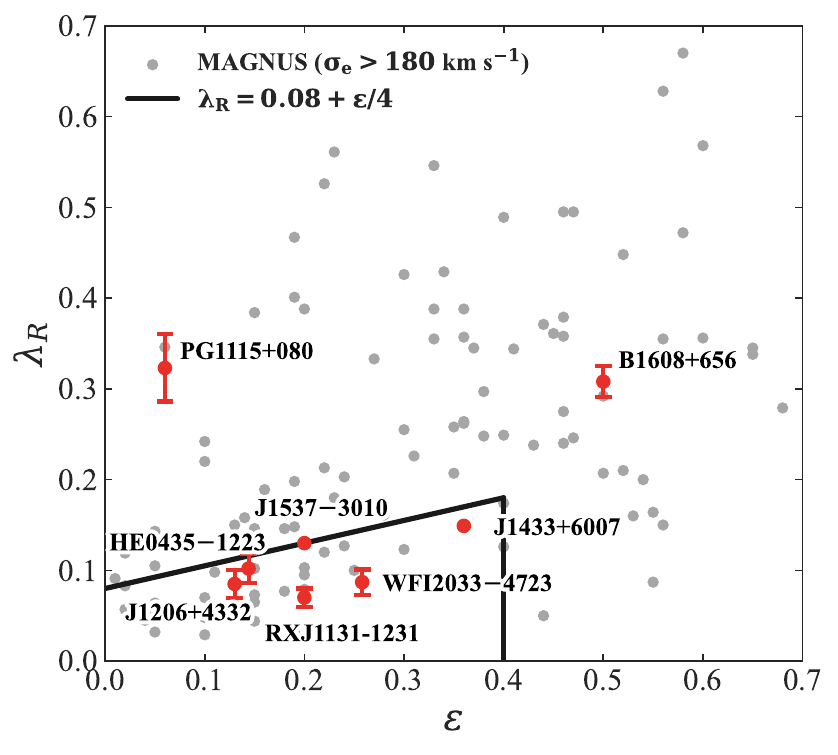}
    \caption{\label{fig:tdcosmo_slow_fast_rot} Kinematic classification of the TDCOSMO lens sample. Spin parameter $\lambda_R$ versus ellipticity $\varepsilon$ for the TDCOSMO deflectors with 2D kinematics (red), compared with early-type galaxies from MAGNUS sample with $\sigma_\mathrm{e} > 180$\,km\,s$^{-1}$ (grey; \citep{Mozumdar_MAGNUS_I}). The black line encloses the slow-rotator region, $\lambda_R < 0.08 + \varepsilon/4$ for $\varepsilon < 0.4$ \citep{Cappellari2016}. Deflectors without error bars have no published uncertainty.} .
\end{figure}

We classify the deflectors in the TDCOSMO sample with 2D kinematics as fast or slow rotators using the stellar spin parameter $\lambda_R = \sum_n F_n R_n |V_n| \, / \sum_n F_n R_n V_{\mathrm{rms},n}$, with $V_{\mathrm{rms}}^2 = V^2 + \sigma^2$ \citep{Emsellem07}, measured within an elliptical aperture. For \Bsixteen, \Jtwelve and \RXJ, we compute $\lambda_R$ from our JWST-NIRSpec IFU kinematics, assigning each spaxel the $V$ and $V_\mathrm{rms}$ of its bin and weighting by the flux of the deflector. Uncertainties are measured using 2000 Monte Carlo realisations drawn from the $V$ and $V_\mathrm{rms}$ covariance matrices. For the remaining deflectors, we take $\lambda_R$ from the corresponding papers (\Jfourteen, \citep{sheu2026}; \Jfifteen, \citep{GalanSchweinfurth:2026}; \HEzero, \PGeleven and \WFItwenty, \citep{Knabel2026}). The aperture used to measure $\lambda_R$ is not uniform across the sample, because the spatial extent of the kinematic data relative to $R_\mathrm{e}$ differs between systems.

Fig.~\ref{fig:tdcosmo_slow_fast_rot} shows the deflectors in the $(\lambda_R, \varepsilon)$ plane, together with the slow-rotator region $\lambda_R < 0.08 + \varepsilon/4$, $\varepsilon < 0.4$ \citep{Cappellari2016}. For comparison, we show MAGNUS early-type galaxies \citep{Mozumdar_MAGNUS_I} with $\sigma_\mathrm{e} > 180$\,km\,s$^{-1}$, which span a redshift ($0.24 < z < 0.75$) and velocity dispersion range similar to those of the TDCOSMO deflectors and use the same $\lambda_R$ definition. Atmospheric seeing smears light and biases $\lambda_R$ low, so a PSF correction \citep{Graham_2018} is important for seeing-limited data such as MAGNUS \citep{Mozumdar_MAGNUS_II}. For the JWST/NIRSpec measurements, the PSF (FWHM $\sim 0.15$\,arcsec) is several times smaller than the deflector effective radii ($0.53$--$1.51$\,arcsec), and we apply no correction. The values for \Jfourteen (KCWI) and \Jfifteen (MUSE), however, come from seeing-limited data and may be biased low. Apart from \Bsixteen and \PGeleven, all deflectors lie in or at the edge of the slow-rotator region.

\section{Cosmographic goodness of fit}
\label{ap:cosmographic_gof}

\begin{figure}[htbp]
    \includegraphics[width=0.5\textwidth]{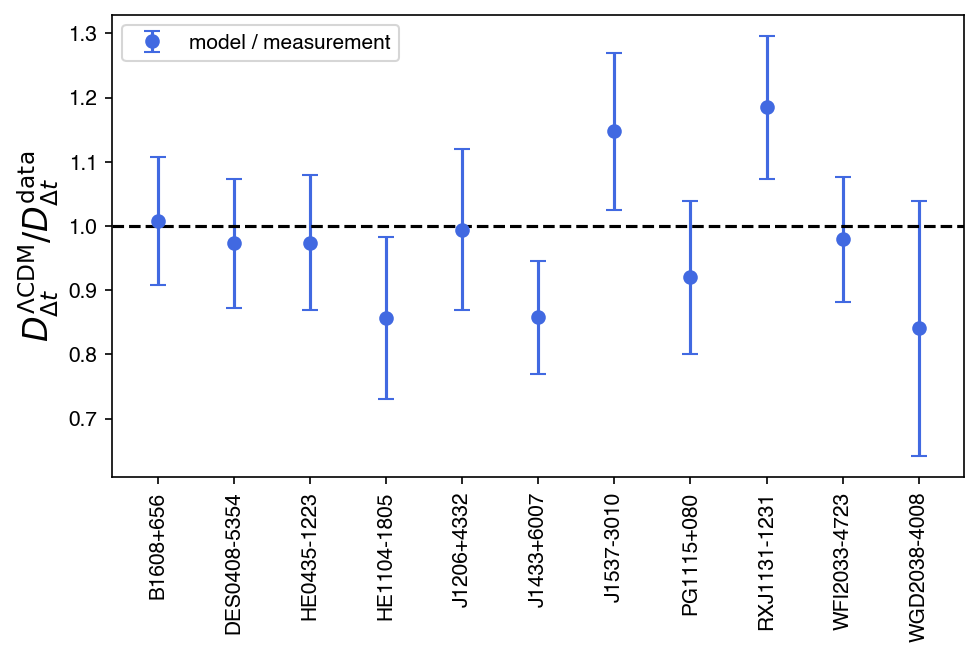}
    \caption{\label{fig:cosmographic_gof} Goodness of fit of the time-delay distances $D_{\Delta t}$ for each individual lens in flat \lcdm cosmology for our baseline configuration (corresponding to run $\Lambda$CDM1a of \tref{tab:results_LCDM}). Points show the ratio of the \Ddt predicted by the maximum-likelihood hierarchical model to the one measured from the lens model and time delays. Error bars combine the measurement and prediction uncertainties in quadrature; the latter includes the intrinsic scatter of $\lint$ and the uncertainty on \kext.}
\end{figure}

We assess how well the cosmological and population model is reproducing the measured time-delay distance of each lens by comparing the \Ddt measured from the lens model and time delays with the one predicted by the maximum-likelihood point of the hierarchical inference ($\Lambda$CDM1a of \tref{tab:results_LCDM}). The prediction combines the cosmological distances with the population-level internal MST, $\lint$, and the external convergence of each line of sight. Its uncertainty therefore includes the intrinsic scatter $\sigma(\lint)$ and the uncertainty on $\kext$, which together amount to 9--12\% per lens and dominate over the 2--10\% precision of most \Ddt measurements. \fref{fig:cosmographic_gof} shows the ratio of the predicted to the measured \Ddt.

All eleven lenses are reproduced within $2\sigma$, with $\chi^2 = 10.0$ for 11 lenses. The largest deviations are for \RXJ, whose measured \Ddt lies $1.9\sigma$ below the prediction, and \Jfourteen, which lies $1.6\sigma$ above it. These two systems pull the population mean of $\lint$ in opposite directions: \RXJ prefers a lower $\lint$ than the sample average, as already noted in \TDCprev, while \Jfourteen prefers a higher one, in line with the individual analysis of \citet{sheu2026}.

\section{Kinematic goodness of fit}
\label{ap:kienatic_gof}

\begin{figure*}[htbp]
    \centering
    \includegraphics[width=0.9\textwidth]{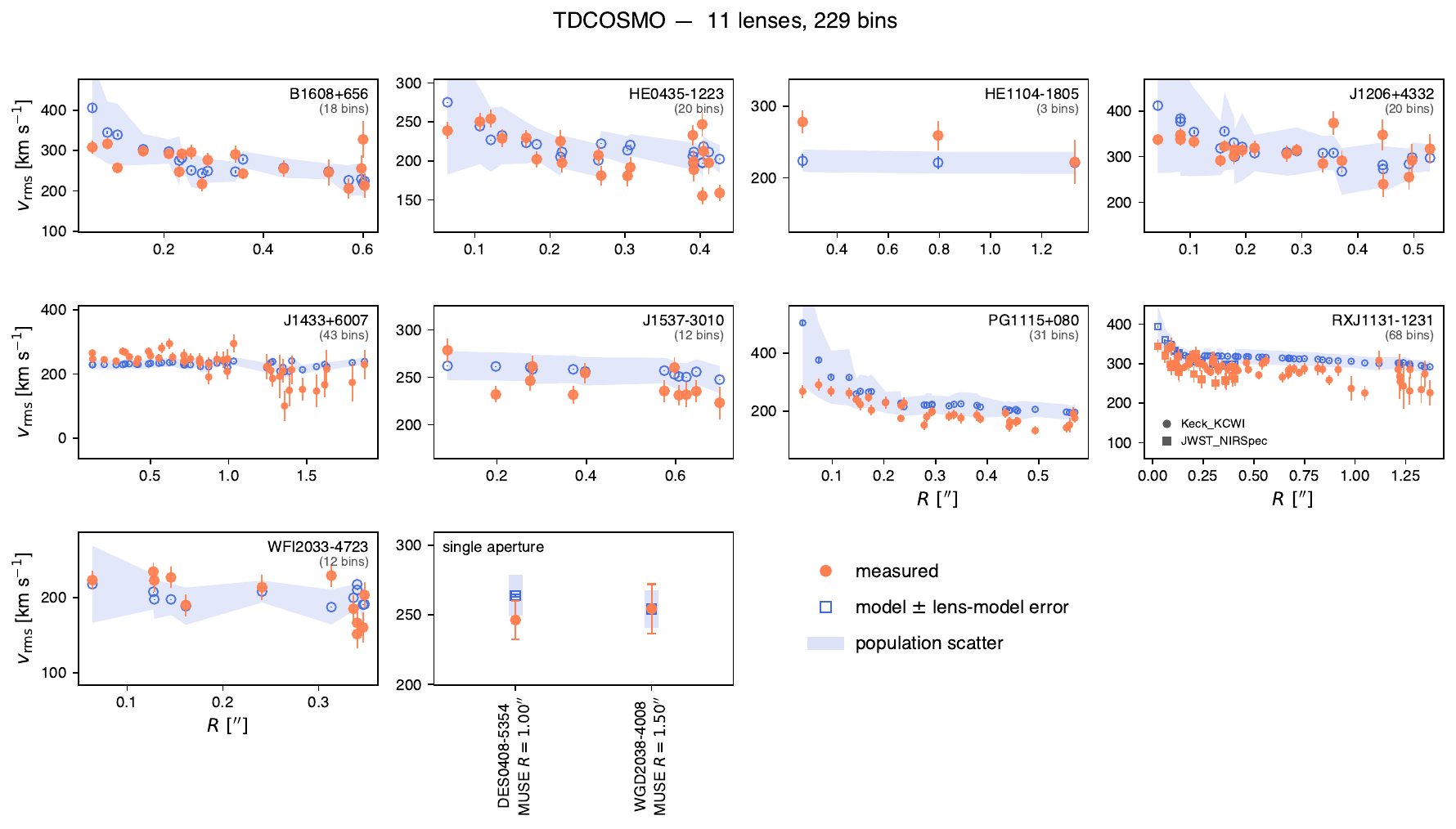}
    \caption{Goodness of fit of the stellar kinematics for the 11 TDCOSMO lenses, for the baseline TDCOSMO + DES-Dovekie inference in flat \lcdm (run $\Lambda$CDM1a). Orange circles are the measured line-of-sight velocity dispersions of each kinematic bin, plotted against the projected radius of the bin center; open blue squares are the predictions of the maximum-likelihood model, with error bars showing the lens-model uncertainty at fixed population parameters. The blue band is the scatter induced by the inferred population distributions of $\lambda_{\rm int}$, the stellar anisotropy and the intrinsic axis ratio. This is representative of the spread in properties of the sample population rather than an uncertainty on the fit itself. spatially resolved kinematics come from JWST/NIRSpec (\Bsixteen, \HEzero, \Jtwelve, \PGeleven, \WFItwenty), Keck/KCWI (\Jfourteen), VLT/MUSE (\HEeleven, \Jfifteen) and both KCWI and NIRSpec for \RXJ. The last panel groups the two lenses with a single MUSE aperture measurement, namely \WGDtwenty and \DESzerofour, labeled with the radius of that aperture.}
    \label{fig:TDCOSMO_kin_gof}
\end{figure*}

\begin{figure*}[htbp]
    \centering
    \includegraphics[width=0.9\textwidth]{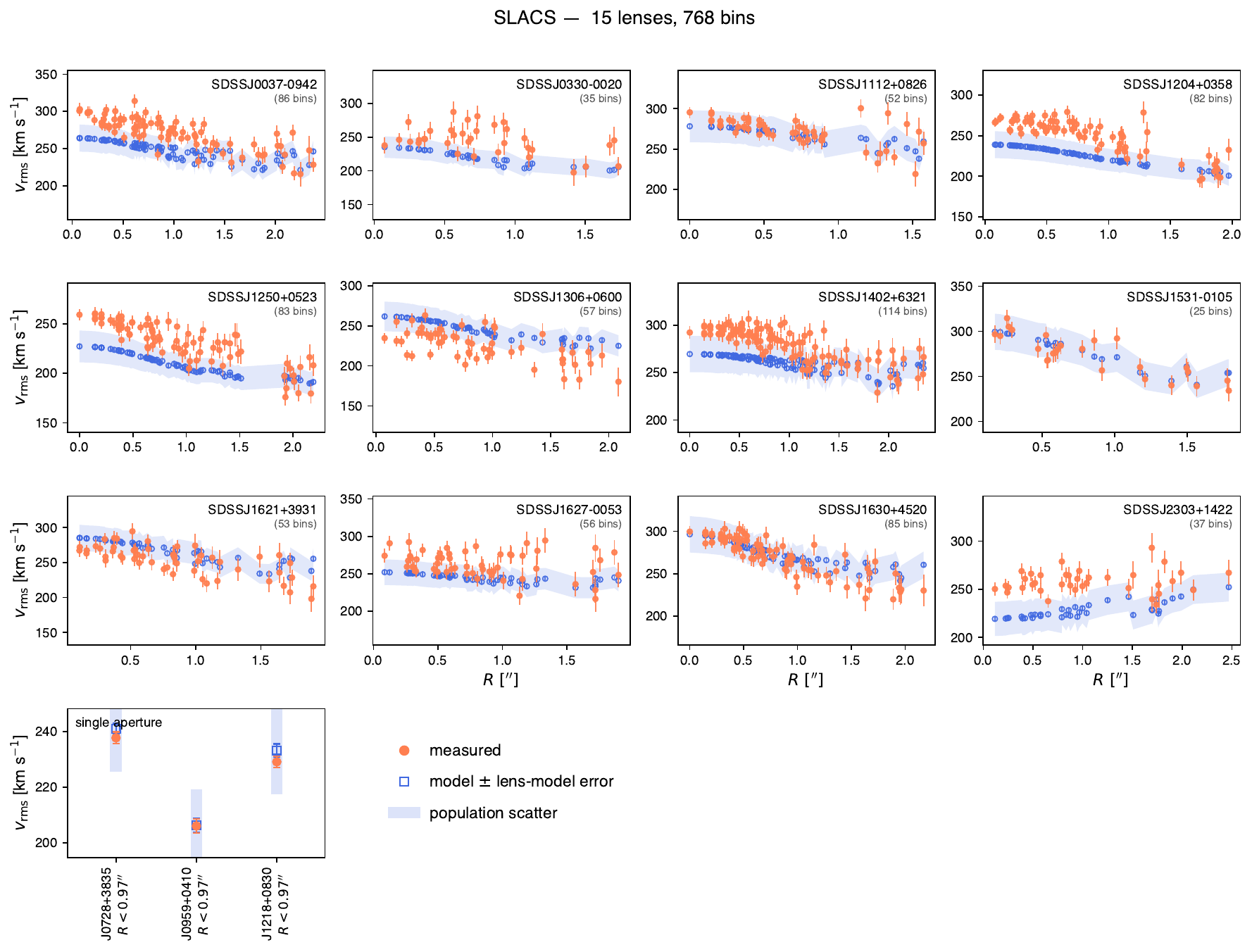}
    \caption{As \fref{fig:TDCOSMO_kin_gof}, for the \nslacstot ~SLACS lenses with Keck/KCWI kinematics entering the joint inference. The last panel groups the three lenses whose KCWI data yield a single bin. Their label gives the outer extent of the spaxel footprint the dispersion is integrated over.}
    \label{fig:SLACS_kin_gof}
\end{figure*}

\begin{figure}[htbp]
    \centering
    \includegraphics[width=0.5\textwidth]{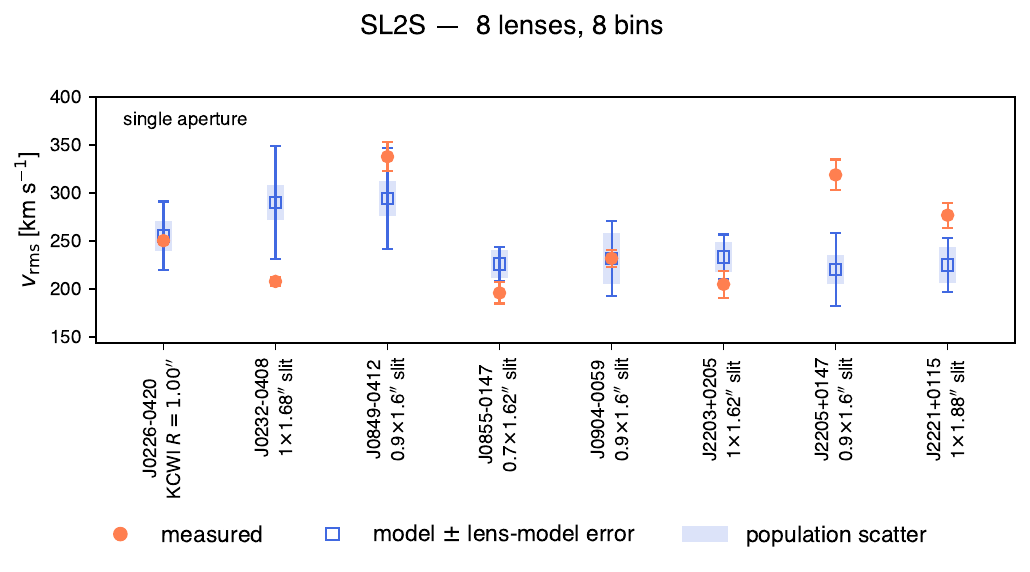}
    \caption{As \fref{fig:SLACS_kin_gof}, for the eight SL2S lenses. All have a single aperture measurement rather than resolved kinematics: a long-slit dispersion for seven of them, whose label gives the slit dimensions, and a Keck/KCWI measurement within a circular aperture of radius $1\arcsec$ for SL2SJ0226$-$0420.}
    \label{fig:SL2S_kin_gof}
\end{figure}

In \fsref{fig:TDCOSMO_kin_gof}, \ref{fig:SLACS_kin_gof} and \ref{fig:SL2S_kin_gof}, we compare the measured $v_{\rm rms}$ with the predictions of the maximum-likelihood of the hierarchical model, for the TDCOSMO lenses ($\Lambda$CDM1a) and for the SLACS and SL2S lenses in the joint inference ($\Lambda$CDM4a). The model errors are split into two parts. The first is the lens-model uncertainty at fixed population parameters, shown as error bars on the model points (in blue). The second is the scatter induced by the population distributions of $\lint$, of the anisotropy and of the intrinsic axis ratio, shown as a blue band. For most lenses, the population scatter dominates the total model uncertainty. The measurement uncertainties are shown as orange error bars. 

\section{Posterior distribution}
\label{app:all_posteriors}
\begin{figure}[htbp]
    \centering
    \includegraphics[width=\linewidth]{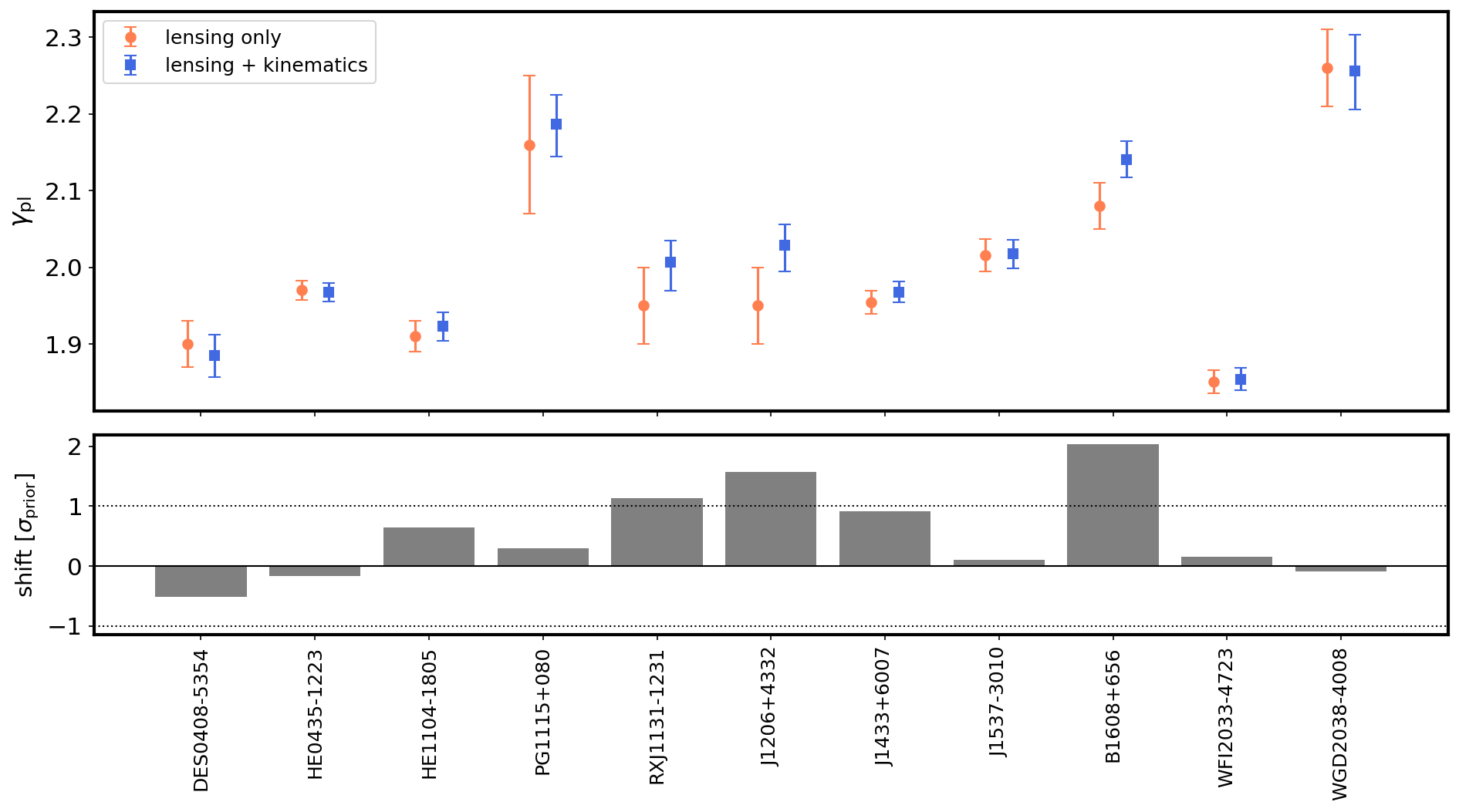}
    \caption{Power-law slope $\gamma_{\rm pl}$ of the 11 \TDCnew lenses. Orange circles show the prior from the lens models (imaging and time delays only); blue squares show the posterior of the baseline hierarchical inference ($\Lambda$CDM1a), in which $\gamma_{\rm pl}$ is sampled for each lens and jointly constrained by the lensing prior and the stellar kinematics. Error bars are 68\% credible intervals. The bottom panel shows the shift of the posterior median with respect to the prior mean, in units of the prior standard deviation.}
    \label{fig:pl_slopes_posterior}
\end{figure}

In \tref{tab:results_LCDM_all_parameters}, we list the constraints on all the cosmological and population-level parameters sampled in the flat \lcdm runs of \tref{tab:results_LCDM}. The individual power-law slopes $\gamma_{{\rm pl},i}$ are also sampled for each lens, but are not listed in this Table. Instead, we compare the $\gamma_{{\rm pl},i}$ posteriors with their lensing-only priors in \fref{fig:pl_slopes_posterior}, which also shows the shift of each posterior induced by the kinematic data, in units of the prior standard deviation. The population parameters are remarkably stable across the different lens samples and the choice of external cosmological dataset. The population means and scatters of $\lambda_{\rm int}$, $\sigma_{\rm t}/\sigma_{\rm r}$ and $q_{\rm int}$, as well as $\alpha_{\lambda}$, are compatible across the different lens samples and choices of \Om prior. Most differences are within $\sim 1\sigma$; the largest, about $1.5\sigma$ for $\alpha_{\lambda}$ and $\sigma(q_{\rm int})$, arise between samples with and without the SLACS lenses.

\renewcommand*\arraystretch{1.5}
\setlength{\tabcolsep}{2pt}
\begin{table*}[htbp]
\caption{\label{tab:results_LCDM_all_parameters}
All cosmological and population parameters in flat \lcdm. The individual power-law slopes $\gamma_{{\rm pl}, i}$ are sampled but not listed.}

\begin{minipage}{\linewidth}
\centering
\footnotesize
\begin{tabular}{lcccccccccc}
\hline \hline
Model ID & $H_{0}$ & $\Om$ & $r_{\rm d}$ [Mpc] & $\lambda_{\rm int,0}$ & $\alpha_{\lambda}$ & $\sigma(\lambda_{\rm int})$ & $\langle \sigma_{\rm t}/\sigma_{\rm r} \rangle$ & $\sigma(\sigma_{\rm t}/\sigma_{\rm r})$ & $\langle q_{\rm int} \rangle$ & $\sigma(q_{\rm int})$ \\
\hline
U$\Lambda$CDM1 & $70.6^{+3.9}_{-3.1}$ & $0.356^{+0.103}_{-0.135}$ & -- & $0.935^{+0.060}_{-0.055}$ & $-0.045^{+0.078}_{-0.081}$ & $0.092^{+0.037}_{-0.025}$ & $0.984^{+0.084}_{-0.072}$ & $0.330^{+0.380}_{-0.186}$ & $0.788^{+0.097}_{-0.116}$ & $0.177^{+0.078}_{-0.069}$ \\
U$\Lambda$CDM2 & $71.9^{+4.3}_{-3.4}$ & $0.306^{+0.133}_{-0.157}$ & -- & $0.976^{+0.058}_{-0.050}$ & $0.047^{+0.054}_{-0.064}$ & $0.127^{+0.027}_{-0.024}$ & $0.934^{+0.101}_{-0.049}$ & $0.438^{+0.384}_{-0.248}$ & $0.861^{+0.049}_{-0.062}$ & $0.082^{+0.038}_{-0.027}$ \\
U$\Lambda$CDM3 & $71.0^{+3.7}_{-3.0}$ & $0.374^{+0.090}_{-0.135}$ & -- & $0.953^{+0.055}_{-0.048}$ & $-0.090^{+0.072}_{-0.078}$ & $0.100^{+0.038}_{-0.028}$ & $0.986^{+0.084}_{-0.072}$ & $0.285^{+0.348}_{-0.160}$ & $0.820^{+0.098}_{-0.131}$ & $0.192^{+0.071}_{-0.067}$ \\
U$\Lambda$CDM4 & $72.1^{+4.1}_{-3.4}$ & $0.323^{+0.118}_{-0.144}$ & -- & $1.002^{+0.050}_{-0.046}$ & $0.005^{+0.053}_{-0.053}$ & $0.132^{+0.028}_{-0.024}$ & $0.937^{+0.101}_{-0.050}$ & $0.494^{+0.339}_{-0.287}$ & $0.901^{+0.044}_{-0.066}$ & $0.093^{+0.040}_{-0.030}$ \\ \hline
$\Lambda$CDM1a & $71.1^{+2.4}_{-2.3}$ & $0.335^{+0.016}_{-0.016}$ & -- & $0.936^{+0.050}_{-0.049}$ & $-0.038^{+0.082}_{-0.088}$ & $0.093^{+0.038}_{-0.026}$ & $0.987^{+0.082}_{-0.073}$ & $0.311^{+0.366}_{-0.175}$ & $0.782^{+0.100}_{-0.119}$ & $0.182^{+0.075}_{-0.071}$ \\
$\Lambda$CDM2 & $71.2^{+2.2}_{-2.2}$ & $0.332^{+0.016}_{-0.015}$ & -- & $0.969^{+0.042}_{-0.043}$ & $0.047^{+0.056}_{-0.057}$ & $0.127^{+0.029}_{-0.024}$ & $0.935^{+0.100}_{-0.049}$ & $0.461^{+0.355}_{-0.253}$ & $0.864^{+0.046}_{-0.059}$ & $0.084^{+0.038}_{-0.028}$ \\
$\Lambda$CDM3 & $71.5^{+2.4}_{-2.3}$ & $0.334^{+0.015}_{-0.016}$ & -- & $0.956^{+0.046}_{-0.042}$ & $-0.090^{+0.072}_{-0.082}$ & $0.099^{+0.040}_{-0.028}$ & $0.980^{+0.086}_{-0.068}$ & $0.323^{+0.373}_{-0.189}$ & $0.814^{+0.102}_{-0.126}$ & $0.192^{+0.070}_{-0.065}$ \\
$\Lambda$CDM4a & $71.3^{+2.2}_{-2.1}$ & $0.333^{+0.016}_{-0.015}$ & -- & $0.993^{+0.042}_{-0.041}$ & $0.016^{+0.051}_{-0.056}$ & $0.131^{+0.029}_{-0.022}$ & $0.928^{+0.100}_{-0.044}$ & $0.419^{+0.387}_{-0.246}$ & $0.901^{+0.044}_{-0.065}$ & $0.093^{+0.043}_{-0.031}$ \\ \hline
$\Lambda$CDM1b & $71.7^{+2.5}_{-2.3}$ & $0.298^{+0.009}_{-0.009}$ & $141.6^{+4.7}_{-4.8}$ & $0.942^{+0.051}_{-0.047}$ & $-0.043^{+0.085}_{-0.086}$ & $0.095^{+0.039}_{-0.028}$ & $0.991^{+0.080}_{-0.074}$ & $0.310^{+0.386}_{-0.170}$ & $0.783^{+0.099}_{-0.119}$ & $0.177^{+0.077}_{-0.070}$ \\
$\Lambda$CDM4b & $72.2^{+2.2}_{-2.2}$ & $0.298^{+0.009}_{-0.008}$ & $140.7^{+4.3}_{-4.2}$ & $0.999^{+0.038}_{-0.037}$ & $0.011^{+0.054}_{-0.052}$ & $0.134^{+0.031}_{-0.024}$ & $0.934^{+0.108}_{-0.050}$ & $0.461^{+0.378}_{-0.273}$ & $0.897^{+0.045}_{-0.063}$ & $0.085^{+0.038}_{-0.025}$ \\
\hline
\end{tabular}
\end{minipage}
\tablefoot{Reported values are medians, with errors corresponding to the 16th and 84th percentiles. Model IDs as in \tref{tab:results_LCDM}; $r_{\rm d}$ is only sampled in combination with DESI BAO.}
\end{table*}
\setlength{\tabcolsep}{6pt}
\renewcommand*\arraystretch{1.0}

\section{New unpublished spectra}
\label{sec:unpub_spec}

\begin{figure}[htbp]
    \centering
    \includegraphics[width=\linewidth]{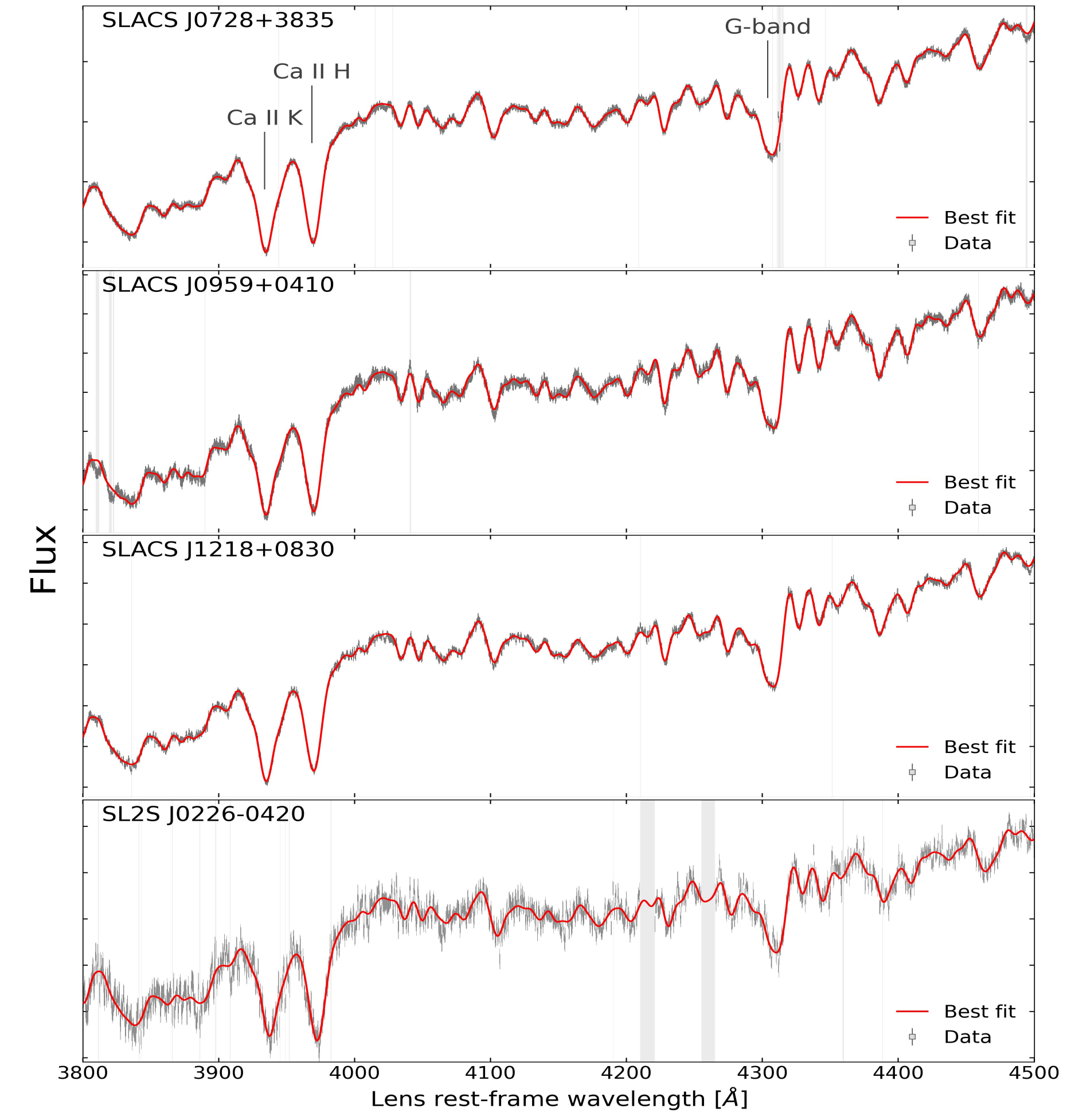}
\caption{Integrated Keck KCWI/KCRM spectra and kinematic fits for the new SLACS and SL2S lenses entering our sample, as described in Appendix~\ref{sec:unpub_spec}.}
    \label{fig:slacs_sl2s_spectra}
\end{figure}

Keck KCWI/KCRM IFU data for three SLACS systems (SLACJ0959+0410, SLACJ1218+0830, and SLACJ0728+3835) and one SL2S system (SL2SJ0226$-$0420) were integrated within circular apertures of 1\arcsec radius from the center of the deflector galaxy. These spectra and their best kinematic model fits with the {\sc squirrel} pipeline are presented in \fref{fig:slacs_sl2s_spectra}. Measured kinematics and their models from the joint inference are presented in \fsref{fig:SLACS_kin_gof} and \ref{fig:SL2S_kin_gof}. Uncertainties are calculated following the methodology of \citet{Knabel25}. spatially resolved kinematics will be presented in a forthcoming paper.

\bibliographystyle{unsrtnat}
\bibliography{references}

\end{document}